\documentclass[11pt]{article}
\usepackage{hyperref}
\usepackage{graphicx}
\usepackage{graphics}
\usepackage{dsfont}
\usepackage{epsfig}
\usepackage{bbm}
\usepackage{bm}
\usepackage{amsmath,amssymb,amsthm,amscd,tabu}
\usepackage{bbold}
\usepackage[sort&compress,square,numbers]{natbib}
\usepackage{float}
\usepackage{braket}
\usepackage{breqn}
\restylefloat{table}
\usepackage{mathdots}
\usepackage{blkarray}
\usepackage{arydshln}

\usepackage{tikz,textcomp}
\usetikzlibrary{decorations.pathreplacing,calc,fadings,fit}
\usetikzlibrary{shapes,arrows,positioning,chains,matrix}
\usepackage{tikz}
\usetikzlibrary{calc,fit,matrix}

\usepackage{listings}

\newcommand*{\ditto}{---\texttt{"}---}

\def\Mgn[#1]#2{{\overline{\cal M}_{#1,#2}}}
\def\pqs[#1,#2]{{\footnotesize{$\left[\begin{array}{c} #1\\#2  \end{array}\right]$}}} 
\def\pqsu[#1,#2]{\left[\begin{array}{c} #1\\#2  \end{array}\right]} 
\def\pqssu[#1,#2]{{\footnotesize{\left[\begin{array}{c} #1\\#2  \end{array}\right]}}} 
\def\pqh[#1,#2]{{\footnotesize{$\left[\begin{array}{c} #1\\#2  \end{array}\right]$}}} 
\def\pqhu[#1,#2]{\left[\begin{array}{c} #1\\#2  \end{array}\right]}

\newcommand{\ba}{\begin{eqnarray*}}
\newcommand{\ea}{\end{eqnarray*}}
\newcommand{\ban}{\begin{eqnarray}}
\newcommand{\ean}{\end{eqnarray}}
\newcommand{\be}{\begin{equation}}
\newcommand{\ee}{\end{equation}}
\newcommand{\ben}{\begin{equation}}
\newcommand{\een}{\end{equation}}
\numberwithin{equation}{section}

\numberwithin{equation}{section}

\begin{document}

\begin{titlepage}
{}~ \hfill\vbox{ \hbox{} }\break

\rightline{
Bonn-TH-2019-05,$\quad$UWThPh-2019-29
}

\vskip 1.5 cm

\centerline{\Large \bf Topological strings on genus one fibered Calabi-Yau 3-folds }  \vskip 0.5 cm
\centerline{\Large \bf  and string dualities }   \vskip 0.1 cm

\renewcommand{\thefootnote}{\fnsymbol{footnote}}
\vskip 10pt \centerline{ 
Cesar Fierro Cota\footnote{fierro@th.physik.uni-bonn.de}, Albrecht Klemm\footnote{aklemm@th.physik.uni-bonn.de}, Thorsten Schimannek\footnote{thorsten.schimannek@univie.ac.at} } \vskip .5cm \vskip 20pt

\begin{center}

{$^{* \dagger}$Bethe Center for Theoretical Physics and $\dagger$Hausdorff Center for Mathematics,}\\ 
	{Universit\"at Bonn, \  D-53115 Bonn, Germany}\vspace{.3cm}\\
{$^\ddagger$Faculty of Physics, University of Vienna}\\{Boltzmanngasse 5, A-1090 Vienna, Austria}\\ [3 mm]
\end{center}

\setcounter{footnote}{0}
\renewcommand{\thefootnote}{\arabic{footnote}}
\vskip 40pt
\begin{abstract}

We calculate the generating functions of BPS indices using their modular properties in Type II and M-theory compactifications 
on compact genus one fibered CY 3-folds with singular fibers and additional rational sections or just $N$-sections,
in order to study string dualities in four and five  dimensions as well as rigid limits in which gravity decouples.
The generating functions are Jacobi-forms of $\Gamma_1(N)$ with the complexified fiber volume as modular parameter.
The string coupling $\lambda$, or the $\epsilon_\pm$ parameters in the rigid limit, as well as the masses of charged 
hypermultiplets and non-Abelian gauge bosons are elliptic parameters. To understand this structure, we show that specific 
auto-equivalences act on the category of topological B-branes on these geometries and generate an action 
of $\Gamma_1(N)$ on the stringy K\"ahler moduli space. We argue that these actions can always be expressed in terms of 
the generic Seidel-Thomas twist with respect to the 6-brane together with shifts of the B-field and are thus monodromies.
This implies the elliptic transformation law that is satisfied by the generating functions. We use Higgs transitions in F-theory 
to extend the ansatz for the modular bootstrap to genus one fibrations with $N$-sections and boundary conditions fix the all 
genus generating functions for small base degrees completely. This allows us to study in depth a wide range of new, 
non-perturbative theories, which are Type II theory duals to the CHL $\mathbb{Z}_N$ orbifolds of the heterotic string 
on $K3\times T_2$. In particular, we compare the BPS degeneracies in the large base limit to the perturbative heterotic 
one-loop amplitude with  $R_+^2 F_+^{2g-2}$ insertions for many new Type II geometries.
In the rigid limit we can refine the ansatz and obtain the elliptic genus of superconformal theories in 5d.
\end{abstract}

\end{titlepage}
\vfill \eject

\newpage

\baselineskip=16pt    

\tableofcontents

\section{Introduction}

In this paper we solve the all-genus topological string partition function $Z_{\text{top.}}$  on compact genus one fibered Calabi-Yau 3 folds $M$ in a large base expansion,
extending the approach of~\cite{Huang:2015sta,HKKprep} to elliptic fibrations with reducible fibers and in particular to geometries that do not exhibit a section but only $N$-sections. 

An elliptic curve is a genus one curve with a marked point which is the zero ${\cal O}$ in the additive group law on the elliptic curve~\cite{MR3363545}. 
An elliptic fibration  over a base $B$ is accordingly defined to be a genus one fibration that has a global section which can be taken to define the zero ${\cal O}$ 
on each fiber and is called the zero section.
The more general term genus one fibration refers to a geometry in which each fiber is a genus one curve, but no assumptions about the presence of a section is made.  

Elliptic fibrations can have additional rational sections and they generate the Modell-Weil group $\text{MW}_M$ which is an Abelian group of finite rank that can have torsion parts.
If a rational section corresponds to a torsion point ${\cal P}$ with $N {\cal P}={\cal O}$ on each fiber one refers to it as an $N$-torsional section and the corresponding geometric group law on the fiber requires
a specialized Weierstrass form of the Jacobian fibration~\cite{Aspinwall:1998xj}. 
On the other hand, independent non-torsional sections correspond to independent elements in the cohomology $H^2(M,\mathbb{Z})$. 
In the F-theory context it is well known that independent rational sections lead to an Abelian gauge group $U(1)^{r-1}$ with $r=\text{rk}\left(\text{MW}_M\right)$~\cite{Morrison:1996pp}.
Explicit geometries were first discussed in~\cite{Klemm:1996hh} to resolve, in compact Calabi-Yau 3-folds, the BPS degeneracies of the $E$-string with respect to the flavour fugacities by breaking the flavour group $E_8$ on the $\frac{1}{2} K3$ with $N-1$ global rational sections into $G$, where $U(1)^{N-1}\times G\subset E_8$ and $G=\{E_7,E_6,D_5,\ldots\}$ for $N=2,3,4,\ldots $.

An important class of fibrations that is discussed in this paper has no section but just $N$-sections.
An $N$-section is defined by $N$ points that can be  identified on each fiber and which are transformed into each other by monodromies in the base.
Accordingly, an $N$-section corresponds to one element in the cohomology $H^2(M,\mathbb{Z})$.
Genus one-fibrations that have no section but only $N$-sections lead to discrete $\mathbb{Z}_N$ gauge symmetries in F-theory.
The corresponding effective theories can be obtained via Higgsing from vacua with $U(1)$ factors and matter multiplets of charge $N$, a 
mechanism first described geometrically in~\cite{deBoer:2001wca,Morrison:2014era}.
Again, $N$-sections that are in an appropriate sense independent lead to Abelian gauge bosons.
When we say that $M$ is a genus one fibration with $N$-sections we will imply that it does not have a section and also no $N'$-sections with $N'<N$.

In addition to rational sections or $N$-sections, a genus one fibration can exhibit fibral divisors that resolve $ADE$-singularities that are fibered over curves in the base. 
There can be also global monodromies that identify nodes of the $ADE$-singularities and lead to the reduced number of fibral divisors of non-simply 
laced gauge groups~\cite{Bershadsky:1996nh}. The corresponding harmonic forms in $H^2(M,\mathbb{Z})$ lead to the Abelian gauge bosons in the Cartan of the corresponding gauge group while, 
from the M-theory perspective, wrapped 2-branes correspond to the $W$-bosons. After compactifying the F-theory vacuum on a circle, the volumes of components of reducible fibers can be 
identified with scalar fields in vector multiplets of the corresponding ${\cal N}=2$ theories. In the effective theory they parametrize the masses of charged hypermultiplets and of the non-Abelian $W$-bosons.
We therefore denote these parameters generically by $\underline{m}$ and, for reasons that we outline below, we call them \textit{geometric elliptic parameters}.

In~\cite{Huang:2015sta,HKKprep} it was assumed that $M$ is an elliptic fibration which has just the zero section and exhibits no fiber singularities more severe than $I_1$ in Kodaira's classifications. 
In particular, all fibers were irreducible and no fibral divisors have been present.
It was then shown that the expansion coefficients\footnote{See~\eqref{eqn:topexp} (without the mass parameters $\underline m$) for the 
precise definition of the expansion.} $Z_\beta(\tau, \lambda)$ of the topological string  partition 
function $Z_{\text{top.}}$  on elliptic Calabi-Yau 3-folds for fixed classes $\beta\in H_2(B,\mathbb{Z})$ in the base are meromorphic Jacobi forms of weight zero, where the elliptic argument $\tau$ is related to the complexified  
volume of the elliptic fiber, while the string coupling constant $\lambda$  appears as elliptic argument.
Moreover, $Z_\beta(\tau, \lambda)$  has  an index  that is given by intersections on the base as  $r^\beta_\lambda = \frac{1}{2} \beta \cdot  (\beta - c_1(B))$. 
The former fact can be argued using invariance of $Z_{\text{top.}}$  under an ${\rm SL}(2,\mathbb{Z})$ action that is 
embedded into  the symplectic monodromy group $\Gamma_W\subset {\sl Sp}(b_3(W),\mathbb{Z})$ acting on 
the integral symplectic basis of periods of the mirror $W$ of the elliptic fibration $M$.
The index follows by combining this action with the background independence equation of $Z_{\text{top.}}$ as a wave 
function~\cite{Witten:1993ed,Klemm:2012sx,Huang:2015sta}, which is equivalent to the holomorphic anomaly equation. Using 
the pole structure of  $Z_\beta$ imposed by the integral BPS expansion of $Z_{\text{top.}}$ one can argue that $Z_\beta$ has a unique denominator and the numerator is a weak Jacobi form. 
The ring of weak Jacobi forms is finitely generated, see \cite{Huang:2015sta} and in particular reference~\cite{Dabholkar:2012nd}.
Therefore one can fix  $Z_\beta$ for low base degree from the vanishing of BPS invariants for high genus that follow from Castelnuovo bounds.\footnote{If $\beta^2<0$ 
one can argue that one can always fix $Z_\beta$~\cite{Huang:2015sta} and the method is quite effective.  
For example the E-string could be solved to seven base windings in~\cite{HKKprep}.} 

A crucial step in generalizing the analysis of \cite{Huang:2015sta,HKKprep} is  to modify and extend these monodromy arguments to genus one fibrations that do not have a section
and also to include those parameters that correspond to the volumes of fibral curves and, after circle compactification, to the vaccum expectation values of scalars in vector multiplets.
The action of the corresponding monodromies for elliptic fibrations with reducible fibers has been calculated in~\cite{Schimannek:2019ijf} and we extend the argument to fibrations without sections.
Using this generalization, we establish that for genus one fibrations with $N$-sections and $N\le 4$, the ${\rm SL}(2,\mathbb{Z})$ part is broken to the finite index subgroup $\Gamma_1(N)$ as well as the fact 
that the complexified  K\"ahler  parameters that correspond to rational fibral curves become elliptic parameters of the coefficients $Z_\beta(\tau, \lambda, {\underline m})$~\footnote{In the rigid limits 
the masses appear as fugacities in the elliptic genus of the 2d gauged linear quiver $\sigma$ model and are naturally identified as elliptic 
parameters~\cite{Haghighat:2013gba,Haghighat:2014vxa,Haghighat:2015ega,DelZotto:2016pvm,Gu:2017ccq,DelZotto:2017mee,DelZotto:2018tcj}. In global threefold cases with multiple sections, a similar behaviour has been observed in~\cite{Lee:2018urn,Lee:2018spm}.}.  This analysis enables us to construct the $Z_\beta(\tau, \lambda, {\underline m})$ in a far more general setting as meromorphic higher degree Jacobi form of $\Gamma_1(N)$ with further 
elliptic parameters, where the numerators are now generated by the larger rings of weak Jacobi-forms under $\Gamma_1(N)$, which exhibit Weyl-invariance in the ${\underline m}$ parameters similar to the cases discussed~\cite{DelZotto:2017mee}. 
The monodromies allow us to determine the precise index matrix with respect to the geometric elliptic parameters $\underline{m}$ and the elliptic transformation law then follows for these parameters already from invariance of the topological string partition function under the corresponding monodromies.
Although the rings of weak higher degree Jacobi forms for $\Gamma_1(N)$ are larger, we show in examples that one can fix~$Z_\beta(\tau, \lambda, {\underline m})$ for small $\beta$ and expect that the Castelnuovo bounds are again sufficient to solve completely for classes with  $\beta^2<0$.

Calculating the monodromy group  $\Gamma_W \in {\rm Sp}(b_3(W),\mathbb{Z})$ in the B-model on $W$ using the complete Picard Fuchs differential ideal and a global integral symplectic basis 
on the resolved complex structure moduli space ${\cal M}_{cs}(W)$ is technically feasible only for models with few complex moduli $h_{2,1}(W)$.
Instead we work on the stringy K\"ahler moduli space ${\cal M}_{ks}(M)$ of $M$ directly.
Extending the method of~\cite{Cota:2017aal,Schimannek:2019ijf,HKKprep}, our strategy is to identify the K\"ahler moduli as central charges of 2-branes and to study auto-equivalences of the category of branes that generate an action of $\Gamma_1(N)$. We will then relate those auto-equivalences to the generic monodromies that correspond to the 
boundary of the geometric cone and to the large volume limiting points and thus show that the $\Gamma_1(N)$ action corresponds indeed to monodromies 
in the stringy K\"ahler moduli space.This allows us to use the known automorphic properties of $Z_{\text{top.}}$ under 
monodromies~\cite{Aganagic:2006wq,Gunaydin:2006bz} and to derive the elliptic transformation law with respect to the geometric elliptic parameters and thus also to
identify their index matrix.

For elliptic fibrations, the most characteristic auto-equivalence is induced by the Fourier-Mukai kernel which is the ideal sheaf ${\cal I}_{\Delta_B}$ of the relative diagonal 
$\Delta_B$ in $M\times_B M$~\cite{bridgeland1997fourier,Andreas:2000sj,Andreas:2001ve,Andreas:2004uf,ruiperez2006relative} and it acts as $\tau \rightarrow \frac{\tau}{N \tau +1}$  on the  $\tau$ parameter~\cite{Schimannek:2019ijf}.
We will call this the $U$-transformation or the relative Conifold transformation.
Furthermore, tensoring objects in the derived category with the line bundle ${\cal O}(D)$, where $D$ is an effective Cartier divisor, also induces an auto-equivalence and corresponds to a shift of the $B$ field by $D$.
In partiular, when $D$ is choosen to be the ``zero'' $N$-section, it leads to the $T$ transformation $\tau\rightarrow\tau+1$.
The Weyl symmetries in the geometric elliptic parameters ${\underline  m}$ arise from monodromies around points where the Calabi-Yau develops a singularity from divisors collapsing to curves~\cite{Klemm:1996kv,Katz:1996ht}.
The corresponding Fourier-Mukai kernels have been indentified in~\cite{Horja:2001cp,Szendroi:2002rf,2002math.....10121S}, see also~\cite{Aspinwall:2001zq}.

After general explanations of the geometric structure  of the fibration at the beginning of 
Section~\ref{sec:branes},  the central charges and the symplectic pairing of the topological $B$-branes are explained in Subsection~\ref{ssec:topologicalbranes}.
The Fourier-Mukai kernels and the corresponding actions on the brane charges are described and their action on the brane charges is calculated in the Subsections~\ref{ssec:fourier-mukai},~\ref{ssec:relativeconi}.
The calculation of the $U$-transformation generalizes the results from~\cite{Schimannek:2019ijf} to genus one fibrations with $N$-sections and relies on a realization of the
Calabi-Yau as a complete intersection inside a compatibly fibered toric ambient space.
In Subsection~\ref{ssec:ez-trans} we argue that the $U$-transformation can always be expressed in terms of the generic Seidel-Thomas twist with respect to the 6-brane together with shifts of the B-field and is thus a monodromy.
For fibrations with bases $B=\mathbb{P}^2$ and $\mathbb{F}_n,\,n\in\mathbb{N}$ we provide the relations and check them via an explicit calculation of the corresponding products of monodromies.
This calculation does not rely on toric geometry and applies to all genus one fibered Calabi-Yau threefolds over the corresponding bases.
At the end of Subsection~\ref{ssec:ez-trans} we also discuss a beautiful connection of this structure to the swampland distance and the emergence conjectures.
The Weyl monodromies that emerge when divisors collapse to curves are discussed in Subsection~\ref{ssec:weyl}.
In Subsection~\ref{ssec:monaut} we then combine the monodromies with the automorphic properties of the topological string partition function $Z_{\text{top.}}$ and derive the elliptic transformation law.
A derivation of the full modular transformation law is given for genus one fibrations without additional $N$-sections or fibral divisors, and where geometric elliptic parameters are thus absent.

We explain in Section~\ref{sec:modularbootstrap} how to use the modular and elliptic transformations of weight $k=0$ and index $r^\beta_\lambda$  under $\Gamma_1(N)$ as well as the pole behaviour to reconstruct $Z_\beta$.
For convenience of  the reader we include in Subsection~\ref{ssec:ringsofforms} the definitions of the rings of holomorphic modular forms of $\Gamma_1(N)$ and discuss the special class of higher degree Jacobi forms that will appear in the topological
string partition function.
Subsection~\ref{ssec:modularbootstrap} contains a review of the main results about the modular bootstrap on elliptic fibrations.
In Subsection~\ref{ssec:closedexpressions} we then discuss the base degree zero part $Z_0$ of the partition function, both on elliptic fibrations with reducible fibers and on genus one fibrations with $N$-sections.
We find that the corresponding free energies $F_{g\ge 0,\beta=0}$ are Jacobi forms of weight $2g-2$ and index $0$.
For particular examples we derive the closed expression at all genera and the derivation can easily be adapted to other genus one fibrations.
In particular, the matter content of the corresponding F-theory vaccum including multiplicities is entirely encoded in $Z_0$~\footnote{Such observations have been independently 
made in the elliptic case by Amir Kashani-Poor~\cite{AKP} and they will also  expanded  in~\cite{OS}.}.
Subsection~\ref{ssec:modan} contains the derivation of the index with respect to the topological string coupling constant $\lambda$, using Witten's form of the wave function equation.
We then obtain the correct Ansatz for general $Z_\beta$ on general genus one fibered Calabi-Yau threefolds with $N$-section from the corresponding Ansatz for elliptic fibrations by considering Higgs transitions in F-theory, see Subsection~\ref{ssec:modgenusone}. We summarize the equations relevant for all modular Ans\"atze in Subsection~\ref{ssec:summary}.

Of course, the full topological partition function contains much richer BPS information than just the massless spectrum of the corresponding F-theory vacuum.
A beautiful application to Type II/heterotic duality for Calabi-Yau 3-folds with $N$-section geometries in four dimensional ${\cal N}=2$ theories is described in Section~\ref{sec:dualityCHL}.
On the heterotic side one expects \cite{Kachru:1997bz,Datta:2015hza} the dual to be  a CHL~\cite{Chaudhuri:1995fk,Chaudhuri:1995bf} 
compactifictation on  $(K3\times T_2)/\mathbb{Z}_N$. This follows from the observation that self-dualities on the heterotic side are identified with 
monodromies in the moduli space of the Calabi-Yau target space on the Type II side~\cite{Klemm:1995tj,Kachru:1995fv}, together with our general 
discussion of the monodromies of genus one fibrations. The $\mathbb{Z}_2$ CHL string exists already for $S^1$ compactifications of 
the $E_8\times E_8$ heterotic string and the corresponding  $\mathbb{Z}_2$ exchanges the two $E_8$ factors while acting as 
a half shift  $x_k\rightarrow x_k+ \pi R$ on the circle. Using  a maximal supersymmetric dual pair in 6d between a Type II 
$\mathbb{Z}_2$ orbifold  on $K3$ and a heterotic  CHL $\mathbb{Z}_2$ orbifold on $T^4$\cite{Schwarz:1995bj}, as well as an adiabatic  argument on the Type 
II side, the $\mathbb{Z}_2$ action on a dual elliptically fibered Calabi-Yau 3-fold was identified in~\cite{Kachru:1997bz}. 
Indeed, this leads to a $2$-section geometry with a compatible $K3$ fibration. 

The automorphisms of $K3$ surfaces have been classified~\cite{MukaiK3} and correspond to conjugacy classes of the Mathieu group $M_{23}$.
More generally, one can classify the discrete symmetries of non-linear sigma models on $K3$, a task completed in~\cite{Gaberdiel:2011fg,Cheng:2016org}.
Using this information in further $T^2$ compactifictions one can fully classify orbifolds of Type II compactifications on $K3\times T^2$ which  yield ${\cal N}=4$  
supergravities in four dimensions~\footnote{This setting is very well studied  in order to understand the microscopic entropy of  ${\cal N}=4$ black holes and because the twining elliptic genera are given by 
Jacobi forms, whose coefficients decompose in a  simple way  into the dimensions of the  largest exceptional discrete groups, a phenomenon known as moonshine.}~\cite{Paquette:2017gmb}.
One can also classify CHL orbifolds of heterotic strings on $K3\times T^2$ which lead to ${\cal N}=2$ effective supergravity actions~\cite{Chattopadhyaya:2016xpa,Chattopadhyaya:2017zul}. 
For the corresponding Type II theories on Calabi-Yau 3-folds with $N$-sections and compatible $K3$ fibration one should be able to 
identify the all-genus BPS amplitudes with a  heterotic one loop integral that contains insertions of the self-dual parts of curvature 
and graviphoton field strength of the form $R^2_+ F_+^{2g-2}$~\cite{Antoniadis:1993ze,Antoniadis:1995zn,Marino:1998pg}.
The corresponding perturbative heterotic amplitudes can be calculated using the Borcherds lift as in~\cite{Marino:1998pg,Klemm:2005pd}.
Similar calculations have been performed for some CHL models in~\cite{Chattopadhyaya:2016xpa,Chattopadhyaya:2017zul}.
In Subsection~\ref{ssec:comparison} we identify novel Type II compactifications which are dual to heterotic compactifications on CHL orbifolds that 
correspond to the conjugacy class $2A$ in $M_{23}$ with non-standard embedding of the gauge connection. We show that the corresponding one-loop 
amplitudes that were obtained in~\cite{Chattopadhyaya:2017zul} match our results from the modular bootstrap on genus one fibrations 
in the large base limit of the compatible $K3$ fibration and thereby provide a strong all genus tests of Type II/CHL duality. Since the heterotic 
dilaton $\Phi_{het}$ is identified  with the K\"ahler parameter of the base of the $K3$ fibration, instanton corrections with non-vanishing base 
degree contribute like $q_{b} = e^{-8\pi^2 \Phi_{het}}$ and predict non-perturbative corrections to the CHL string. 
Type II theory on $N$-section Calabi-Yau 3-folds should  give therefore the full  non-perturbative description of the corresponding CHL string.

An analysis of superconcormal representation theory reveals, that six is the maximal dimension in which superconformal field theories can exist~\cite{Nahm:1977tg}. 
F-theory on fibrations with a contractable configuration of curves in a non-compact base have been used to study~\cite{Morrison:1996pp,Klemm:1996hh} and classify $N=(2,0)$~\cite{Heckman:2013pva} and $N=(1,0)$~\cite{Heckman:2015bfa} superconformal  
field theories. Building blocks in this classifications are  minimal SCFTs studied in~\cite{Haghighat:2014vxa} and among the latter is a particularly interesting one, the so called $E$-string.
This is geometrically engineered as the rigid F-theory limit on an elliptic Calabi-Yau 3-fold with a rational curve of self-intersection $-1$ in the base..
The limit decouples gravity and emerges geometrically when the normal direction of the rational curve is decompactified, so that the compact part of the geometry is an  $\frac{1}{2}K3$, i.e. an elliptic surface with 12 
$I_1$ fibers.
The latter can also be obtained as a nine-fold blowup of $\mathbb{P}^2$~\cite{Morrison:1996pp,Klemm:1996hh}.    
The topological string partition function encodes the elliptic genera of tensionless strings that arise in the superconformal limit~\cite{Haghighat:2014vxa}.
 
 In Section~\ref{sec:decoupling} we study corresponding gravity decoupling limits now in the context of genus one fibrations with $N$-sections.
 To this end we use toric geometry to construct genus one fibrations over the Hirzebruch  surface $\mathbb{F}_1$.
 The latter is a  rational fibration  with fiber $F\sim \mathbb{P}^1$ with $F^2=0$ and base $S\sim \mathbb{P}^1$,  a section with $S^2=-1$. 
 The decoupling limit corresponds hence to the limit of large fiber $F$, or equivalently to contracting the base $S$.
Taking this limit on genus one fibrations with $N$-sections~\footnote{We also consider 
 geometries with pseudo $N$-sections that nevertheless exhibit $\Gamma_1(N)$ modularity. Those occur when the K\"ahler 
 moduli of the one or two rational sections considered for the  $E_7\times U(1)$ and $E_6\times U(1)^2$ splitting of the $E$-string~\cite{Klemm:1996hh} are frozen to zero value by the toric embedding.}
 over $\mathbb{F}_1$, we find for $N=2,3,4$ that, due to the additional $U(1)_R$ symmetry in the local limit, the $Z_{k\cdot S}(\tau, \lambda)$ can be refined for $k$-th multi wrapping $\beta =k\cdot S$  
 of the base class $S$ to  $Z_{k\cdot S}(\tau, \epsilon_\pm)$.
 Adapting  the refined modular bootstrap  approach with $SL(2,\mathbb{Z})$ Jacobi forms with two elliptic parameters $\epsilon_\pm$  described in~\cite{Gu:2017ccq}  
 to  similar Jacobi forms  of  $\Gamma_1(N)$  we can extract the refined BPS invariants $N_{j_L,J_R}^\beta$  in the full 5d spin representations of the little group 
 $SU(2)_L\times SU(2)_R$ as decribed in Subsection~\ref{ssec:wilsonline}. In particular, to  determine the elliptic index with respect to 
 the refined parameters $\epsilon_\pm$, we use the relation of the latter to the anomaly polynomials for the chiral $6d$ space time -- or quiver worldsheet 
 theory~\cite{DelZotto:2016pvm, DelZotto:2017mee}.   
 The BPS spectrum can be explained by discrete Wilson lines on the $S^1$ that compactifies from six to five dimension.
 In particular  for $N=2$ we find the refined BPS invariants of the 5d theory  with an $\mathbb{Z}_2$ discrete Wilson line that was used in~\cite{Kim:2014dza}.
 The  latter was used to obtain  the $E$-string  spectrum, up to the effect of the Wilson line, from the  the elliptic genus of a 2d quiver gauge theory  with 
 $SO(16)$ gauge theory, which has a brane description as $k$ $D2$ branes stretched between a stack of one $O8$-- and $8$ $D8$ branes and one NS5-brane.
  
Our toric construction of genus one fibered Calabi-Yau 3-folds over various bases with $N$-sections  as well as with Abelian and non-Abelian gauge symmetry enhancements is illustrated with typical examples in Section~\ref{sec:examples}. There we apply the modular bootstrap in detail and also discuss the manifestation of field theoretic Higgs transitions at the level of the geometry and the partition function.
We provide the toric data related to the ambient space, study the Mori cone in the relevant phases and obtain the Picard-Fuchs differential ideals as well as the discriminants. Longer
expressions are relegated to Appendix \ref{app:localgeometries} and \ref{app:Pfdisc}.  From the toric data mirror symmetry is 
manifest~\cite{batyrev1993dual,Batyrev:1994pg} and the periods at large radius as well as the genus zero BPS invariants can be calculated from the GKZ solutions as in~\cite{Hosono:1993qy,Hosono:1994ax}.
If all discriminant components are known, the genus one BPS invariants follow as in~\cite{Bershadsky:1993ta}.

We provide auxiliary data for our discussion in the Appendices. Moreover, Appendix~\ref{app:ftheory} contains a very brief review of the F-theory dictionary that we frequently
employ to describe geometric structures in terms of the physics in the corresponding F-theory vacua.
In Appendix~\ref{app:characteristic} we prove an identity between charateristic classes that is crucial to obtain the auto-equivalences and monodoromies in Subsection~\ref{ssec:relativeconi} as well
as in the derivation of the modular anomaly equation in Subsection~\ref{ssec:modan}.\\\vspace{.3cm}

\noindent
{\bf Acknowledgement:} We would like to thank Min-xin Huang and Sheldon Katz for discussions and the collaboration on compact 
elliptic fibrations using the modular bootstrap and  Michelle del Zotto,  Jie Gu, Amir Kashani-Poor,  Gugielmo Lockhart  and 
Cumrun Vafa  for sharing insights into the non-compact cases. Special thanks goes to Xin Wang for checking the compatibility of 
the refined local limits of the $N$-section cases with the blow up equations.         
We also thank Paul Oehlmann for insightful discussions regarding the physics of compactifications on genus one fibrations.
We thank Johanna Knapp and David Erkinger for valuable discussions about D-brane monodromies and their realizations in the gauged linear sigma model, as well as the swampland distance program in the context of fibrations.
The work of Thorsten Schimannek is supported by the Austrian Science Fund (FWF):P30904-N27. Cesar Fierro Cota would like to thank the financial support from
the fellowship  ``Regierungsstipendiaten CONACYT-DAAD mit Mexiko'' under the grant number 2014 (50015952) and the Bonn-Cologne Graduate School of Physics and Astronomy for their generous support.

\section{The geometry of elliptic and genus one fibrations}
\label{sec:geometry}
 
Before we begin, let us briefly review the geometry of an elliptic or genus one fibered Calabi-Yau threefold.
Note that we follow the convention from the F-theory literature that elliptic fibrations have at least one section and use the word genus one fibration for torus fibrations that do not necessarily have a section.
We will always assume that the fibration is flat (i.e. the dimension of the fiber does not jump) and that there are no multiple fibers.

The Shioda-Tate-Wazir theorem \cite{TateWazir} states that the homology group $H_4(M)$ of an elliptically fibered threefold  is generated by three types of divisors.
These are \textit{vertical divisors} $D_i=\pi^{-1}\tilde{D}_i$ where $\tilde{D}_i\in H_{2}(B)$ is a divisor in the base $B$ of the fibration, \textit{fibral divisors} that consist of rational curves fibered over a divisor in $B$ and
\textit{sections}.
It is useful to distinguish between holomorphic sections, that intersect every fiber in a point, and rational sections, that intersect every smooth irreducible fiber in a point.
By convention an elliptic fibration has at least one section, holomorphic or rational, and we can choose any section to be the \textit{zero section}.
Irreducible fibers can then be canonically identified with $\mathbb{C}/(\mathbb{Z}+\tau\mathbb{Z})$ where the origin is the point that corresponds to the intersection with the zero section.
Addition of points defines a group law on the fiber that can be extended to the set of rational sections.
This leads to the \textit{Mordell-Weil group} $MW(M)$.
If we only consider the zero section and sections that are linearly independent in the Mordell-Weil group $MW(M)$ together with a basis of vertical divisors and fibral divisors we obtain a basis of $H_4(M)$.

In our convention a genus one fibration might not have a section but only $k$-sections that intersect the generic fiber $k$ times.
The points experience monodromy along loops in the base which distinguishes a $k$-section from a union of $k$ sections.
However, a genus one fibration has an associated Jacobian fibration where every fiber is replaced by its moduli space of degree $0$ line bundles.
Note that an elliptic fibration is birationally equivalent to it's associated Jacobian fibration and on smooth fibers the zero section can be identified with the trivial line bundle.
The Mordell-Weil group law then locally amounts to taking tensor products.
In the case of genus one fibrations one can identify the $k$-sections with degree $k$ line bundles on the (smooth) fibers.
It is then clear that the Mordell-Weil group of the Jacobian fibration acts on the set of $k$-sections.

The Shioda-Tate-Wazir theorem has subsequently been generalized to genus one fibered threefolds, i.e. fibrations that do not admit a section but only multi-sections \cite{Braun:2014oya}.
The only difference is that one considers $k$-sections (for minimal $k$) instead of sections. To obtain a basis one picks a ``zero $k$-section'' and acts on it with the free generators of the Mordell-Weil group of the Jacobian fibration.

The homology of an elliptic or genus-one fibered Calabi-Yau manifold, including the intersection structure, is encoded in the effective theory that is associated to the Calabi-Yau via F-theory.
This provides a concise way to describe e.g. the group of divisors in terms of the corresponding physical gauge group.
We will frequently make use of this dictionary and a brief review of the relevant entries can be found in Appendix \ref{app:ftheory}.

For elliptically fibered Calabi-Yau manifolds there is a homomorphism from the Mordell-Weil group to $H_4(M,\mathbb{Z})$ called the \textit{Shioda map}
\begin{align}
	\sigma:\,MW(M)\rightarrow H_4(M,\mathbb{Z})\,.
\end{align}
The explicit form of this map is reviewed in Appendix~\eqref{app:ftheory} but one can uniquely define it in terms of it's intersection properties~\cite{Park:2011ji}.
This definition generalizes to genus one fibrations with $N$-sections where the domain is of course just a set.
Let us define the inner product
\begin{align}
	\langle\,,\,\rangle:\,H_4(M)\times H_4(M)\rightarrow H_2(B)\,,\quad (S,S')\mapsto-\pi(S\cdot S')\,.
	\label{eqn:inner}
\end{align}
For an $N$-section $E$ of a Calabi-Yau threefold we then define $\sigma(E)=E+D$ where $D$ is the unique linear combination of the zero-$N$-section, vertical divisors and fibral divisors
such that $\sigma(E)$ is orthogonal with respect to $\langle\cdot,\cdot\rangle$ to the subspace spanned by those divisors in $H_4(M)$.

Irreducible curves in $M$ can either arise from curves in the base, from rational curves that are fibers of fibral divisors or from isolated rational curves over points of the base.
We will denote the latter two collectively as \textit{fibral curves}.
It is easy to see that the image of an $N$-section under the Shioda map $\sigma$ will have non-zero intersection only with isolated fibral curves.
The class of the generic fiber is irreducible if there are no reducible fibers.
We can therefore expand the topological string partition function $Z=\exp(\sum_{g=0}^\infty \lambda^{2g-2}F_{g})$ as
\begin{align}
Z(\tau,\underline{m},\underline{t},\lambda)=Z_0(\tau,\lambda)\left(1+\sum\limits_{\beta\in H^{1,1}(B,\mathbb{Z})}Z_\beta(\tau,\underline{m},\lambda)Q^\beta\right)\,,
	\label{eqn:topexp}
\end{align}
where we find that for an elliptic or genus one fibration with $N$-sections for $N\in\{1,2,3,4\}$, the K\"ahler modulus $\tau$ should be choosen such that $N\tau$ is the complexified volume of the generic fiber (this is discussed in Section~\ref{ssec:modgenusone}).
Moreover, $\underline{m}$ are complexified volumes of fibral curves and $Q^\beta=\exp(2\pi i\sum_i \beta^i t_i)$ where $t_i,\,i=1,...,h^{1,1}(B)$ are shifted volumes of curves in the base.
The shift is linear in $\tau$ and its necessity for elliptic fibrations has first been observed by~\cite{Klemm:2012sx,Alim:2012ss}.
It can also be derived from the action of the relative conifold transformation~\cite{Schimannek:2019ijf}.
We will extend this derivation to genus one fibrations in Section~\ref{ssec:relativeconi}.
The topological string coupling constant is denoted by $\lambda$.

\section{Branes, derived equivalences and monodromies}
\label{sec:branes}
In this paper we adopt the philosophy that modular properties of the topological string partition function are consequences of the general transformation behaviour under monodromies in the \textit{stringy} K\"ahler moduli space.
Here the attribute ``stringy'' indicates two important differences to the classical K\"ahler moduli space of a Calabi-Yau.
First, the K\"ahler form is combined with the B-fields into the \textit{complexified K\"ahler form}.
Second, the complexified K\"ahler cone is extended with the K\"ahler cones of other geometries and cones that do not admit a geometric interpretation~\cite{Witten:1993yc,Aspinwall:1993nu}.
A canonical example for such a non-geometric ``phase'' is the moduli cone of a Landau-Ginzburg model.
Together these cones form the so-called \textit{fully enlarged K\"ahler moduli space}.
The third difference is, that the duality group of the string compactification is quotiented out.

At least for hypersurfaces in toric ambient spaces it is relatively easy to construct the fully enlarged K\"ahler moduli space.
It is much harder to decide whether two points correspond to dual compactifications.
However, according to the homological mirror symmetry conjecture the action of monodromies in the stringy K\"ahler moduli space, where we choose the large volume limit of a Calabi-Yau $M$ as the base point, can be lifted to autoequivalences of the category of topological B-branes on $M$~\cite{Kontsevich:1994dn}.
This allows us to use the machinery of Fourier-Mukai transformations and thus to calculate the action of certain generic autoequivalences on the complexified K\"ahler moduli.

Our main interest will be in what we call the \textit{relative conifold transformation} on elliptic and genus-one fibered Calabi-Yau threefolds.
For fibrations over $\mathbb{P}^2$ we show that the action on the brane charges can be expressed in terms of transformations that are related to monodromies in the complex structure moduli space of the mirror.
This implies that the corresponding points in the K\"ahler moduli space are identified by a duality.
In the case of elliptic fibration this duality is closely related to T-dualizing along both cycles of the elliptic fiber.
When the base is a Hirzebruch surface $\mathbb{F}_i,\,i=0,1$ we observe analogous relations for all geometries that we study in this paper.
Having thus established that relative Conifold transformations correspond to monodromies in the stringy K\"ahler moduli space, we relate it to the modular properties of the topological string partition function.

We start this section with a cursory review of topological B-branes and Fourier-Mukai transformations where we will also introduce our conventions.
A pedagogical review can be found in~\cite{Aspinwall:2004jr}.

\subsection{A brief introduction to topological B-branes}
\label{ssec:topologicalbranes}
Topological B-branes on a Calabi-Yau manifold $n$-fold $M$ are objects in the bounded derived category of quasicoherent sheaves $D^b(M)$ on $M$~\cite{Douglas:2000gi}.
An object $\mathcal{F}^\bullet$ in this category is a bounded complex of quasicoherent sheaves and quasi-isomorphic complexes are identified.
This implies (non-trivially) that every object can be represented by a bounded complex of locally free sheaves
\begin{align}
	\mathcal{F}^\bullet=0\rightarrow\mathcal{E}^{m}\rightarrow\mathcal{E}^{m+1}\rightarrow\dots\rightarrow 0\,,
	\label{eqn:fbullet}
\end{align}
where the superscript indicates the position in the complex.

Locally free sheaves are equivalent to vector bundles and the complex~\ref{eqn:fbullet} can be thought of as a stack of $n$-branes (at even positions) and anti-$n$-branes (at odd positions in the complex).
The identification of objects in the derived category implements the fact that via brane/anti-brane annihilation many stacks are identified in the infrared limit.
In particular, branes that do not wrap the Calabi-Yau correspond to quasicoherent sheaves that are not locally free and can be realized by annihilating brane/anti-brane pairs with lower-dimensional branes dissolved on their world volumes.

The RR charges of B-branes are classified by $K$-theory or, since we are only interested in geometries without torsion, by the vertical cohomology on $M$.
For a B-brane $\mathcal{F}^\bullet$ the asymptotic central charge can be calculated using the Gamma class formula~\cite{Iritani:2009}
\begin{align}
	\Pi_{\text{asy}}(\mathcal{F}^\bullet)=\int\limits_M e^{\omega}\Gamma_{\mathbb{C}}(M)(\text{ch}\,\mathcal{F}^\bullet)^\vee\,,
	\label{eqn:bbranecharge}
\end{align}
where in terms of the Chern classes $c_2,c_3$ of $M$
\begin{align}
	\Gamma_{\mathbb{C}}(M)=1+\frac{1}{24}c_2+\frac{\zeta(3)}{(2\pi i)^3}c_3\,.
	\label{eqn:gammaclass}
\end{align}
The action of the operator $\vee:\oplus_k H^{k,k}(M)\rightarrow\oplus_k H^{k,k}(M)$ is linear and determined by $\delta^\vee\mapsto(-1)^i\delta$ for $\delta\in H^{i,i}(M)$.

It has been conjectured by Kontsevich that mirror symmetry relates B-branes on $M$ to A-branes on the mirror $W$~\cite{Kontsevich:1994dn}.
The latter are calibrated Lagrangian $k$-cycles $L$ and the central charge is determined by the holomorphic $(3,0)$-form $\Omega$ as
\begin{align}
	\Pi(L)=\int\limits_L\Omega\,.
	\label{eqn:abranecharge}
\end{align}
The mirror map then identifies the asymptotic central charge of $\mathcal{F}^\bullet$ \eqref{eqn:bbranecharge} with the logarithmic terms in the central charge \eqref{eqn:abranecharge} of the mirror brane $L$.
Furthermore, the intersection between two A-branes $L_1\cap L_2$ is equal to the open string index of the mirror branes
\begin{align}
	\chi(\mathcal{E}^\bullet,\mathcal{F}^\bullet)=\int\limits_M\text{Td}(M)\text{ch}(\mathcal{E}^\bullet)^\vee\text{ch}(\mathcal{F}^\bullet)\,,
	\label{eqn:openindex}
\end{align}
where the \textit{Todd class} $\text{Td}(M)$ can be expressed in terms of the Chern classes via
\begin{align}
	\text{Td}(M)=1+\frac12c_1(M)+\frac{1}{12}\left(c_1(M)^2+c_2(M)\right)+\frac{1}{24}c_1(M)c_2(M)+\dots\,.
\end{align}
Of course, for a Calabi-Yau manifold $c_1(M)=0$ and  most of the terms vanish.

We will now introduce a set of branes that generate the charge lattice.
The structure sheaf $\mathcal{O}_M$ corresponds to a $6$-brane, while the skyscraper sheaf $\mathcal{O}_{p}$ with support on some point $p\in M$ corresponds to a $0$-brane.
The short exact sequence
\begin{align}
	0\rightarrow \mathcal{O}_M(-D)\rightarrow\mathcal{O}_M\rightarrow\mathcal{O}_D\rightarrow0\,,
\end{align}
implies a quasi-isomorphism between a $4$-brane $\mathcal{O}_D$ with support on the effective Cartier divisor $D$ and the complex of vector bundles
\begin{align}
	\mathcal{F}^\bullet_{D}\equiv0\rightarrow\mathcal{O}_{M}(-D)\rightarrow\mathcal{O}_M\rightarrow 0\,.
\end{align}
The position of the brane is encoded in the maps of the complex but will not be important to us.
Two-branes with support on a curve $C$ can be obtained as K-theoretic push-forwards
\begin{align}
	\mathcal{C}^\bullet=\iota_! \mathcal{O}_{C}\left(K_{C}^{1/2}\right)\,.
\end{align}

Let us assume a basis of the K\"ahler cone is given by $J_i,\,i=1,...,h^{1,1}(M)$ and denote the dual curves by $C_i=1,\,...,h^{1,1}(M)$.
We expand the K\"ahler form $\omega$ as
\begin{align}
	\omega=J_i t^i\,,
\end{align}
and, using the Gamma class formula \eqref{eqn:gammaclass}, calculate the asymptotic central charges~(see e.g.~\cite{Gerhardus:2016iot} for details of the calculation)
\begin{align}
	\begin{split}
	\Pi_{\text{asy}}(\mathcal{O}_M)=&\frac16c_{ijk}t^it^jt^k+b_it^i+\frac{\zeta(3)}{(2\pi i)^3}\chi\,,\\
	\Pi_{\text{asy}}(\mathcal{F}_{J_i}^\bullet)=&-\frac12 c_{ijk}t^jt^k-\frac12 c_{iij}t^j-\frac16 c_{iii}-b_i\,,\\
		\Pi_{\text{asy}}(\mathcal{C}_i^\bullet)=&t^i\,,\quad \Pi_{\text{asy}}(\mathcal{O}_{p})=-1\,,
	\end{split}
	\label{eqn:branebasis}
\end{align}
where $\chi$ is the Euler characteristic of $M$ and we introduced
\begin{align}
	c_{ijk}=\int_MJ_iJ_jJ_k\quad\text{and}\quad b_i=\frac{1}{24}\int_M c_2 J_i\,.
	\label{eqn:topdef}
\end{align}
In the remainder of this paper we will refer to the central charges of this basis as
\begin{align}
	\vec{\Pi}=\left(\Pi^{(6)},\Pi^{(4)}_i,\Pi^{(2)}_i,\Pi^{(0)}\right)^t\,,
\end{align}
and the Calabi-Yau $M$ as well as the choice of basis for the K\"ahler cone will be clear from the context.

\subsection{Fourier-Mukai transformations and monodromies}
\label{ssec:fourier-mukai}
The periods of the holomorphic $(k,0$)-form can be calculated to arbitrary order in the complex structure parameters of $W$ and it is well known that they experience monodromy when transported along non-contractible loops
in the complex structure moduli space.
The homological mirror symmetry conjecture implies that the B-model monodromies lift, in the A-model, to auto-equivalences of the category of B-branes.
On the other hand, a theorem by Orlov~\cite{Orlov:1996} states that equivalences of derived categories $D^b(X),\,D^b(Y)$ are always expressible as Fourier-Mukai transformations
\begin{align}
	\Phi_{\mathcal{E}}:\,\mathcal{F}^\bullet\mapsto R\pi_{1*}\left(\mathcal{E}\otimes_L L\pi_2^*\mathcal{F}^\bullet\right)\,,
	\label{eqn:fmtransform}
\end{align}
where $\pi_i,i=1,2$ are the projections from $Y\times X$ to the $i$-th factor and the so-called Fourier-Mukai kernel $\mathcal{E}$ is a quasicoherent sheaf on $Y\times X$. The letters $L$ and $R$ indicate that the corresponding derived version of a functor is to be taken.

It is, in general, difficult to evaluate this expression and it is often easier to obtain the corresponding action on the brane charges.
This can be done with the Grothendieck-Riemann-Roch formula which for $f:X\times Y\rightarrow Y$ states that
\begin{align}
	\text{ch}(f_*\mathcal{F}^\bullet)=f_*\left[\text{ch}(\mathcal{F}^\bullet)\cdot f^*\text{Td}(X)\right]\,.
	\label{eqn:grr}
\end{align}
Here we already assume that $X\times Y$ and $Y$ are smooth and neither of the associated K-groups contains torsion such that we can work directly in cohomology.

For several important B-model monodromies the corresponding Fourier-Mukai kernel is known to be of a generic form which allows us to calculate the corresponding action on the brane charges.
We will now discuss the transformations that are most relevant for our discussion.

\paragraph{Large volume monodromies}
Let us denote the embedding of the diagonal $\Delta=M$ into $M\times M$ by $j:M\rightarrow M\times M$.
The Fourier-Mukai kernel that corresponds to the large volume monodromy around the divisor in K\"ahler moduli space where we move to infinity in the direction of a divisor $D$ inside the K\"ahler cone is given by
$j_*\mathcal{O}(D)$~\cite{Horja:1999}. The general formula \eqref{eqn:fmtransform} can be evaluated (see e.g. \cite{Aspinwall:2001zq,Distler:2002ym}) and one finds that the transformation maps
\begin{align}
	\mathcal{F}^\bullet\mapsto\mathcal{F}^\bullet \otimes \mathcal{O}(D)\,.
\end{align}
At the level of the central charges this transformation just leads to a shift in the classical/logarithmic terms.
In particular, the charges of 2-branes that wrap a curve $C$ are shifted by $D\cdot C$.
Of course this is nothing but a shift of the $B$-fields.
It is easy to see, that for large volume transformations with respect to $D=J_a$ the brane charges~\eqref{eqn:branebasis} transform as $\vec{\Pi}\mapsto M_a\cdot\vec{\Pi}$, with
\begin{align}
	M_{a,IJ}=&\left(\begin{array}{cccc}
	1&-\delta_{aj}&0&0\\
		0&\delta_{ij}&-c_{aij}&c_{ai}^+\\
		0&0&\delta_{ij}&-\delta_{ai}\\
		0&0&0&1
	\end{array}\right)\,,\quad c^\pm_{ai}=\frac12\left(c_{aai}\pm c_{aii}\right)\,,
	\label{eqn:lrmon}
\end{align}
where $I,J=1,...,2+2h^{1,1}$ and $i,j=1,...,h^{1,1}$ while $c_{ijk}$ was defined in~\eqref{eqn:topdef}.

\paragraph{Seidel-Thomas twists and conifold transformations}
For loops around a component of the discriminant where the Calabi-Yau $M$ itself or a divisor in $M$ collapses to a point, the corresponding transformation of B-branes is a Seidel-Thomas twist~\cite{Seidel:2000ia,Aspinwall:2001dz,Horja:2001cp}.
Here we will only be interested in the case where $M$ itself collapses which is conjectured to correspond to the principal component of the discriminant~\cite{Horja:1999,Morrison:2000bt}.
We will denote the corresponding transformation as a \textit{conifold transformation} or \textit{conifold monodromy}.

The Fourier-Mukai kernel is given by the ideal sheaf $\mathcal{I}_\Delta$ of the diagonal in $M\times M$.
In the derived category this is quasi-isomorphic to the complex
\begin{align}
	0\rightarrow\mathcal{O}_{M\times M}\rightarrow \mathcal{O}_\Delta\rightarrow 0\,.
	\label{eqn:quasiiso}
\end{align}
One can use the Grothendieck-Riemann-Roch theorem~\eqref{eqn:grr} to translate \eqref{eqn:fmtransform} into an action on the Chern characters of branes
\begin{align}
	\text{ch}(\Phi_{\mathcal{E}}(\mathcal{F}^\bullet))=\pi_{1*}\left(\text{ch}(\mathcal{E})\cdot\pi_2^*\left[\text{ch}(\mathcal{F}^\bullet)\text{Td}(M)\right]\right).
	\label{eqn:fmkcharges}
\end{align}
Using \eqref{eqn:quasiiso} and $\Phi_{\mathcal{O}_\Delta}(\mathcal{F}^\bullet)=\mathcal{F}^\bullet$ (see~\cite{huybrechts2006fourier}, p. 114) this leads to
\begin{align}
	\text{ch}(\Phi_{\mathcal{I}_\Delta}(\mathcal{F}^\bullet))=\text{ch}(\mathcal{F}^\bullet)-\pi_{1*}\pi_2^*\left(\text{ch}(\mathcal{F}^\bullet)\text{Td}(M)\right)\,.
	\label{eqn:fmconi}
\end{align}
With the definition of the open string index \eqref{eqn:openindex} the action on the central charges then takes the simple form
\begin{align}
	\Pi(\mathcal{F}^\bullet)\mapsto\Pi(\mathcal{F}^\bullet)-\chi(\mathcal{F}^\bullet,\mathcal{O}_M)\Pi(\mathcal{O}_M)\,.
	\label{eqn:fmstconifold}
\end{align}

It is perhaps instructive to give some geometric intuition for this action.
Note that multiplication of the charge $\text{ch}(\mathcal{F}^\bullet)$ with the Todd-class $\text{Td}(M)=1+\dots$ does not change the support of the corresponding brane.
Now the effect of the pull-back and push-forward operations in the second term is to make any brane $\mathcal{F}^\bullet$ ``wrap'' the first factor of $M\times M$ and to project out branes that do not have point-like support on the second factor.
The only brane that survives this operation is the zero-brane corresponding to a skyscraper sheaf $\mathcal{O}_{pt.}$ which is transformed into the structure sheaf $\mathcal{O}_M$.
Combining this with the first term in \eqref{eqn:fmstconifold} one can see that the zero brane is transformed into a bound state of a zero brane and an anti six-brane while branes that do not contain embedded zero branes remain unaffected.

Of course, this pictures receives corrections due to the presence of the Todd-class and it is easy to evaluate \eqref{eqn:fmstconifold} exactly.
In particular, we find that our basis of brane charges~\eqref{eqn:branebasis} transforms as $\vec{\Pi}\mapsto M_C\cdot\vec{\Pi}$, with
\begin{align}
	M_{C,IJ}=&\left(\begin{array}{cccc}
		1&0&0&0\\
		-\kappa_i&\delta_{ij}&0&0\\
		0&0&\delta_{ij}&0\\
		-1&0&0&1
	\end{array}\right)\,,\quad\text{where}\quad \kappa_i=\frac16 c_{iii}+2b_i\,.
	\label{eqn:cmon}
\end{align}
The definition of $b_i$ has been given in~\eqref{eqn:topdef}.

\subsection{Relative conifold transformations and $\Gamma_1(N)$}
\label{ssec:relativeconi}
Given a family of complex elliptic curves, the lattice of brane charges is generated by the zero brane $\mathcal{O}_{pt.}$ and the two-brane $\mathcal{O}_C$.
The large volume monodromy $T$ and the conifold monodromy $U$ respectively act on the vector of central charges $\vec{\Pi}=\left(\Pi(\mathcal{O}_C),\,\Pi(\mathcal{O}_{pt.})\right)^t$ as
\begin{align}
	T:\,\vec{\Pi}\mapsto\left(\begin{array}{rr}1&-1\\0&1\end{array}\right)\cdot\vec{\Pi}\,,\qquad U:\,\vec{\Pi}\mapsto\left(\begin{array}{rr}0&1\\-1&1\end{array}\right)\cdot\vec{\Pi}\,.
\end{align}
Note that the Todd-class is trivial and therefore the geometric intuition that we outlined above applies without further modification.
It is easy to see that the corresponding matrices, which we will also denote by $T$ and $U$, generate the modular group $SL(2,\mathbb{Z})$.
In particular, the normalized volume of the curve
\begin{align}
	\tau =-\frac{\Pi(\mathcal{O}_{C})}{\Pi(\mathcal{O}_{pt.})}\,,
\end{align}
transforms as
\begin{align}
	T:\,\tau\mapsto\tau+1\,,\qquad U:\,\tau\mapsto\frac{\tau}{1+\tau}\,.
\end{align}

It turns out that on an elliptic or genus-one fibration this transformation can be performed fiberwise.
Perhaps not surprisingly, the relevant Fourier-Mukai kernel is the ideal sheaf of the relative diagonal $\mathcal{I}_{\Delta_B}$ in the relative fiber product $M\times_B M$~\cite{ruiperez2006relative}.
Following~\cite{Schimannek:2019ijf} we will therefore call this a \textit{relative conifold transformation}.
Using the singular Riemann-Roch theorem an analogous formula to~\eqref{eqn:fmconi} can be derived and reads
\begin{align}
	\text{ch}(\Phi_{\mathcal{I}_{\Delta_B}}(\mathcal{F}^\bullet))=\text{ch}(\mathcal{F}^\bullet)-\pi_{1*}\pi_2^*\left(\text{ch}(\mathcal{F}^\bullet)\text{Td}_{M/B}\right)\,,
\end{align}
where $\text{Td}_{M/B}$ is the Todd-class of the so-called \textit{virtual relative tangent bundle}.

To define the latter, one needs a \textit{local complete intersection} (l.c.i.) morphism $i:M\rightarrow V$ that embeds $M$ into a smooth ambient space $V$ such that $V$ is a bundle $\pi':V\rightarrow B$ over $B$ and $\pi=\pi'\circ i$~\cite{fulton1984intersection}.
When $M$ is a hypersurface or complete intersection in a toric ambient space $V$, the toric ambient space itself often exhibits a compatible toric fibration structure and the inclusion of $M$ is an l.c.i. morphism~\cite{Schimannek:2019ijf}.
In the rest of the paper we will assume this to be the case.
The leading behavior of the Todd-class was then found to generically be
\begin{align}
	\text{Td}_{M/B}=1-\frac12 c_1(B)+\dots\,.
	\label{eqn:relativetodd}
\end{align}

Let us now assume that $M$ is an elliptic or genus one fibration and parametrize the K\"ahler form as
\begin{align}
	\omega=\tau\cdot (E_0+D)+\sum\limits_{i=1}^rm_i\cdot\sigma(E_i)+\sum_{i=r+1}^{\text{rk}(G)}m_i\cdot D_{f,i-r}+\sum_{i=1}^{b_2(B)}\tilde{t}_i\cdot D_i'\,.
	\label{eqn:kaehlerexpansion}
\end{align}
where the vertical divisors $D_i'$ are dual to the curves $C_i=\frac1N E_0\cdot D_i$ and $D_i=\pi^{-1}\tilde{D}_i,\,i=1,...,b_2(B)$ is a basis of vertical divisors.
Moreover, $E_i,\,i=0,...,r$ are independent $N$-sections and $D$ is a vertical divisor such that $\tilde{E}_0=E_0+D$ is orthogonal to all of these curves~\footnote{Recall that independence for $N$-sections
means that they cannot be related via the action of the Jacobian fibration.}.
We will also assume that the fibral divisors $D_{f,i}$ are choosen such that they are orthogonal to the zero-section.

Using the generic behavior of the Todd-class $\text{Td}_{M/B}$~\eqref{eqn:relativetodd} one can calculate the action of the relative Conifold transformation $R$ on the K\"ahler parameters~\eqref{eqn:relativetodd}.
The action on the normalized volumes of curves in the base $\tilde{t}_i,\,i=1,...,b_2(B)$ is
\begin{align}
	\begin{split}
		\tilde{t}_i\mapsto& \tilde{t}_i+\frac{1}{1+N\tau}\left(\frac12\tilde{a}_i\tau^2+\frac{1}{24}\int_Mc_2(M)D_i+\frac N2 a_i\tau-\frac{N}{2}m^am^bC^i_{ab}\right)\\
		=&\tilde{t}_i+\frac{\tilde{a}_i}{2N}\tau-\frac{\tilde{a}_i}{2N}\frac{\tau}{1+N\tau}+\frac{1}{24}\int_M c_2(M)D_i-\frac{1}{1+N\tau}\cdot \frac{N}{2} m^am^b C^i_{ab}+A\,,
	\end{split}
\end{align}
where
\begin{align}
	A=\frac12\frac{N\tau}{N\tau+1}\left(a_i-\frac{1}{12}\int_M c_2(M)D_i\right)\,,\quad\tilde{a}_i=\int\limits_M \tilde{E}_0^2\cdot D_i\,,\quad a_i=\int\limits_B c_1(B) D_i\,,
	\label{eqn:defA}
\end{align}
and
\begin{align}
	C^i_{ab}=\frac{1}{N}\cdot\left\{\begin{array}{cl}
				-\pi_*\left(\sigma(E_a)\cdot\sigma(E_b)\right)\cdot C_\beta&\text{ for }1\le a,b\le r\\
				-\pi_*\left(D_{f,a}\cdot D_{f,b}\right)\cdot C_\beta&\text{ for }r<a,b\le\text{rk}(G)\\
				0&\text{ otherwise}
			\end{array}\right.\,,
	\label{eqn:defC}
\end{align}
It was shown in~\cite{Esole:2018bmf} that $A$ vanishes for generic elliptic fibrations.
A proof that this result extends at least to the classes of genus one fibrations studied in~\cite{Klevers:2014bqa} can be found in Appendix~\ref{app:characteristic}.
We can then introduce the shifted K\"ahler parameters
\begin{align}
	t_i=\tilde{t}_i+\frac{\tilde{a}_i}{2N}\tau\,,
	\label{eqn:kaehlershift}
\end{align}
and find the action
\begin{align}
	U:\,\left\{\begin{array}{rcl}
		\tau&\mapsto&\tau/(1+N\tau)\\
		m_i&\mapsto&m_i/(1+N\tau)\,,\quad i=1,...,\text{rk}(G)\\
		Q_i&\mapsto&(-1)^{a_i}\exp\left(-\frac{N}{1+N\tau}\cdot \frac12m^am^b C^i_{ab}+\mathcal{O}(Q_i)\right)Q_i
	\end{array}\right.\,,
	\label{eqn:raction}
\end{align}
where $Q_i=\exp(2\pi i t_i)$ and the action on $Q_i$ receives corrections that are double exponentially surpressed in the large base limit.
Note that the large volume transformation $T$ that shifts $\tau\mapsto\tau +1$ while leaving $m_i,\,i=1,...,\text{rk}(G)$ invariant acts as
\begin{align}
	T:\,\left\{\begin{array}{rcl}
		\tau&\mapsto&\tau+1\\
		m_i&\mapsto& m_i\,,\quad i=1,...,\text{rk}(G)\\
		Q_i&\mapsto&(-1)^{\frac{\tilde{a}_i}{2N}}Q_i
	\end{array}\right.\,.
\end{align}

At this point we should review the congruence subgroups $\Gamma_0(N)$ and $\Gamma_1(N)$ of $\Gamma=SL(2,\mathbb{Z})$.
The \textit{Hecke congruence subgroup} of level $n$ is defined as
\begin{align}
	\Gamma_0(N)=\left\{\left(\begin{array}{cc}a&b\\c&d\end{array}\right)\in\Gamma\,:\,c\equiv 0\,(\text{mod }n)\right\}\,,
\end{align}
while
\begin{align}
	\Gamma_1(N)=\left\{\left(\begin{array}{cc}a&b\\c&d\end{array}\right)\in\Gamma\,:\,a,d\equiv 1\,(\text{mod }n)\,,\quad c\equiv 0\,(\text{mod }n)\right\}\,,
\end{align}
and it is clear that $\Gamma_1(N)\subseteq\Gamma_0(N)\subseteq\Gamma$.
Both groups have an interesting moduli problem associated to it which will be relevant to us later.
Namely, $\Gamma_0(N)$ acts on the complex structure parameter $\tau$ of an elliptic curve such that a cyclic subgroup of order $N$ is preserved while
$\Gamma_1(N)$ preserves also the generator of this group~\cite{Silverman:1338326}.

Let us now discuss generators for $\Gamma_0(N)$ and $\Gamma_1(N)$ with $N\le4$.
Using the elements
\begin{align}
	\tilde{T}=\left(\begin{array}{cc}1&1\\0&1\end{array}\right)\,,\quad \tilde{U}_N=\left(\begin{array}{cc}1&0\\N&1\end{array}\right)\,
\end{align}
we can write $\Gamma_0(N)=\langle \tilde{T},\tilde{U}_N,-\mathbb{1}\rangle$ and $\Gamma_1(N)=\langle \tilde{T},\tilde{U}_N\rangle$.
Note that the set of generators is not always minimal and in particular one finds $\Gamma_0(2)=\Gamma_1(2)$.
Starting with $N=5$ the generating sets become more complicated and we will restrict ourselves to $N\le4$.
Because $\Gamma_0(N)$ is then obtained from $\Gamma_1(N)$ by adjoining the matrix $-\mathbb{1}$ it is clear that the action on $\tau$ and therefore also the rings of modular forms for both groups are identical.

With these definitions we see that for $N\le 4$ the monodromies $U$ and $T$ generate an action of the congruence subgroup $\Gamma_1(N)\subseteq SL(2,\mathbb{Z})$, or equivalently $\Gamma_0(N)$, on the K\"ahler parameters.
Under this action $\tau$ transforms like a modular parameter, $m_i$ transform as elliptic parameters and the exponentiated K\"ahler parameters $Q_i$ transform, up to a multiplier system,
like lattice Jacobi forms of weight $0$ and with index matrix given by $C^i_{ab}$.

Note that on the enlarged moduli space, where one does not normalize the 0-brane charge, $U$ and $T$ generate an action of $\Gamma_1(N)$ and, for $N>2$, not of $\Gamma_0(N)$.
In the following we will therefore talk about modularity with respect to $\Gamma_1(N)$ although the rings of modular forms do not distinguish between the groups.

\subsection{EZ-transformations and wall monodromies}
\label{ssec:ez-trans}
The Seidel-Thomas twists admit several generalizations.
Horja constructed so-called \textit{EZ-transformations} that arise from loci where a subvariety $E\hookrightarrow M$ collapses onto a subvariety $Z$~\cite{horja1999hypergeometric,Horja:2001cp}.
The conjectured physical interpretation is that there is a whole category of branes becoming massless at these loci.
Essentially the massless objects arise from pullbacks of branes on $Z$ and the induced action on any brane $B$ binds a particular subset of massless objects that depends on $B$.
The explicit form of the corresponding monodromy action is significantly more complicated than for the Seidel-Thomas twist and we refer to \cite{Aspinwall:2002nw} for a discussion that is aimed towards physicists.
However, it is clear that the relative conifold transformation should correspond to an EZ-transformation that arises from the monodromy around a locus in the K\"ahler moduli space where the generic fiber of the fibration
collapses to a point.

More generally, EZ-transformations are conjectured to be realized in the A-model as \textit{wall monodromies} around phase boundaries. 
For any Calabi-Yau variety that is a fibration we expect that, when all K\"ahler parameters but the volume $\tau$ of the fiber are deep inside the K\"ahler cone, the locus $\tau=0$ marks a boundary between the geometric cone and a hybrid phase in
the stringy K\"ahler moduli space.
The hybrid phase can be thought of as a Landau-Ginzberg model that is fibered over a non-linear sigma model.
In general, when the volume of the fiber of a fibration vanishes, the category of massless branes is conjectured to be generated by pull-backs of branes in the derived category of the base along the morphism that defines the fibration~\cite{Aspinwall:2002nw}.
One of the branes that becomes massless is therefore in particular the 6-brane that is the pre-image of the base itself.
However, the locus in the moduli space where the volume of the 6-brane vanishes is conjectured to correspond to the principal component of the discriminant.
The generic monodromy around this locus is just given by a Seidel-Thomas twist with respect to the 6-brane~\eqref{eqn:fmstconifold}.
This implies that in the complex structure moduli space of the mirror the point $\tau=0$ with all other K\"ahler moduli sent to infinity cannot be a normal crossing but corresponds to a tangency between the principal component of the discriminant and
the union of the large complex structure divisors.
It was shown in~\cite{Aspinwall:2001zq} (and we review below) that such tangencies imply a relation between the Seidel-Thomas twist with respect to the 6-brane and the large volume monodromies.

We will illustrate these somewhat technical statements with an example.
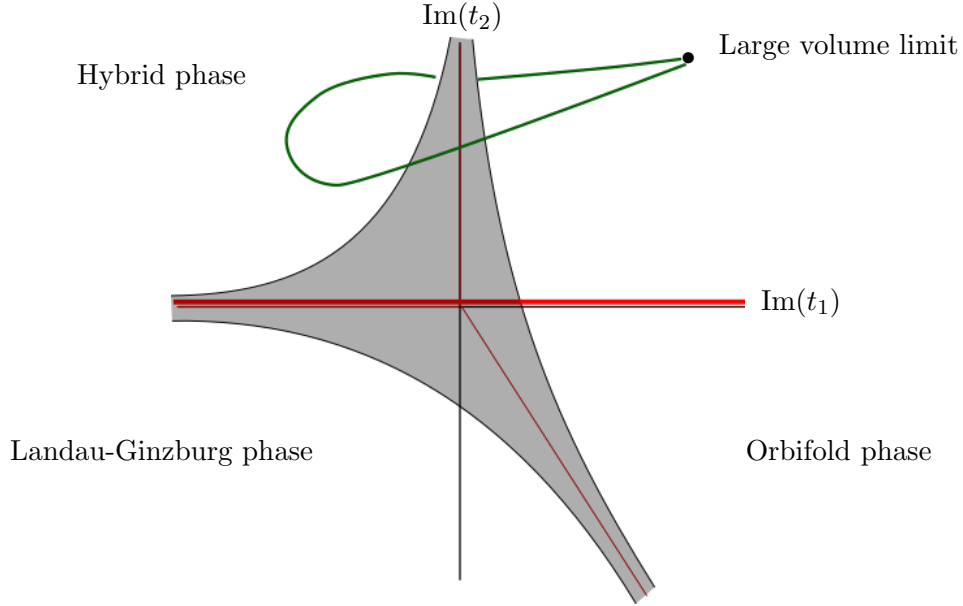
\begin{figure}[h!]
	\begin{tikzpicture}[remember picture,overlay,node distance=4mm]
		\node[] at (7,7.25) {$\bullet$};
		\node[] at (9,7.4) {Large volume limit};
		\node[] at (9,2) {Orbifold phase};
		\node[] at (0,2) {Landau-Ginzburg phase};
		\node[] at (0,7) {Hybrid phase};
		\node[] at (8.5,4) {$\text{Im}(t_1)$};
		\node[] at (4,7.8) {$\text{Im}(t_2)$};
	\end{tikzpicture}
	\centering
	\includegraphics[width=.5\linewidth]{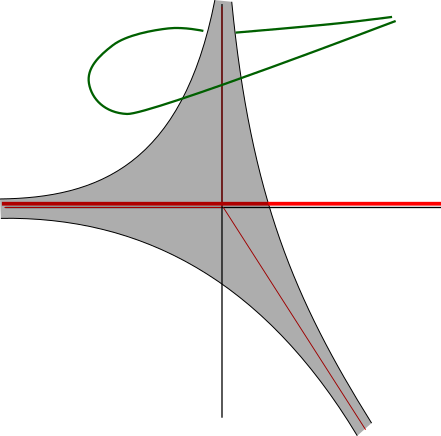}
	\caption{Diagram of the FI-parameter space of the gauged linear sigma model that realizes the non-linear sigma model into the degree $18$ hypersurface inside $\mathbb{P}(1,1,1,6,9)$.
	The amoeba of the principal component of the discriminant, which corresponds to a non-compact Coulomb branch, is indicated in grey. Two geometric phases on the right are seperated by a mixed Coulomb-Higgs-branch.}
	\label{fig:amoeba}
\end{figure}
Consider the well-studied Calabi-Yau threefold $X_{18}$ that corresponds to a generic degree $18$ hypersurface in the weighted projective space $\mathbb{P}(1,1,1,6,9)$.
It is elliptically fibered over the base $\mathbb{P}^2$ and the group of divisors is generated by a section $E_0$ and one vertical divisor $D_b$.
A basis of the K\"ahler cone is generated by $J_1=E_0+3D_b$ and $J_2=D_b$ and we expand the K\"ahler form as
\begin{align}
	\omega=t^1J_1+t^2J_2\,.
\end{align}
The topological invariants~\eqref{eqn:topdef} are
\begin{align}
	c_{111}=9\,,\quad c_{112}=3\,,\quad c_{122}=1\,,\quad c_{222}=0\,,\quad \vec{b}=\left(17/4,\,3/2\right)\,.
\end{align}
In Batyrev coordinates the discriminant consists of two components
\begin{align}
	\Delta_1=(1-432z_1)^3-432^3\cdot27\cdot z_1^3z_2\,,\quad \Delta_2=1+27z_2\,.
\end{align}
The mirror maps are given by
\begin{align}
	t_1=\frac{1}{2\pi i}\log(z_1)+\mathcal{O}(z)\,,\quad t_2=\frac{1}{2\pi i}\log(z_2)+\mathcal{O}(z)\,,
\end{align}
where $t_1$ is the complexified volume of the generic fiber and $t_2$ the complexified volume of a degree one curve in the base.
Note that there is a triple tangency between $\Delta_1=0$ and $z_3=0$ at $z_1=1$.

At leading order the complexified K\"ahler parameters $t_1,t_2$ can be identified with FI-parameters of a GLSM that realizes the Calabi-Yau as a vacuum manifold.
The FI-parameter space of such a GLSM is depicted in figure~\ref{fig:amoeba}.
A geometric phase that realizes $X_{18}$ can be found for $\text{Im}(t_1)\gg0,\,\text{Im}(t_2)\gg0$ while there is a hybrid phase around $\text{Im}(t_1)\ll0,\text{Im}(t_2)\gg0$.
The hybrid phase essentially corresponds to the Landau-Ginzburg model of the elliptic fiber that is fibered over a non-linear sigma model with target space $\mathbb{P}^2$.
The geometric and the hybrid phase are seperated by a ``tentacle'' of the amoeba of $\Delta_1=0$.

There is a wall monodromy in the A-model that corresponds to a loop around the boundary between the two phases that is taken deep inside the limit of large base volume, i.e. $\text{Im}(t_2)\gg 0$.
To obtain the mirror transformation in the B-model we have to move around the discriminant $\Delta_1=0$ close to the plane $z_1=0$.
This has been done for $X_{18}$ in \cite{Aspinwall:2001zq} and the procedure is as follows.

\begin{figure}[h!]
	\centering
	\begin{minipage}[b]{.4\textwidth}
	\includegraphics[width=\linewidth]{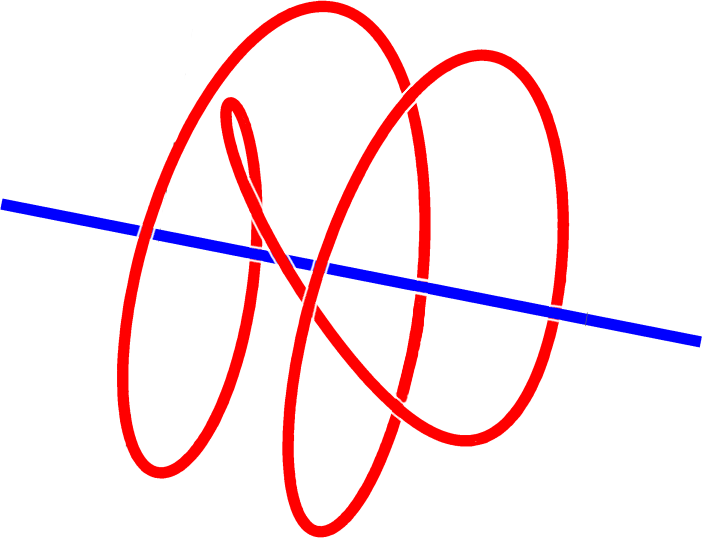}
	\end{minipage}
	\hspace{1cm}
	\begin{minipage}[b]{.4\textwidth}
	\includegraphics[width=\linewidth]{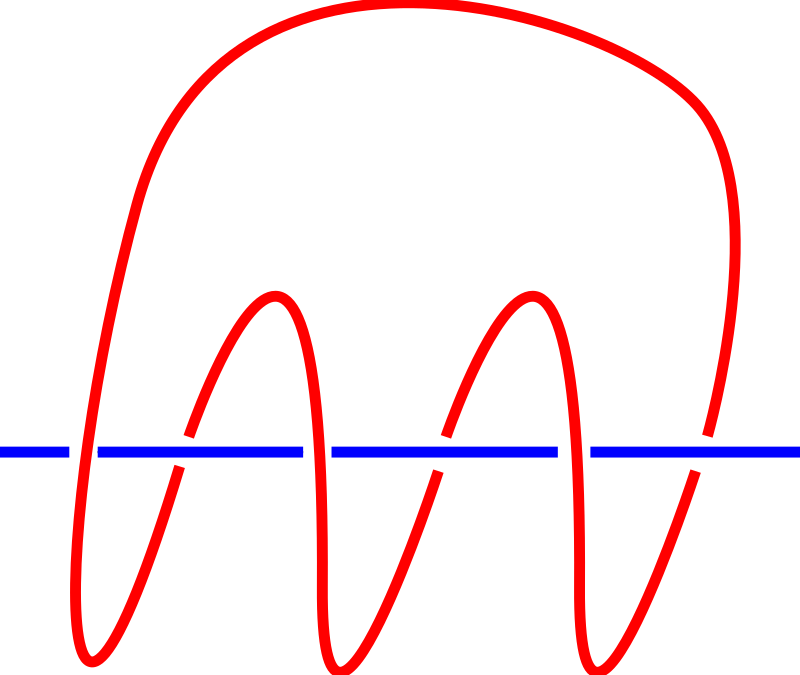}
	\end{minipage}
	\caption{The left image shows a plot of the link that is formed by $\Delta_1=0$ (red) and $z_3=0$ (blue) inside a small 3-sphere around the point $z_1=1,z_3=0$ in the complex structure moduli space of the mirror of $X_{18}$. We use stereographic coordinates on the 3-sphere and the blue line closes at infinity. The right image shows an equivalent link.}
	\label{fig:x18link}
\end{figure}

One considers a small 3-sphere $S_\epsilon$ around $z_1=1,\,z_2=0$. The intersections $L_1=S_\epsilon\cap\{\Delta_1=0\}$ and $L_2=S_\epsilon\cap\{z_1=1\}$ are both unknots inside $S_\epsilon$ that form a non-trivial link.
The shape of this link can be seen directly by taking stereographic coordinates on $S_\epsilon$.
A plot as well as a schematic depiction of the link is given in figure \ref{fig:x18link}.
In the sketch on the right the large volume monodromy corresonds to a loop around the blue line taken outside the link while the generic monodromy around the conifold can be obtained by following a loop around the ``handle'' of the red line.
We now claim that the wall monodromy $M_{\text{W}}$ is related to the generic conifold monodromy $M_{C}$ and to the large base volume monodromy $M_{b}$ is given by
\begin{align}
	M_{\text{W}}=M_b^{-1}\cdot M_C\cdot M_b^{-1}\cdot M_C\cdot M_b^{-1}\cdot M_C\cdot M_b^3\,,
	\label{eqn:wall}
\end{align}
while referring the reader to \cite{Aspinwall:2001zq} for a detailed discussion of the appropriate paths.

Now how does the wall monodromy $M_{\text{W}}$ act on the brane charges of $X_{18}$?
Using~\eqref{eqn:lrmon} and~\eqref{eqn:cmon} one can immediately calculate the action of $M_W$ on our period basis~\eqref{eqn:branebasis}.
It turns out that~\eqref{eqn:wall} \textit{exactly} reproduces the action of the Fourier-Mukai transformation that is associated to the ideal sheaf of the relative diagonal.

We are now going to show that this relation is generic for any elliptic or genus-one fibration over $\mathbb{P}^2$ and derive an analogous relation over Hirzebruch surfaces $\mathbb{F}_n$.
Let us assume that $J_b$ is an effective Cartier divisor on \textit{any} Calabi-Yau threefold $M$ such that
\begin{align}
	J_b^3=0\,,\quad\text{and}\quad \int_Mc_2(M)\cdot J_b=36\,.
	\label{eqn:p2cond}
\end{align}
We can then use~\eqref{eqn:lrmon} and~\eqref{eqn:cmon} to obtain
\begin{align}
	\begin{split}
		&\left(M_b^{-1}\cdot M_C\right)^3\cdot M_b^3=\left(\begin{array}{cccc}
			1&-3\delta_{bj}&3c_{bbj}&0\\
			-c_{bbi}&x_{2,ij}&x_{3,ij}&0\\
			0&-\delta_{bj}\delta_{bi}&\delta_{ji}+c_{bbj}\delta_{bi}&0\\
			0&0&-c_{bbj}&1
	\end{array}\right)\,,\\
		&x_{2,ij}=\left(2c_{bbi}-c_{bi}^-\right)\delta_{bj}+\delta_{ji}\,,\quad x_{3,ij}=c_{bbj}(c_{bi}^--c_{bbi}-\kappa_i)\,.
	\end{split}
	\label{eqn:p2wallmon}
\end{align}
This exactly reproduces the action of the $U$-transformation~\eqref{eqn:raction} on the 2-brane charges of an elliptic or genus-one fibration over $\mathbb{P}^2$.

Let us now assume instead that there are two effective Cartier divisors $J_1,J_2$ and some $n\in\mathbb{N}$ such that
\begin{align}
	J_2^2=J_1^3=J_1^2J_2=J_1J_2^2=0\,,\quad \int_Mc_2(M)\cdot J_1=12(2+n)\,,\quad \int_Mc_2(M)\cdot J_2=24\,.
	\label{eqn:fncond}
\end{align}
This is satisfied when $M$ is an elliptic or genus-one fibration over the Hirzebruch surface $\mathbb{F}_n$.
More precisely, if we denote the base and the fiber of $\mathbb{F}_n$ respectively by $B$ and $F$ then $J_1=\pi^{-1}(B+F)$ and $J_2=\pi^{-1}(F)$.
We can then calculate
\begin{align}
\begin{split}
	&\left(M_1^{-1}\cdot M_C\cdot M_2^{-1}\cdot M_C\right)^2\\
	=&\left(\begin{array}{cccc}
		1&n\cdot\delta_{2j}&c_{11j}-n\cdot c_{12j}&0\\
		-c_{12i}&y_{22,ij}&y_{23,ij}&2c_{1i}^--c_{2ii}+4c_{12i}+c_{11i}\\
		0&y_{32,ij}&y_{33,ij}&2(\delta_{1i}+\delta_{2i})\\
		0&0&-c_{12j}&1
	\end{array}\right)\,,\\
	y_{22,ij}=&\delta_{2j}\left((n-1)c_{12i}-\frac{n}{2}c_{2ii}-c_{1i}^--c_{11i}\right)+\delta_{1j}\left(\frac12c_{2ii}-c_{12i}\right)+\delta_{ji}\,,\\
	y_{23,ij}=&(c_{12j}+c_{11j})\left(\frac12c_{2ii}-c_{12i}\right)-(c_{1i}^-+\kappa_i)c_{12j}+2\left(c_{2ji}+c_{1ji}\right)\,,\\
	y_{32,ij}=&-\delta_{2i}\delta_{1j}-\delta_{1j}\delta_{2i}+n\delta_{2i}\delta_{2j}\,,\\
	y_{33,ij}=&-\delta_{2i}(c_{12j}+c_{11j})-\delta_{1i}c_{12j}+\delta_{ji}\,,
\end{split}
\end{align}
and find that
\begin{align}
M_W=\left(M_1^{-1}\cdot M_C\cdot M_2^{-1}\cdot M_C\right)^2\cdot M_1^2\cdot M_2^2\,,
	\label{eqn:fnwallmon}
\end{align}
reproduces the action of the $U$-transformation~\eqref{eqn:raction}.

For the sake of completeness let us also discuss the case of a $K3$ fibration over $\mathbb{P}^1$.
Assume that there is an effective Cartier divisor $J_b$ on any Calabi-Yau threefold $M$ such that
\begin{align}
	J_b^2=0\,,\quad \int_Mc_2(M)\cdot J_b=24\,.
	\label{eqn:p1cond}
\end{align}
Then from $J_b^2=0$ it follows that $M$ is fibered over $\mathbb{P}^1$ and the intersection with $c_2(M)$ implies that the generic fiber is a $K3$~\cite{oguiso1993}.
We calculate
\begin{align}
	\begin{split}
		\left(M_b^{-1}\cdot M_C\right)^2 =&\left(\begin{array}{cccc}
			-1&0&0&0\\
			\frac12 c_{bjj}&\delta_{ij}+\left(c_{bjj}-\kappa_j\right)\delta_{bi}&2c_{bij}&-c_{bjj}\\
			\delta_{bj}&-\delta_{bj}\delta_{bi}&\delta_{ij}&2\delta_{bj}\\
			0&-\delta_{bi}&0&1
		\end{array}\right)\,,
	\end{split}
\end{align}
and see that the wall monodromy
\begin{align}
	M_W=\left(M_b^{-1}\cdot M_C\right)^2\cdot M_b^2\,,
\end{align}
corresponds, in Batyrev coordinates, to a double tangency between the principal component of the discriminant and the large base divisor where the $K3$ fiber collapses to a point.

The generic form of the monodromies~\eqref{eqn:p2wallmon} and~\eqref{eqn:fnwallmon} suggests that the conditions~\eqref{eqn:p2cond} and~\eqref{eqn:fncond} are also sufficient for a Calabi-Yau threefold to be elliptic or genus one fibered respectively over $\mathbb{P}^2$ or $\mathbb{F}_n$.
This can possibly derived from the more general criteria by Oguiso~\cite{oguiso1993} but we are not aware of any previous discussion of these criteria.

\paragraph{A comment on the swampland}
Let us briefly mention that there is a beautiful connection of this story to the swampland distance conjectures, which posit that around loci that are at inifinite distance in the moduli space an infinite tower of states becomes massless.
Moreover, according to the emergence conjecture, integrating out the tower of states is what generates the infinite distance in the first place.
The swampland distance conjecture in the context of the stringy K\"ahler moduli space has been discussed in~\cite{Blumenhagen:2018nts,Lee:2018urn,Lee:2018spm,Lee:2019tst,Erkinger:2019umg,Joshi:2019nzi} and for a recent review of the swampland program see~\cite{Palti:2019pca}.
As we discussed in great detail, the boundary between the hybrid phase and the geometric cone corresponds to a tangency between the discriminant and the large complex structure divisors in the complex structure moduli space of the mirror.
This implies that it is at infinite distance and it is natural to ask what the towers of states are that become massless.

The massless states can either correspond to wrapped D-branes or to Kaluza-Klein modes and the wrapped D-branes can become massless on their own or only relative to the Planck scale.
The states that arise from wrapped D-branes are called Ramond-Ramond states.
At the phase boundaries where the fiber of a fibration collapses we can identify a tower of light Kaluza-Klein modes that arise from the large volume of the base.
As we discussed above, the massless branes are supported on the restriction of the fibration to cycles in the base.

On a K3-fibered Calabi-Yau threefold there is only a finite number of Ramond-Ramond states that become massless independently of the Planck scale, because neither multi-wrappings of the Calabi-Yau itself nor branes that wrap the fiber over multiple points lead to independent states in the effective theory.
In particular, one can consider the multi-scaling limit where the towers from Kaluza-Klein modes and those branes that only become massless relative to the Planck scale decouple, i.e. one sends the Planck scale to infinity.
The remaining massless states then yield the $W$- and $Z$-bosons of an 4d $N=2$ gauge theory~\cite{Aspinwall:2002hg}.
When the Planck scale is fixed at a finite value, one expects that NS5-branes that wrap the $K3$ fiber lead to a heterotic string where the tension corresponds to the volume of the fiber~\cite{Lee:2019tst}.

However, for a genus one fibration there is a tower of states from branes that wrap the restriction of the fibration to irreducible curves in the base~\cite{Witten:1996qb}.
For this reason there does not appear to be a well-defined field theory limit when the Planck scale is sent to infinity and when it is kept at a finite value, as expected from 2-fold T-duality, another Type II string emerges~\cite{Corvilain:2018lgw}.

An independent study of the inifinite distance limits from boundaries of the K\"ahler cone where the fiber of a fibration collapses has been performed in the work~\cite{lee2019emergent} that appeared during the final stage of the preparation of this paper.
\subsection{When divisors collapse to curves}
\label{ssec:weyl}
Another example of an EZ-twist arises when a fibral divisor collapses to the curve in the base of the fibration~\cite{horja1999hypergeometric}.
It is well known that the monodromies around the corresponding locus in K\"ahler moduli space generate the action of the Weyl group on the K\"ahler moduli that parametrize
the volumes of the components of the familiy of reducible fibers~\cite{Katz:1996ht,2002math.....10121S,Szendroi:2002rf}.
Let us assume that a rationally fibered fibral divisor $D_f$ collapses to a curve of genus $g$ and denote the generic fiber of $D_f$ by $C$. Then for Calabi-Yau threefolds the corresponding action on brane charges reads~\cite{Aspinwall:2001zq}
\begin{align}
	\Pi\left(\mathcal{F}^\bullet\right)\mapsto\Pi\left(\mathcal{F}^\bullet\right)-\chi\left(\mathcal{O}_{D_f}-(1-g)\mathcal{O}_C,\mathcal{F}^\bullet\right)\cdot\Pi\left(\mathcal{O}_C\right)+\chi\left(\mathcal{O}_C,\mathcal{F}^\bullet\right)\cdot\Pi\left(\mathcal{O}_{D_f}\right)\,.
	\label{eqn:fmweyl}
\end{align}
For the examples with fibral divisors that we consider in this paper we verified this formula using matrix factorization techniques in the corresponding gauged linear sigma models~\cite{Herbst:2008jq,Erkinger:2017aaa}.

\subsection{Monodromies and automorphic properties of $Z_{\text{top}}$}
\label{ssec:monaut}
Having discussed the monodromies that generate the modular transformations of the K\"ahler moduli we now want to relate this to the modular properties of the topological string partition function $Z_{\text{top}}$.

The automorphic properties under general monodromies can be derived from the wave function interpretation of $Z_{\text{top}}$~\cite{Witten:1993ed}.
This approach was explored e.g. in~\cite{Aganagic:2006wq,Gunaydin:2006bz}.
The corresponding Hilbert space arises from quantizing the symplectic vector space $H^3(W)$ where $W$ is again the mirror of $M$.
Of course, one can equally well consider the symplectic vector space structure on the quantum cohomology ring of $M$.
If we choose the {real polarization}, our positions in this phase space correspond to the central charges of a basis of $0$- and $2$-branes and the conjugate momenta are central charges of $4$- and $6$-branes.

Let us first consider the special case that a monodromy does not mix position and momenta or, in other words, it transforms the lattice of $0$- and $2$-brane charges into itself.
In that case the topological string partition function is expected to be invariant under the corresponding action on the flat coordinates.
This happens to be the case for some of the monodromies that we discussed above:
The large volume monodromies act as $t\mapsto t+1$ on the flat coordinates and the $q$-expansion of $Z_{\text{top}}$ can be interpreted as a consequence.
Less trivial are the monodromies that generate the action of the Weyl group.
They transform the volumes of fibral curves into each other and $Z_{\text{top}}$ has to be invariant under this action.

We will now show that the elliptic transformation law of the topological string partition function with respect to the geometric elliptic parameters can also be derived in this way.
The large volume transformation that shifts $\tau\mapsto\tau+1$ will again be denoted by $T$ and for every volume $m_i$ of a rational fibral curve there is a corresponding large volume transformation $M_i$ that acts as $m_i\mapsto m_i+1$.
The inverse $U^{-1}$ of the transformation~\eqref{eqn:raction} acts as
\begin{align}
	U^{-1}:\,\left\{\begin{array}{rcl}
		\tau&\mapsto&\tau/(1-N\tau)\\
		m_i&\mapsto&m_i/(1-N\tau)\,,\quad i=1,...,\text{rk}(G)\\
		Q_i&\mapsto&(-1)^{a_i}\exp\left(\frac{N}{1-N\tau}\cdot \frac12m^am^b C^i_{ab}+\mathcal{O}(Q_i)\right)Q_i
	\end{array}\right.\,,
	\label{eqn:ractioni}
\end{align}
where the definitions of $Q_i,\,i=1,...,b_2(B)$, $a_i$ and $C^i_{ab}$ are as in~\eqref{eqn:raction},~\eqref{eqn:defA} and~\eqref{eqn:defC}.
Then the combination $E_a=M_a\cdot U^{-1}\cdot M_a^{-1}\cdot U$ acts as
\begin{align}
	E_a:\,\left\{\begin{array}{rcl}
		\tau&\mapsto&\tau\\
		m_i&\mapsto&m_i\,,\quad i=1,...,\text{rk}(G),\,i\ne a\\
		m_a&\mapsto&m_a+N\cdot \tau\,,\\
		Q_i&\mapsto&\exp\left(\frac{N^2\cdot\tau}{2}C^i_{aa}+NC^i_{(ab)}m^b\right)Q_i
	\end{array}\right.\,.
	\label{eqn:eaction}
\end{align}
Note that \eqref{eqn:eaction} is exact even away from the large base limit.
It is then clear that invariance of $Q^\beta Z_\beta(\tau,\vec{m},\lambda)$ under the action of $M_i$ and $E_i$ implies that
\begin{align}
	\begin{split}
	&Z_{\beta}(\tau,m_1,...,m_a+\kappa N\tau+\rho,...,m_{\text{rk}(G)},\lambda)\\
		=&\exp\left[-\frac{\beta_i}{2}\left(C^i_{aa}\kappa^2N^2\tau+ 2C^i_{(ab)}N\kappa m^b\right)\right]Z_\beta(\tau,\vec{m},\lambda)\,,
	\end{split}
\end{align}
for all $\kappa,\rho\in\mathbb{Z}$ and $\beta\in H_2(B,\mathbb{Z})$.
Therefore $Z_\beta(\tau,\vec{m},\lambda)$ satisfies the elliptic transformation law for the geometric elliptic parameters $m_i,\,i=1,...,\text{rk}(G)$ and the index matrix is given by $C^i_{ab}$.

It should be possible to make a similar argument that relates the full modular transformation law to the $U$-monodromy, although the mixing of positions and momenta requires a careful treatment of the transformation of $Z_{\text{top}}$.
The elliptic transformation law with respect to the topological string coupling constant would then be implied by the modular transformation law.
However, this proves to be surprisingly subtle.
We will not solve this problem here but to highlight the difficulties it is instructive to review the situation for an elliptic fibration that leads to a trivial gauge group.

As we already discussed above, the Calabi-Yau threefold $M=X_{18}$ is elliptically fibered over $B=\mathbb{P}^2$.
In particular, following the discussion in~\ref{ssec:relativeconi}, we can introduce K\"ahler parameters $\tau,t$ such that
\begin{align}
	T:\,\left\{\begin{array}{rcl}
		\tau&\mapsto&\tau+1\\
		t&\mapsto&t+\frac32
	\end{array}\right.\,,\quad
	U:\,\left\{\begin{array}{rcl}
		\tau&\mapsto&\frac{\tau}{1+\tau}\\
		t&\mapsto&t+\frac32
	\end{array}\right.\,,
\end{align}
where in the $U$ transformation we have suppressed terms that are exponentially surpressed in the large base limit.
On the other hand, our choice of gauge for the holomorphic 3-form $\Omega$ implies that it transforms like a modular form of weight $-1$, i.e.
\begin{align}
	T:\,\Omega\mapsto\pm\Omega\,,\quad U:\,\Omega\mapsto\frac{\pm1}{1+\tau}\Omega\,,
\end{align}
where the sign can be fixed by studying the action of the monodromy in the complex structure moduli space~\cite{Huang:2015sta}. 
But $\Omega$ is a section of a line bundle $\mathcal{L}$ on the complex structure moduli space of the mirror and $\lambda$ is a section of $\mathcal{L}$ as well.
Therefore the topological string coupling also transforms like a modular form of weight $-1$.
This is also clear from the action of the monodromies on the enlarged moduli space.

We now want to relate this to the modular properties of the topological string partition function on $X_{18}$.
If we consider the string partition function in holomorphic polarization it is invariant under monodromies~\cite{Aganagic:2006wq}~\footnote{Possibly up to an overall factor due to the change of gauge which is irrelevant for our discussion.}.
The expansion 
\begin{align}
	Z(\tau,t,\lambda)=\exp\left(\sum\limits_{g=0}^\infty\lambda^{2g-2}F_g\right)\,,
\end{align}
then implies that the free energies $F_g$ transform like modular forms of weight $2g-2$~\footnote{Following \cite{Aganagic:2006wq} this sum should actually start at genus two. That the genus zero free energy of $X_{18}$ also admits
an expansion in terms of quasi-modular forms hints towards the existence of an anholomorphic free energy at genus zero.}.
It also follows from the discussion in~\cite{Aganagic:2006wq} that if we take the limit where the imaginary parts of all K\"ahler parameters except for $\tau$ go to infinity, then the anholomorphic free energies will be a polynomial in $(\text{Im}\tau)^{-1}$.
However, it is easy to see that in this limit the partition function has to remain invariant under the action of $U$ and under the large complex structure monodromies.
We can conclude that the anholomorphic free energies $F_g$ are almost holomorphic modular forms of weight $2g-2$.

Note that this implies that the coefficients $Z_d(\tau,\lambda)$ in the expansion
\begin{align}
	Z(\tau,t,\lambda)=Z_0(\tau,\lambda)\left(1+\sum\limits_{d=1}^\infty Z_d(\tau,\lambda)Q^d\right)\,,
\end{align}
are, up to a multiplier system, almost holomorphic modular forms of weight $0$.
The multiplier system is a consequence of the non-trivial transformation of the base parameter under $U$ and $T$.
It turns out that Witten's wave function equation then implies that $Z_d(\tau,\lambda)$ is a weak Jacobi form of weight $0$ and index $m=d(d-3)/2$~\cite{Huang:2015sta,Katz:2016SM}.
This can be shown as follows.

A weak Jacobi form $\phi_{k,m}(\tau,z)$ (see~\ref{ssec:ringsofforms} for the definition) can be written as a power series $\phi_{k,m}(\tau,z)\in \widetilde{M}_\bullet[[z]]$, where $\widetilde{M}_\bullet=\mathbb{C}[E_2,E_4,E_6]$ is the ring of quasi-modular forms.
It is clear that
\begin{align}
	\tilde{\phi}(\tau,z)=\exp\left(\frac{\pi^2}{3}mz^2E_2(\tau)\right)\phi_{k,m}(\tau,z)\,,
	\label{eqn:e2phi}
\end{align}
transforms like a weak Jacobi form of index $0$ under modular transformations of $\tau$.
This implies that $\phi_{k,m}(\tau,z)$ satisfies the differential equation
\begin{align}
	\left(\frac{\partial}{\partial E_2}+\frac{(2\pi i)^2}{12}mz^2\right)\phi_{k,m}(\tau,z)=0\,.
\end{align}
On the other hand, let us denote the ring of almost holomorphic modular forms by $\widehat{M}_\bullet=\mathbb{C}[\hat{E}_2,E_4,E_6]$ and assume that an element $\hat{f}(\tau,z)\in \widehat{M}_\bullet[[z]]$ satisfies the differential equation
\begin{align}
	\left(\frac{\partial}{\partial \hat{E}_2}+\frac{(2\pi i)^2}{12}mz^2\right)\hat{f}(\tau,z)=0\,.
\end{align}
Then we know that
\begin{align}
	\frac{\partial}{\partial\hat{E}_2}\exp\left(\frac{\pi^2}{3}mz^2\hat{E}_2(\tau)\right)\hat{f}(\tau,z)=0\,,
\end{align}
and the limit
\begin{align}
	f(\tau,z)=\lim_{\text{Im}\tau\rightarrow0}\hat{f}(\tau,z)\,,
\end{align}
satisfies the modular transformation law of a weak Jacobi form.

Let us introduce $z=\frac{\lambda}{2\pi i}$ where $\lambda$ is again the topological string coupling.
The wave function equation for $X_{18}$ then reads in holomorphic polarization~\cite{Klemm:2012sx,Alim:2012ss} 
\begin{align}
	\left(\frac{\partial}{\partial \hat{E}_2}+\frac{(2\pi i)^2}{12}\frac{d(d-3)}{2}z^2\right) Z_d(\tau,z)=0\,.
	\label{eqn:difE2}
\end{align}
In the holomorphic limit $Z_d(\tau,z)$ therefore satisfies the modular transformation law of a weak Jacobi form.
Furthermore, the Gopakumar-Vafa formula \eqref{eqn:gvexpansion} implies that the partition function is invariant under shifts $z\rightarrow z+1$ and therefore admits an expansion in terms of $y=\exp(2\pi i z)$.
The elliptic transformation law for shifts of $z$ by $\tau$ follows by combining the modular transformation law and invariance under constant shifts.
Assuming validity of the wave function equation we have therefore proven that the coefficients $Z_d(\tau,z)$ are weak Jacobi forms of weight $0$ and index $d(d-3)/2$ with a multiplier system.
An analogous argument can be made for other elliptic and genus one fibrations that do not exhibit any fibral divisors or additional (multi-)sections.
The deriviation of the corresponding modular anomaly equations will be performed in Section~\ref{ssec:modan}.

For elliptic and genus-one fibrations with reducible fibers the exponentiated volumes of curves in the base transform like lattice Jacobi forms with non-trivial index matrix.
It is not clear to us how the above argument that relates the relative conifold transformation to the modular properties of the partition function on $X_{18}$ can be generalized.
However, it was found that generalizations of the Huang-Katz-Klemm conjecture hold for geometries with fibral divisors~\cite{DelZotto:2017mee}, multiple sections~\cite{Lee:2018urn,Lee:2018spm} and also for the refined topological string~\cite{Gu:2017ccq,DelZotto:2017mee}.
In this paper we study many more examples, including genus-one fibrations with $N$-sections.
We find that
\begin{align}
	Z_\beta'=Q^\beta\cdot Z_\beta(\tau,\vec{m},\lambda)\,,
\end{align}
always transforms like a lattice Jacobi form under $\Gamma_1(N)$.
The weight is generically zero and the index with respect to the topological string coupling $\lambda$ is $\frac12\beta\cdot(\beta-c_1(B))$.
The index matrix of $Z_\beta'$ with respect to the geometric elliptic parameters $\vec{m}$ is zero.

We conjecture that if the gauge group is fully Higgsable then the topological string partition function is of the form given in~\eqref{eqn:jacAns1}.
The ansatz~\eqref{eqn:jacAns2} for genus-one fibrations with $N$-sections where $N\in\{2,3,4\}$ then follows from the argument that we outline below.

\section{The modular bootstrap for elliptic and genus one fibrations}
\label{sec:modularbootstrap}
In this section we want to generalize the modular bootstrap that has been developed for elliptic fibrations to genus fibered Calabi-Yau threefolds.
Based on the results from the previous section it is already clear that instead of modular forms for $SL(2,\mathbb{Z})$ we will need to consider congruence subgroups $\Gamma_1(N)$.
We start with a review of modular forms, Jacobi forms and the modular bootstrap for elliptic fibrations.
By considering Higgs transitions and the corresponding relations among topological string partition functions we are then going to obtain the modular ansatz for genus one fibrations.
We will then analyse the base degree zero contributions and find closed expressions for genus one fibrations but also for elliptic fibrations with reducible fibers.
Finally, we will study the modular anomaly equations.

\subsection{Rings of modular forms and Jacobi forms for $\Gamma_1(N)$}
\label{ssec:ringsofforms}
Much of the following will be well known to the reader.
Nevertheless, in the case of modular and Jacobi forms for congruence subgroups some details will be crucial for our discussion
and for this reason we do not relegate this section to the Appendix.

We recall the definition of the congruence subgroup $\Gamma_1(N)\subseteq SL(2,\mathbb{Z})$ from~\eqref{ssec:relativeconi}
\begin{align}
	\Gamma_1(N)=\left\{\left(\begin{array}{cc}a&b\\c&d\end{array}\right)\in\Gamma\,:\,a,d\equiv 1\,(\text{mod }n)\,,\quad c\equiv 0\,(\text{mod }n)\right\}\,,
\end{align}
and note that $\Gamma_1(1)=SL(2,\mathbb{Z})$.
A modular form $f$ of weight $k$ for the congruence subgroup $\Gamma\subset SL(2,\mathbb{Z})$ is a holomorphic function on the upper half-plane $\mathbb{H}=\{\tau\in\mathbb{C},\,\text{Im}(\tau)>0\}$ that is holomorphic at $\tau\rightarrow i\infty$
and satisfies
\begin{align}
	f\left(\frac{a\tau+b}{c\tau+d}\right)=(c\tau+d)^kf(\tau)\,,\quad\text{for}\,\left(\begin{array}{cc}a&b\\c&d\end{array}\right)\in\Gamma\,.
\end{align}
We denote the vector space of modular forms of weight $k$ for a group $\Gamma$ by $M_k(\Gamma)$ and use the short-hand $M_k(N)$ when $\Gamma=\Gamma_1(N)$.
The corresponding rings of modular forms will be denoted by $M_*(\Gamma)$ or $M_*(N)$.

The Eisenstein series $E_{2k}(\tau),\,k>1$ are modular forms for $SL(2,\mathbb{Z})$ of weight $2k$ and can be written as
\begin{align}
	E_{2k}(\tau)=1+\frac{2}{\zeta(1-2k)}\sum\limits_{n=1}^\infty\frac{n^{2k-1}q^n}{1-q^n}\,.
\end{align}
For $k=1$ one obtains the quasi modular Eisenstein series $E_2(\tau)$ of weight $2$.
The Dedekind $\eta$-function
\begin{align}
	\eta(\tau)=q^{\frac{1}{24}}\prod\limits_{i=1}^\infty (1-q^i)\,,
\end{align}
is also not quite modular but satisfies
\begin{align}
	\eta(\tau+1)=e^{\frac{\pi i}{12}}\eta(\tau)\,,\quad\eta(-1/\tau)=\sqrt{-i\tau}\eta(\tau)\,.
\end{align}
However, $\Delta_{12}(\tau)=\eta(\tau)^{24}$ is a modular form of weight 12 for $SL(2,\mathbb{Z})$ that vanishes as $\tau\rightarrow i\infty$.
To generate the rings of modular forms for congruence subgroups $\Gamma_1(N)$ let us also introduce
\begin{align}
	E_2^{(N)}(\tau)=-\frac{1}{N-1}\partial_\tau\log\left(\frac{\eta(\tau)}{\eta(N\tau)}\right)\,,
	\label{eqn:gene2}
\end{align}
which for any $N$ is a modular form for $\Gamma_0(N)$ and therefore in particular a modular form for $\Gamma_1(N)$.
We can then generate the rings of modular forms for $\Gamma_1(N)$ and $N\in\{1,2,3,4\}$ as
\begin{align}
	\begin{split}
	M_*(1)=&\langle E_4(\tau),\,E_6(\tau)\rangle\,,\\
	M_*(2)=&\langle E_2^{(2)}(\tau),\,E_4(\tau)\rangle\,,\\
	M_*(3)=&\langle E_2^{(3)}(\tau),\,E_4(\tau),\,E_6(\tau)\rangle\,,\\
	M_*(4)=&\langle E_2^{(2)}(\tau),\,E_2^{(4)}(\tau),\,E_4(\tau),\,E_6(\tau)\rangle\,.
	\end{split}
	\label{eqn:modgens}
\end{align}
It is important to note, that for $N=1$ a modular form $f\in M_k(1)$ that exhibits a zero of order $n$ at $\tau\rightarrow i\infty$
can be written as
\begin{align}
	f(\tau)=\Delta_{12}(\tau)^n\cdot f'(\tau)\,,
\end{align}
where $f'(\tau)\in M_{k-12\cdot n}(1)$ does not vanish at infinity.
For $N=2$ the corresponding decomposition of $f\in M_k(2)$ is
\begin{align}
	f(\tau)=\Delta_4(\tau)^n\cdot f'(\tau)\,,
	\label{eqn:n2factorization}
\end{align}
with $f'\in M_{k-4\cdot n}(2)$ and we introduced
\begin{align}
	\Delta_{4}(\tau)=\frac{\eta(2\tau)^{16}}{\eta(\tau)^8}=\frac{1}{192}\left(E_4(\tau)-E_2^{(2)}(\tau)^2\right)\,.
\end{align}
Due to the larger set of generators~\eqref{eqn:modgens} an analogous factorization is not possible for a general $f\in M_*(N)$ with $N>2$.

The theory of Jacobi forms has been developed in~\cite{eichler1985theory} and was extended to multiple elliptic parameters and general finite index subgroups of symplectic groups in~\cite{Ziegler1989}.
We will only be interested in the following special case.
A weak Jacobi form of weight $k$ for $\Gamma_1(N)\subseteq SL(2,\mathbb{Z})$ and with index a symmetric matrix $C\in M_{m\times m}(\frac12\mathbb{Z})$ is a holomorphic function $\phi(\tau,z)\equiv\phi(\tau,z_1,...,z_m)$ on $\mathbb{H}\times\mathbb{C}^m$ that
satisfies the \textit{modular transformation law}
\begin{align}
	\phi\left(\frac{a\tau+b}{c\tau+d},\,\frac{z}{c\tau+d}\right)=(c\tau+d)^k\exp\left(2\pi i\frac{cz^t Cz}{c\tau+d}\right)\phi(\tau,z)\,,
\end{align}
for all $a,b,c,d\in\mathbb{Z}$ with
\begin{align}
\left(\begin{array}{cc}a&b\\c&d\end{array}\right)\in\Gamma_1(N)\,,
\end{align}
as well as the \textit{elliptic transformation law}
\begin{align}
	\phi\left(\tau,z+\lambda\tau+\mu\right)=\exp\left(-2\pi i\left[\lambda^tC\lambda\tau+\lambda^tCz+z^tC\lambda\right]\right)\phi(\tau,z)\,,
	\label{eqn:elliptictrans}
\end{align}
for any $\lambda\in \mathbb{Z}^n$ and $\mu\in\mathbb{Z}^n$.
It admits a Fourier expansion
\begin{align}
	\phi(\tau,z_1,...,z_m)=\sum\limits_{n\ge 0}\sum\limits_{r\in\mathbb{Z}^n}c(n,r)q^n\zeta^r\,,
	\label{eqn:jacobifourier}
\end{align}
with $q=\exp(2\pi i\tau)$ and $\zeta^r=\exp\left(2\pi i z\cdot r\right)$.

Of particular importance will be the weak Jacobi forms
\begin{align}
	\begin{split}
		\phi_{-2,1}(\tau,z)=&-\frac{\theta_1(\tau,z)^2}{\eta(\tau)^6}=(2\pi i z)^2+\frac{1}{12}E_2(\tau)(2\pi i z)^4+\mathcal{O}(z^6)\,,\\
		\phi_{0,1}(\tau,z)=&4\left[\frac{\theta_2(\tau,z)^2}{\theta_2(\tau,0)^2}+\frac{\theta_3(\tau,z)^2}{\theta_3(\tau,0)^2}+\frac{\theta_4(\tau,z)^2}{\theta_4(\tau,0)^2}\right]=12+E_2(\tau)(2\pi i z)^2+\mathcal{O}(z^4)\,.
	\end{split}
\end{align}
of index one and respective weight $-2$ and $0$.
Recall that the Jacobi theta functions are defined as
\begin{align}
	\begin{split}
		\theta_1(\tau,z)=&\vartheta_{\frac{1}{2}\frac{1}{2}}(\tau,z)\,,\quad \theta_2(\tau,z)=\vartheta_{\frac{1}{2}0}(\tau,z)\,,\\
		\theta_3(\tau,z)=&\vartheta_{00}(\tau,z)\,,\quad \theta_4(\tau,z)=\vartheta_{0\frac{1}{2}}(\tau,z)\,,\\
		\vartheta_{ab}(\tau,z)=&\sum\limits_{n=-\infty}^\infty e^{\pi i(n+a)^2\tau+2\pi i z(n+a)+2\pi i b(n+a)}\,.
	\end{split}
	\label{eqn:jacobitheta}
\end{align}
If we denote the vector space of weak Jacobi forms of weight $k$ and index $C$ for a congruence subgroup $\Gamma$ by $J^{\text{\textit{weak}}}_{k,C}(\Gamma)$ then
\begin{align}
	J^{\text{\textit{weak}}}_{k,m}\left(SL(2,\mathbb{Z})\right)=\oplus_{j=0}^m M_{k+2j}\left(SL(2,\mathbb{Z})\right)\left[\phi_{-2,1}(\tau,z)^j,\phi_{0,1}(\tau,z)^{m-j}\right]\,.
\end{align}
When we discuss the refinement of the topological string partition function over $(-1)$-curves we will also need the Jacobi form
\begin{align}
	\begin{split}
		\phi_{-1,\frac12}(\tau,z)=&i\frac{\theta_1(\tau,z)}{\eta(\tau)^3}\,,
	\end{split}
\end{align}
of weight $-1$ and index $1/2$.

If a weak Jacobi form $\phi(\tau,z)$ for $\Gamma_1(N)$ satisfies the stronger condition
\begin{align}
	\phi\left(\tau,z+\frac{1}{N}\right)=\phi(\tau,z)\,,
\end{align}
i.e. $c(n,r)$ in~\eqref{eqn:jacobifourier} vanishes for $r\notin N\mathbb{Z}$, then the elliptic transformation law
follows from the modular transformation law.
This can be easily shown from
\begin{align}
	\begin{split}
		&\phi\left(\tau,z+\lambda\tau+\mu\right)=\phi\left(\tau,z+\lambda\tau\right)\\
		=&\phi\left(\frac{\tilde{\tau}}{1+N\tilde{\tau}},\frac{\tilde{z}}{1+N\tilde{\tau}}+\frac{1}{N}\lambda\right)\,,\quad\text{with}\quad \tilde{\tau}=\frac{\tau}{1-N\tau}\,,\,\tilde{z}=\frac{z}{1-N\tau}-\frac{1}{N}\lambda\,.
	\end{split}
\end{align}

A simple example is $\phi(\tau,z)=\phi'(N\tau,Nz)$, where $\phi'(\tau,z)$ is any weak Jacobi form for $SL(2,\mathbb{Z})$.
We will encounter expressions of the form $\phi'(N\tau,z)$ that satisfy the elliptic transformation law~\eqref{eqn:elliptictrans} only for $\lambda\in N\mathbb{Z}^n$.
They satisfy the definition after substituting $z\rightarrow Nz$ and in an abuse of language we will also refer to those objects as Jacobi forms for $\Gamma_1(N)$.
Note that if the index of $\phi'(\tau,z)$ is $m$ then the index of $\phi'(N\tau,z)$ will be $m/N$.

Another important role will be played by objects of the form
\begin{align}
	\phi(\tau,z)=\phi'(N\tau,\tau,z_1,...,z_{m-1})\,,
	\label{eqn:phiNtau}
\end{align}
where $\phi'$ is again a weak Jacobi form for $SL(2,\mathbb{Z})$ of weight $k$ and we assume that the index $C$ is block diagonal such that $C_{1i}=c\cdot\delta_{i,1}$ for $i=1,...,m$.
It is a priori not clear that $\phi(\tau,z)$ in~\eqref{eqn:phiNtau} is regular at $\tau\rightarrow i\infty$.
However, it is easy to see that $q^c\phi(\tau,z)$ transforms like a Jacobi form for $\Gamma_1(N)$ of weight $k$ and with index matrix $C'_{ij}=C_{i+1,k+1}/N$ for $i,j=1,...,m-1$.
A special case are Jacobi forms that depend on a single elliptic parameter.
In particular, one can check that for any $N$ the functions
\begin{align}
	A_N=\frac{\phi_{0,1}(N\tau,\tau)}{\phi_{-2,1}(N\tau,\tau)}\,,\quad B_N=\frac{1}{q\cdot\phi_{-2,1}(N\tau,\tau)^{N}}\,,
\end{align}
are modular forms of respective weights $2$ and $2N$ under the action of $\Gamma_1(N)$ and are holomorphic at $\tau\rightarrow i\infty$.
Making an ansatz and comparing a sufficient number of coefficients we then derive the relations
\begin{align}
	\begin{split}
		\phi_{0,1}(2\tau,\tau)=&\frac{E_2^{(2)}(\tau)}{\left[q\Delta_{4}(\tau)\right]^{\frac12}}\,,\quad\phi_{-2,1}(2\tau,\tau)=-\frac{\phi_{0,1}(2\tau,\tau)}{E_2^{(2)}(\tau)}\,,\\
		\phi_{0,1}(3\tau,\tau)=&\frac{E_2^{(3)}(\tau)}{\left[q\Delta_6(\tau)\right]^{\frac13}}\,,\quad\phi_{-2,1}(3\tau,\tau)=-\frac{\phi_{0,1}(3\tau,\tau)}{E_2^{(3)}(\tau)}\,,
		\label{eqn:modhiggs}
	\end{split}
\end{align}
as well as
\begin{align}
	\begin{split}
		\phi_{0,1}(4\tau,\tau)=&\frac{1}{\left[q\Delta_8(\tau)\right]^{\frac14}}\frac{E_2^{(2)}+3E_2^{(4)}}{4}\,,\quad\phi_{-2,1}(4\tau,\tau)=-4\frac{\phi_{0,1}(4\tau,\tau)}{E_2^{(2)}+3E_2^{(4)}}\,,
		\label{eqn:modhiggs1}
	\end{split}
\end{align}
where we introduced $\Delta_6\in M_6(3)$ and $\Delta_8\in M_8(4)$ with
{\small
\begin{align}
	\begin{split}
		\Delta_6(\tau)=&\frac{\eta(3\tau)^{18}}{\eta(\tau)^6}=q^2+6q^3+\dots =\frac{1}{2^4\cdot 3^6}\left[7\left(E_2^{(3)}\right)^3-5E_2^{(3)}E_4-2E_6\right]\,,\\
		\Delta_8(\tau)=&\frac{\eta(2\tau)^8\eta(4\tau)^{16}}{\eta(\tau)^8}=q^3+8q^4+\dots\\
		=&\frac{1}{2^{17}\cdot 3^2\cdot 17}\left[187 \left(E_2^{(2)}\right)^4 - 144 \left(E_2^{(4)}\right)^4- 33 E_4^2 -E_6\left( 154 E_2^{(2)} - 144 E_2^{(4)} \right)\right] \,.
	\end{split}
\end{align}
}
The denominators of $\phi_{0,1}(N\tau,\tau)$ in~\eqref{eqn:modhiggs} and~\eqref{eqn:modhiggs1} can be expanded into Eisenstein-like series
\begin{align}
	\left[q\Delta_{2N}(\tau)\right]^{\frac{1}{N}}=\frac{1}{\sigma_1(N-1)}\sum\limits_{k=1}^\infty \sigma_1(N\cdot k-1)q^k\,,
\end{align}
for $N=2,3,4$ where $\sigma_k(d)$ is the divisor function.

\subsection{The modular bootstrap for elliptic fibrations}
\label{ssec:modularbootstrap}
Before we consider genus one fibrations let us review the modular bootstrap for elliptic fibrations.
The K\"ahler form can be expanded as
\begin{align}
	\omega=\tau\cdot (E_0+c_1(B))+\sum_{i=1}^{r}m_i\cdot\sigma(E_i)+\sum_{i=r+1}^{\text{rk}(G)}m_i\cdot D_{f,i-r}+\sum_{i=1}^{b_2(B)}\tilde{t}_i\cdot D_i'\,,
	\label{eqn:kaehlerform}
\end{align}
where $E_0$ is the class of the zero-section, the divisors $E_i,\,i=1,...,r$ correspond to the sections that generate the Mordell-Weil group, $r$ is the rank of the Mordell-Weil group and $\sigma$ is the Shioda map~\eqref{eqn:shioda}.
The fibral divisors are denoted by $D_{f,i},\,i=1,...,\text{rk}(G)$ and the vertical divisors $D_i',\,i=1,...,b_2(B)$ are dual to the curves $C_i=E_0\cdot D_i,\,i=1,...,b_2(B)$.
From the discussion in~\ref{ssec:relativeconi} it is clear that to see the modular structure we need to introduce shifted K\"ahler parameters $t_i,\,i=1,...,h^{1,1}(B)$ that are defined as
\begin{align}
	t_i=\tilde{t}_i+\frac{\tilde{a}_i}{2}\tau\,,\quad\text{with}\quad \tilde{a}_i=\int_Bc_1(B)\cdot \pi(D_i)\,.
	\label{eqn:tshift}
\end{align}
We assume that $M$ is an elliptically fibered Calabi-Yau threefold such that on a generic point on the Coulomb branch the gauge group is Abelian.
In other words, at a generic point in the complex structure moduli space of $M$ there are no fibral divisors.

Then, if we expand the topological string partition function as in~\eqref{eqn:topexp}, the coefficients take the form
\begin{align}
	Z_\beta(\tau,\underline{m},\lambda)=\frac{1}{\eta(\tau)^{12\cdot c_1(B)\cdot\beta}}\frac{\phi_\beta(\tau,\underline{m},\lambda)}{\prod_{l=1}^{b_2(B)}\prod_{s=1}^{\beta_l}\phi_{-2,1}(\tau,s\lambda)}\,.
	\label{eqn:jacAns1}
\end{align}

The numerators $\phi_\beta(\tau,\underline{m},\lambda)$ are Jacobi forms of weight
\begin{align}
	w=6c_1(B)\cdot\beta-\sum_l\beta_l\,,
	\label{eqn:jacobiweight}
\end{align}
and index
\begin{align}
	r^\beta_\lambda=\frac12\beta\cdot(\beta-c_1(B))+\sum_l\frac{\beta_l(\beta_l+1)(2\beta_l+1)}{6}\,,
	\label{eqn:jacobiindex1}
\end{align}
with respect to the elliptic parameter $\lambda$.
The index matrix with respect to the \textit{geometric} elliptic parameters $\underline{m}$ is
\begin{align}
	r^\beta_{ij}=\left\{\begin{array}{cl}
		-\frac12\pi_*\left(\sigma(E_i)\cdot\sigma(E_j)\right)\cdot\beta&\text{ for }1\le i,j\le r\\
		-\frac12\pi_*\left(D_{f,i}\cdot D_{f,j}\right)\cdot \beta&\text{ for }r<i,j\le\text{rk}(G)\\
		0&\text{ otherwise}
	\end{array}\right.\,,
	\label{eqn:jacobiindex2}
\end{align}
where $\beta$ is a curve of volume $t_i$.
Note that
\begin{align}
	b_{ij}=-\pi_*\left(\sigma(E_i)\cdot\sigma(E_j)\right)\,,
\end{align}
is the so-called height pairing of the sections $E_i$ and $E_j$.

The ansatz~\eqref{eqn:jacAns1} is based on the results from~\cite{Huang:2015sta,DelZotto:2017mee,Lee:2018urn,Lee:2018spm} and we provide additional evidence in this paper.
The Dedekind $\eta$-function in the denominator is exactly cancelling the zero at $\tau\rightarrow i\infty$ that comes from $Q^\beta$ due to the shift~\eqref{eqn:tshift}.
Moreover, the product $\prod_{l=1}^{b_2(B)}\prod_{s=1}^{\beta_l}\phi_{-2,1}(\tau,s\lambda)$ is such that for a given degree the associated Gopakumar-Vafa invariants vanish when the genus is greater than some highest value~\cite{Huang:2015sta}.
Assuming that $Z_\beta$ is a meromorphic Jacobi form of the given weight and index therefore seems to imply this ansatz at least modulo some assumptions on the vanishing of enumerative invariants.
Note that a more complicated form of the denominator occured in~\cite{DelZotto:2017mee} but is excluded by our assumption on the Coulomb branch.

It can be further constrained if there are additional dualities for which we know the transformation behaviour of the topological string partition function.
This is usually the case when a monodromy does not act on the 0-brane charge and does not transform 2-branes into higher dimensional branes.
A prime example is the action of the affine Weyl group (see Section~\ref{ssec:weyl}) as was first explored in~\cite{Katz:1996ht}. 
In the context of the modular bootstrap on  non-compact Calabi-Yau that engineer 6d SCFTs, this symmetry was understood as a consequence of properties of elliptic genera~\cite{DelZotto:2017mee}.

More generally, in the complex structure moduli space of an elliptically or genus one fibered Calabi-Yau $M$ there is a sublocus such that $M$ remains non-singular and the gauge group $G'$ associated to $M$ is ``maximally non-Abelian''.
On this locus the vacuum expectation values of all hypermultiplets in the adjoint representation of non-Abelian factors of $G'$ are set to zero.
The affine Weyl groups of the non-Abelian factors of $G'$ act on the geometric elliptic parameters $\underline{m}$ and $Z_\beta(\tau,\underline{m},\lambda)$ is invariant under this action.

\subsection{Closed expressions for $Z_{\beta=0}$}
\label{ssec:closedexpressions}
We will now discuss the base degree zero contributions.
For the readers convencience let us recall that the Gopakumar-Vafa formula expresses the sum of the free energies in terms of integer invariants $n_\beta^g$ via
\begin{align}
	\log(Z)=\sum\limits_{g=0}^\infty \lambda^{2g-2}F_g=\sum\limits_{\beta\in H_2(M,\mathbb{Z})}\sum\limits_{g=0}^\infty\sum\limits_{m=1}^\infty \frac{n_\beta^g}{m}\left(2\sin\left(\frac{m\lambda}{2}\right)\right)^{2g-2}q^{\beta m}\,,
	\label{eqn:gvexpansion}
\end{align}
where we have omitted the classical terms.
The crucial observation by~\cite{Huang:2015sta} is that the only non-vanishing Gopakumar-Vafa invariants with degree zero in the base are at genus zero and genus one.
All the free energies for $g\ge 2$ are therefore entirely determined by multi-covering contributions from the genus zero curves.
The corresponding sum can be evaluated using
\begin{align}
	\sum\limits_{m=1}^\infty\frac{1}{m}\frac{q^m}{\left(2\sin\left(\frac{m\lambda}{2}\right)\right)^2}=\lambda^{-2}\cdot\text{Li}_3(q)+\sum\limits_{g=1}^\infty\lambda^{2g-2}(-1)^{g+1}\frac{B_{2g}}{2g[(2g-2)!]}\text{Li}_{3-2g}(q)\,.
	\label{eqn:multi}
\end{align}
Moreover, the contributions of constant maps are given by
\begin{align}
	F_g^{const}=(-1)^{g}\chi\frac{B_{2g}B_{2g-2}}{4g(2g-2)[(2g-2)!]}\,,\quad\text{for}\quad g\ge 2\,.
\end{align}

An analysis for elliptic fibrations without reducible fibers has been performed in~\cite{Huang:2015sta} and we will start with a review of this simpler situation.
The instanton contribution to the genus zero free energy then takes the form
\begin{align}
	F_0^{inst}=-\chi\cdot\sum\limits_{i=1}^\infty\text{Li}_3(q^i)+\mathcal{O}(Q)\,,
\end{align}
where we have used $Q$ to denote K\"ahler moduli of curves in the base and using
\begin{align}
\begin{split}
		E_{2g-2}(\tau)
		=&1-\frac{2(2g-2)}{B_{2g-2}}\sum\limits_{i=1}^\infty \text{Li}_{3-2g}(q^i)=1+\mathcal{O}(q)
\end{split}
\end{align}
it follows that the higher genus free energies are
\begin{align}
	F_{g\ge2}=(-1)^g\chi\frac{B_{2g}B_{2g-2}}{4g(2g-2)(2g-2)!}E_{2g-2}(q)+\mathcal{O}(Q)\,.
\end{align}
The only non-vanishing Gopakumar-Vafa invariants at genus 1 are in multiples of the class of the generic fiber and given by the Euler characteristic of the base $n^1_{T^2}=\chi_B$.
However, the corresponding free energy also receives multi-covering contributions~\eqref{eqn:multi} and reads
\begin{align}
	F_1^{inst}=\left(\chi_B-\frac{\chi}{12}\right)\cdot\sum\limits_{i=1}^\infty\text{Li}_1(q^i)\,.
\end{align}

The closed modular expressions for the base degree zero contributions to the higher genus free energies of geometries without reducible fibers rely on the somewhat miraculous interplay between geometric and number theoretic formulas.
It is therefore particularly interesting to see how the generalization of these expressions can be derived for fibrations with reducible fibers that lead to additional elliptic parameters.

\paragraph{Geometries with reducible fibers}
In our more general setup the only non-vanishing Gopakumar-Vafa invariants for base degree zero still arise at genus zero and genus one.
Instead of aiming for full generality we illustrate the situation for geometries with reducible fibers at the example of $M_2^{(2)}=\left(F_{6}\rightarrow\mathbb{P}^2\right)\left[U(1)\right]_3^{-216}$ (see Section~\ref{sec:m22} for a more detailed discussion of this geometry).
The fibration has a holomorphic zero-section $s_0$ and a rational section $s_1$ that generates the Mordell-Weil group.
Here one finds that the base degree zero contribution $F_{0,\beta=0}$ to the genus zero free energy $F_0=F_{0,\beta=0}+\mathcal{O}(Q)$ takes the form
\begin{align}
	\begin{split}
		F_{0,\beta=0}=&-\chi\cdot\sum\limits_{i=1}^\infty \text{Li}_3(q^i)\\
		&144\cdot\text{Li}_3(y)+144\cdot \sum\limits_{i=1}^\infty\left(\text{Li}_3(q^{i}y^{-1})+\text{Li}_3(q^{i}y)\right)\\
		&18\cdot\text{Li}_3(y^{2})+18\cdot \sum\limits_{i=1}^\infty\left(\text{Li}_3(q^{i}y^{-2})+\text{Li}_3(q^{i}y^2)\right)\,.
	\end{split}
	\label{eqn:f0polylogs}
\end{align}
where $q=\exp(2\pi i\tau),\,y=\exp(2\pi im)$ and $m$ is the volume of the isolated rational fibral curves that lead to charge one hypermultiplets.

More precisely, a contribution of $\text{Li}_3(y)$ arises from curves that intersect $s_1$ transversely but do not intersect $s_0$ while $\text{Li}_3(y^{2})$ stems from curves $C$ that intersect as $C\cdot s_1=-1$ and $C\cdot s_0=0$.
The $U(1)$ charge of a fibral curve is counted by the intersection with $s_1-s_0$ while the Kaluza-Klein charge corresponds to the intersection with $s_0$~\footnote{This statement is slightly modified in the presence of fibral divisors.}.
Therefore the geometry contains $144$ curves that lead to matter with $U(1)$ charge one while another $18$ fibral curves lead to hypermultiplets of charge two. Moreover, $\chi=-216$ is the Euler characteristic of $M_3^{(2)}$. 
The multiplicity of matter representations is thus directly encoded in the enumerative invariants of the geometry.

We will now derive closed expressions for the free energies at genus $g\ge 2$.
To avoid unnecessary prefactors let us also introduce $y=\exp(z)$.
It follows from basic properties of the polylogarithm that
\begin{align}
	\text{Li}_{s}(q^i y^b)=\text{Li}_s(q^i e^{bz})=\sum\limits_{m=0}^\infty \frac{z^m}{m!}\partial_z^m\text{Li}_s(q^i e^{bz})=\sum\limits_{m=0}^\infty \frac{(bz)^m}{m!}\text{Li}_{s-m}(q^i)\,,
\end{align}
and therefore
\begin{align}
	\begin{split}
	\sum\limits_{i=1}^\infty \left(\text{Li}_{3-2k}(q^{i}y^{-b})+\text{Li}_{3-2k}(q^{i}y^{b})\right) =2\sum\limits_{i=1}^\infty\sum\limits_{m=0}^\infty \frac{(bz)^{2m}}{(2m)!}\text{Li}_{3-2k-2m}(q^i)
	\end{split}
	\label{eqn:plidentity1}
\end{align}
We can then use the expansion
\begin{align}
	\text{Li}_s(e^z)=\Gamma(1-s)(-z)^{s-1}+\sum\limits_{k=0}^\infty \frac{\zeta(s-k)}{k!}z^k\,,
\end{align}
which is valid for integer $s\le0$ and $|z|<2\pi$ as well as the fact that
\begin{align}
	\zeta(-n)=(-1)^n\frac{B_{n+1}}{n+1}\,,
\end{align}
for positive integers $n$ to obtain
\begin{align}
	\text{Li}_{3-2k}(e^{bz})=(2k-2)!(bz)^{2-2k}-\sum\limits_{m=0}^\infty\frac{B_{2m+2k-2}}{2m+2k-2}\frac{(bz)^{2m}}{(2m)!}
	\label{eqn:plidentity2}
\end{align}
where we have used that the Bernoulli numbers $B_n$ vanish for odd $n>1$.
Putting \eqref{eqn:plidentity1} and \eqref{eqn:plidentity2} together we find that
\begin{align}
	\begin{split}
		\Phi^b_{2k-2}(\tau,z)\equiv&\text{Li}_{3-2k}( y^b)+\sum\limits_{i=1}^\infty \left(\text{Li}_{3-2k}(q^{i}y^{-b})+\text{Li}_{3-2k}(q^{i}y^{b})\right)\\
		=&(2k-2)!(bz)^{2-2k}-\sum\limits_{m=0}^\infty\frac{B_{2m+2k-2}}{2m+2k-2} E_{2k+2m-2}\frac{(bz)^{2m}}{(2m)!}\,.
	\end{split}
\end{align}
It is easy to see that $\Phi^b_{2k-2}(\tau,z)$ is a weak Jacobi form of weight $2k-2$ and index $0$.
In total the base degree zero parts of the free energies at genus $g\ge 2$ are therefore given by
\begin{align}
	F_{g\ge2,\beta=0}=(-1)^g\chi\frac{B_{2g}B_{2g-2}}{4g(2g-2)(2g-2)!}E_{2g-2}(q)-(-1)^g\frac{B_{2g}}{2g[(2g-2)!]}\sum\limits_{q=1}^\infty n_q\Phi_{2g-2}^q\,,
\end{align}
where $n_q$ is the number of hypermultiplets of charge $q$.

\paragraph{Genus-one fibrations with multi-sections}
Let us now consider the same problem for fibrations without a section. 
Again, the only non-vanishing Gopakumar-Vafa invariants for base degree zero arise at genus zero and genus one.
We will restrict to the case that the gauge group is $G=\mathbb{Z}_N$ which means that there is only one linearly independent $N$-section and there are no fibral divisors.

We start with $N=4$ where the genus zero free energy takes the form
\begin{align}
	F_0= \sum_{i=0}^\infty\left[n_1\cdot\text{Li}_3(q^{1+4i})+n_2\cdot\text{Li}_3(q^{2+4i})+n_3\cdot\text{Li}_3(q^{3+4i})+n_4\cdot\text{Li}_3(q^{4+4i})\right]\,,
\end{align}
where $n_m$ is the number of fibral curves that intersect the $4$-section $m$ times.
Since the generic fiber intersects the $4$-sections four times it is clear that $n_1=n_3$ and $n_4=-\chi$ where $\chi$ is again the Euler-characteristic of the Calabi-Yau.
The expression can therefore be rewritten as
\begin{align}
	F_0=-\chi\cdot\sum\limits_{i=1}^\infty\text{Li}_3(q^{4i})+n_1\cdot\sum\limits_{i=1}^\infty\text{Li}_3(q^i)+(n_2-n_1)\cdot\sum\limits_{i=1}^\infty\text{Li}_3(q^{2i})\,.
\end{align}
It is easy to see that for an $N$-section geometry we have multiplicities $n_i,\,i=1,...,N$ and we expect to be able to write
\begin{align}
	F_0=\sum\limits_{m=1}^Nn_m\sum\limits_{i=0}^\infty\text{Li}_3(q^{N\cdot i+m})=\sum\limits_{m=1}^{N}n_m'\cdot\sum\limits_{i=1}^\infty\text{Li}_3(q^{m i})\,,
	\label{eqn:f0multiidentity}
\end{align}
where the coefficients $n_m',\,m=1,...,N$ are defined recursively via
\begin{align}
	n_m'=n_m-\sum\limits_{k<m,k|m}n_k'\,.
\end{align}
However, if we assume that the independent multiplicities are generic, this is only possible when all $0<k<\lfloor\frac{N}{2}\rfloor$ divide $N$.
This is true for the five cases $N=1,2,3,4,6$.
For other values of $N$ the identity~\eqref{eqn:f0multiidentity} severely constrains the spectrum of charged hypermultiplets.
If it can be satisfied we find that
\begin{align}
	\sum\limits_{m=1}^Nn_m'=n_N\,.
\end{align}
and it is always true that $n_N=-\chi$.
We can than use essentially the same argument that worked for elliptic fibrations without reducible fibers to obtain
\begin{align}
	F_{g\ge2}=(-1)^{g+1}\frac{B_{2g}B_{2g-2}}{4g(2g-2)(2g-2)!}\sum\limits_{m=1}^Nn_m' E_{2g-2}(m\tau)+\mathcal{O}(Q)\,.
\end{align}
It might be that modularity for general $N$ arises from a more complicated relation between polylogarithms and modular forms for $\Gamma_1(N)$.
Another possiblity is that in those cases the base degree zero contribution to the higher genus free energies is not modular at all.
Perhaps the most exciting, although entirely speculative resolution would be that the identity \eqref{eqn:f0multiidentity} puts a genuine constraint on the curves and intersections in those geometries.
\subsection{Modular anomaly equations}
\label{ssec:modan}
In this section our aim is to derive the index $r_\beta^\lambda$ of the topological string partition function $Z_{\text{top}}$ with respect to the topological string coupling constant $\lambda$.
As was explained in Section~\ref{ssec:monaut}, this is captured by a modular anomaly equation of the form~\eqref{eqn:difE2}.

For ordinary elliptic fibrations, modular anomaly equations have also been derived via duality with elliptic genera of strings and the anomaly inflow mechanism in a 6d supergravity theory \cite{Gu:2017ccq,DelZotto:2017mee,DelZotto:2018tcj,Lee:2018urn}.
Here we extend the discussion of~\cite{Klemm:2012sx,Alim:2012ss} and argue that whenever $M$ is elliptic or genus one fibered without fibral divisors or additional (multi-)sections, a modular anomaly equation can be derived from the background independence equations introduced in~\cite{Witten:1993ed}.
Strictly speaking, a modification of the background independence equation is necessary that will be discussed in a seperate paper which is written by a different set of authors~\cite{HKKprep}.
For the reader uninterested in technical details we summarize the main result in the paragraph below.

For a Calabi-Yau threefold $M$ that exhibits a genus one fibration with $N$-section such that $h^{1,1}(M) = h^{1,1}(B)+1$, the coefficient $Z_\beta(\tau,\lambda)$ in~\eqref{eqn:topexp} satisfies
\begin{equation}\label{eq:indextop}
\Bigg(\frac{\partial}{\partial E_2} + \frac{(2\pi i\lambda)^2 }{12}r_\beta^\lambda\Bigg) Z_\beta = 0\,,
\end{equation}
where
\begin{equation}\label{eq:topindex}
r_\beta^\lambda = \frac{1}{2N}\beta\cdot\big(\beta - c_1(B)\big)\,.
\end{equation}
This implies that it is a meromorphic Jacobi form for $\Gamma_1(N)$ of weight $k=0$ and index $r_\beta^\lambda$ with respect to the topological string coupling $\lambda$.
Since the topological string partition function is independent of complex structure deformations, this derivation also applies
to arbitrary genus one fibered Calabi-Yau 3-folds where the gauge group can be completely Higgsed to a discrete group.
Together with the derivation of the index matrix with respect to the geometric elliptic parameters we therefore derive all indices directly from properties of the geometry.

Recall that we denote by $W$ the mirror dual of $M$.
The topological string partition function $\widehat{Z}$~\footnote{We reserve the hat notation for the topological string partition function in holomorphic polarization~\cite{Aganagic:2006wq}.} is identified with a wave function $\Psi$ on a quantum Hilbert space.
The latter is obtained by quantizing the phase space $H_3(W,\mathbb{R})$ with symplectic form $\Sigma$ given by its intersection pairing.
Quantum background independence of $\widehat{Z}$ on the complex moduli space $\mathcal{M}$ of $W$ leads to the heat equation~\cite{Witten:1993ed} (see also~\cite{Verlinde:2004ck,Aganagic:2006wq,Gunaydin:2006bz})
\begin{align}\label{eqn:waveft}
\Bigg(\frac{\partial}{\partial t^{\bar{a}} } - \frac{\lambda^2}{2} C_{\bar{a}}^{bc} \frac{D}{D t^a} \frac{D}{D t^b} \Bigg) \widehat{Z} = 0\,.
\end{align}
Here the coupling $C_{\bar{a}}^{bc}$ is a section of $\mathcal{L}^{-2} \otimes \text{Sym}^2\Big(T^{(1,0)}_{\mathcal{M}} \Big) \otimes T_{\mathcal{M}}^{*(0,1)}$, where $\mathcal{L}$ is the K\"ahler line bundle on $\mathcal{M}$ with fibers $H^{3,0}(W)$.
More precisely,~\eqref{eqn:waveft} was derived as an infinitesimal consequence of the freedom of choice of polarization on $H_3(W,\mathbb{R})$ and a change on $\bar{t}^a$ acts on the wave function $\hat{Z}$ by a Bogoliubov transformation.
A reformulation of the latter transformation is proposed in~\cite{HKKprep}. However, in the limit case we consider, our result matches with this proposal.

In the following we introduce some quantities, which follow from special geometry over the complex moduli space $\mathcal{M}$ of  $W$.
The sections of the K\"ahler line bundle $\mathcal{L}$ are holomorphic three forms $\Omega$ varying holomorphically over $\mathcal{M}$.
Moreover, one can choose a symplectic basis in cohomology $\alpha_I, \beta^I \in H^3(W,\mathbb{Z})$ and a dual homology basis $A^I, B_I \in H_3(W,\mathbb{Z}), I = 0, \ldots, h^{2,1}(W)$, with pairings $\int_W \alpha_I \wedge \beta^J = - \int_W \beta^J\wedge\alpha^I = \int_{A^J} \alpha_I = -\int_{B_I}\beta^J = \delta_I^J$ and $\int_{A^J}\beta^I=\int_{B_I}\alpha^J=0$.
Then for a given $z \in \mathcal{M}$ the holomorphic three-form can be expanded as
\begin{align}
\Omega(z) = X^I(z) \alpha_I - F_I(z) \beta^I \,.
\end{align}
As a consequence of special geometry one can write the periods as
\begin{align}
\vec{\Pi} = \begin{pmatrix} X^I  \\ F_I \end{pmatrix} = X^0 \begin{pmatrix} 1 \\ t^k \\ 2 \mathcal{F}^{(0)} -t^k \frac{\partial}{\partial t^k} \mathcal{F}^{(0)} \\ \frac{\partial}{\partial t^k} \mathcal{F}^{(0)} \end{pmatrix} \,,
\end{align}
where $\mathcal{F}^{(0)}$ is the holomorphic prepotential $F(X) = (X^0)^2 \mathcal{F}^{(0)}(\underline{t})$ and one can introduce the flat coordinates
\begin{align}
t^a = \frac{X^a}{X^0} = \frac{1}{2\pi i} \log(z^a) +\mathcal{O}(\underline{z}) \,.
\end{align}
The K\"ahler potential $K$ is related to the holomorphic $3$-form $\Omega$ via
\begin{align}\label{eq:Kahler}
e^{-K} = i \int_W \Omega \wedge \bar{\Omega} \,, 
\end{align}
which can be expressed in terms of the periods as
\begin{align}
	e^{-K}= i\vec{\Pi}^\dagger \Sigma \vec{\Pi} = 4\vert X^0 \vert^2 \left[ \text{Im} \left(\mathcal{F}^{(0)}\right) -t_2^a \text{Re} \left( \partial_{t^a} \mathcal{F}^{(0)} \right) \right] \,.
\end{align}
Here $t_2^a := \text{Im} (t^a)$. A straightforward calculation using~\eqref{eq:Kahler} provides the leading terms
\begin{equation}\label{eq:classKahler}
e^{-K} = \frac{4}{3} c_{abc} t_2^a t_2^b t_2^c + \frac{\zeta(3) \chi (M)}{4 \pi^3} +\mathcal{O}(\underline{Q}, \underline{\bar{Q}})\,,
\end{equation}
where $c_{abc}$ denotes the classical intersection numbers on $M$ and $Q^a=\exp(2\pi i t^a)$. The Weil-Petersson metric follows from the K\"ahler potential and reads $G_{a\bar{b}}=\partial_a \partial_{\bar{b}} K$. Moreover, the coupling $C_{\bar{a}}^{bc}$ is related to the anti-holomorphic Yuakawa coupling $\bar{C}_{\bar{a}\bar{b}\bar{c}} \in \bar{\mathcal{L}}^2\otimes \text{Sym}^3\Big(T_{\mathcal{M}}^{*(0,1)}\Big)$, the inversion of the Weil-Petersson metric and the K\"ahler potential via
\begin{align}\label{eqn:3coup}
C_{\bar{a}}^{bc} = e^{2K} \bar{C}_{\bar{a}\bar{b}\bar{c}} G^{a\bar{a}} G^{\bar{b}b}\,.
\end{align}

In the remainder of this section we will assume that $M$ is a genus-one fibered Calabi-Yau threefold without fibral divisors or additional (multi-)sections.
In particular, we use the parametrization of the K\"ahler form~\eqref{eqn:kaehlerexpansion}
\begin{align}
	\omega=\tau\cdot\left(\tilde{E}_0-\frac{1}{2N}\sum_i \tilde{a}_iD'_i\right)+t^iD_i'
\end{align}
where $\tilde{E}_0$ and $D'_i,\,i=1,...,h^{1,1}(B)$ have been defined in Section~\ref{ssec:relativeconi}.
The only complexified K\"ahler moduli are $\tau$ and $t^i,\,i=1,...,h^{1,1}(B)$ where $\tau$ parametrizes the volumes of isolated fibral curves while the shifted K\"ahler parameters $t^i$ parametrize the volumes of curves in the base.

As pointed out in Section~\ref{ssec:monaut}, we are interested in the limit  where $\widehat{Z}$ exhibits its anholomorphic behaviour exclusively due to polynomials in $(\text{Im}\,\tau)^{-1}$.
In addition to that, we consider the base parameter limit  $\text{Im}\,{t}^i \sim \frac{1}{h} \rightarrow \infty$,  where $h$ is some real order parameter close to zero, while keeping the fiber parameter $\text{Im}\,\tau$ finite.
We refer to the latter limit as the \textit{small fiber limit}\footnote{Recall from section \ref{ssec:modgenusone} that only in the small fiber limit the exponentiated K\"ahler parameters of the base transform like Jacobi forms under relative Conifold transformation.}.
At the end of the day, the quantity of main interest will be the topological string partition function in the \textit{holomorphic limit}
\begin{align}
	Z(\tau,\underline{t},\lambda) =\lim_{\text{Im}\,\underline{t},\tau \to \infty} \widehat{Z}(\tau,\bar{\tau},\underline{t},\underline{\bar{t}},\lambda)\,.
\end{align}

Let us define
\begin{align}
	\hat{E}_0=\tilde{E}_0-\frac{1}{2N}\sum_i \tilde{a}_iD'_i\,,
\end{align}
and denote with $c_{\tau a b}$ the classical intersection matrix given by the intersections $\hat{E}_0\cdot J_a\cdot J_b$ where $J_a\in\left\{\hat{E}_0,D'_{i=1,...,h^{1,1}(B)}\right\}$.
We find that
 \begin{align}\label{eq:classinttau}
c_{\tau ab} &=\left(
    \begin{array}{c;{2pt/2pt}c}
      \alpha & \vec{0}^T \\ \hdashline[2pt/2pt]
      \vec{0}     & Nc_{ij}   \end{array}
\right)\,.
\end{align}
Here $c_{ij} = \tilde{D}'_i\cdot \tilde{D}'_j$ is the intersection form on the base $B$ with $D'_i=\pi^{-1}\tilde{D}'_i$.
The $N$ factor arises due to the intersection with the zero-$N$-section.
Moreover, $c_{\tau\tau\tau} = \alpha$ is a constant that will drop out from the calculation.
With the information obtained from (\ref{eq:classinttau}), (\ref{eq:classKahler}) and taking the appropriate inversions of $G_{a\bar{a}}$, we are able to compute the coupling $C_{\bar\tau}^{ab}$~\eqref{eqn:3coup} in the small fiber limit.
Putting everything together, the result reads
\begin{align}
C_{\bar\tau}^{ab} 
&=\left(
    \begin{array}{c;{2pt/2pt}c}
       \alpha\frac{\tau_2^2}{V^2} h^4 + \mathcal{O}(h^5) & \frac{1}{6V^2 \tau_2^2}(2\alpha \tau_2^3 - \widehat{\chi})Nt_2^j h^3 +\mathcal{O}(h^5) \\ \hdashline[2pt/2pt]
        \frac{1}{6V^2 \tau_2^2}(2\alpha \tau_2^3 - \widehat{\chi})Nt_2^i h^3 +\mathcal{O}(h^5)          & \frac{1}{4N \tau_2^2 }c^{ij}  + \mathcal{O}(h)  \end{array}
\right)\,.
\end{align}
Here $V = N c_{ij}t^i t^j$, $\widehat{\chi} =  \frac{\zeta(3) \chi (M)}{4 \pi^3}$, and $c^{ij}$ is the inverse of $c_{ij}$.

With this information at our disposal, we proceed applying the small fiber limit on the wavefunction equation~\eqref{eqn:waveft}. 
Denote by $\mathcal{L}_W$ the differential operator in (\ref{eqn:waveft}). In the small fiber limit, Witten's wave function equation reduces to 
\begin{equation}
\lim_{h \to 0} \mathcal{L}_W  = \frac{\partial}{\partial\hat{E}_2} + \frac{\lambda^2}{24 N }c^{ij}\frac{\partial}{\partial t^i }\frac{\partial}{\partial t^j}\,,
\end{equation}
where we used the derivative relation $\partial_{\bar{\tau}} =- \frac{3}{2\pi i \tau_2^2}\partial_{\hat{E}_2}$.
Considering the Fourier expansion of $\widehat{Z}$ in the base parameters around the small fiber limit
\begin{equation}
\lim_{h \to 0} \widehat{Z} = \widehat{Z}_0\Bigg( 1 + \sum_{\beta \in H_2(B,\mathbb{Z}) } \widehat{Z}_\beta(\tau,\bar{\tau}) Q^\beta\Bigg)\,.
\end{equation}
we find that
\begin{equation}\label{eq:intermediate}
\lim_{h \to 0} \mathcal{L}_W\Big( \widehat{Z}_0 Q^\beta \Big)=\Bigg[\frac{1}{2N} \Bigg(\beta - \frac{c_1(B)}{2}\Bigg)^2 - \frac{1}{5760}\sum_{m = 1}^N \frac{n_m}{N}\Bigg] Q^\beta\,.
\end{equation}

The first term on the right hand side of~\eqref{eq:intermediate} comes from the base derivatives in $\mathcal{L}_W$ acting on $Q^{\beta-\frac{c_1(B)}{2}}$.
The shift $c_1(B)$ appears due to the classical contributions from $F_{g=1}$ within $\widehat{Z}_0$ which read
\begin{equation}
	F_{g=1} \Big\vert_{\text{class}} = -\frac{1}{24} \sum_a  t^a \int_M c_2(M) \cdot D_a' =-\frac12\sum_a t^a\int_Bc_1(B)\tilde{D}_a'\,.
\end{equation}

The second term in~\eqref{eq:intermediate} appears due to the action of $\partial_{\hat{E}_2}$ on the base degree zero contribution from $F_{g=2}$
which we derived in the previous section.
It can be separated into a purely modular part and a quasi-modular part. The latter reads
\begin{equation}\label{eq:g2quasi}
F_{g=2,\beta=0} \Big\vert_{\text{quasi}} = - \frac{1}{5760} \sum_{n=1}^N \frac{n_m}{N} E_2(\tau)\,. 
\end{equation}
Performing the map $E_2 \mapsto \hat{E_2}$  we recover the non-holomorphic counterpart of~\eqref{eq:g2quasi} in $\widehat{Z}_0$.
Using the pure gravitational anomaly constraints for an effective supergravity in a 6d theory, which read
\begin{align}
\begin{split}
H-V + 29T = 273 \,, \quad 9-T& = c_1^2(B)\,,
\end{split}
\end{align}
 we are able to verify
\begin{equation}
\sum_{m=1}^N n_m = 60 c_1^2(B) \,.
\label{eqn:piece2}
\end{equation}

Finally, joining the pieces (\ref{eq:intermediate}) and (\ref{eqn:piece2}) we take the coefficient $Q^\beta$  of $\lim_{h\to 0} \mathcal{L}_W \widehat{Z}$. The result reads
\begin{equation}
\Bigg( \frac{\partial}{\partial \hat{E}_2} + \frac{(2\pi i \lambda)^2}{24 N } \beta\cdot\big(\beta - c_1(B)\big) \Bigg) \widehat{Z}_\beta (\tau,\bar{\tau})= 0\,.
\end{equation}
Taking the holomorphic limit $\text{Im}\,\tau \rightarrow \infty$, we obtain~\ref{eq:indextop}.

\subsection{The modular ansatz for genus one fibrations with $N$-sections}
\label{ssec:modgenusone}
We will now derive the modular ansatz for the topological string partition function on genus one fibered Calabi-Yau threefolds that do not have a section but only $N$-sections.
For reasons that have been described in Section~\ref{ssec:relativeconi} we restrict ourselves to $N\in\{2,3,4\}$.
Then our analysis of the monodromies in the stringy K\"ahler moduli space led to the conjecture that $Q^\beta Z_\beta$ transforms like a weak Jacobi form of weight zero and index $\frac{1}{2N}\beta\cdot(\beta-c_1(B))$ under the action of $\Gamma_1(N)$.

We could now try to derive the denominator of the ansatz following the argument for elliptic fibrations to justify~\eqref{eqn:jacAns1}.
However, this is complicated by the fact that the rings of modular forms for $\Gamma_1(N)$ in general contain multiple irreducible elements that vanish at $\tau\rightarrow i\infty$.
Let us therefore follow a different route and derive the ansatz by considering Higgs transitions.

We will first recall the geometric transition that corresponds to Higgsing an Abelian gauge group into a discrete gauge group within F-theory.
This will allow us to identify a good set of K\"ahler parameters on the multi-section geometry and we see that the relative conifold monodromy and a large volume monodromy generate an action of a congruence subgroup $\Gamma_1(N)$ of $PSL(2,\mathbb{Z})$ on the moduli.
We will then use the Higgs transition and the previously conjectured modular properties of the topological string partition function on geometries with multiple sections
to show that the topological string partition function on fibrations that only have multi-sections can be expressed in terms of Jacobi forms for $\Gamma_1(N)$.

It is well known that via F-theory multi-section geometries lead to discrete gauge symmetries.
Moreover, general arguments of quantum gravity imply that discrete gauge symmetries are always the remnant of a Higgsed continuous gauge symmetry.
It is therefore expected that for every multi-section geometry $X'$ there exists a geometry $X$ with multiple sections such that $X'$ can be obtained
from $X$ via a conifold transition.
More precisely, the geometric transition relates the five dimensional theories that are obtained after compactifying the F-theory vacua on a circle.
First the volume of an isolated fibral curve shrinks to zero such that the corresponding hypermultiplets of $U(1)$ charge $q$ become massless.
Note that the volumes of fibral curves correspond to gauge fluxes along the $S^1$.
A subsequent complex structure deformation can be interpreted as giving a vacuum expectation value to the massless scalar fields in the hypermultiplets.

\begin{figure}[h!]
	\centering
	\begin{tikzpicture}[remember picture,overlay,node distance=4mm]
		\node[] at (0.3,.2) {$C_{-1}$};
		\node[] at (4,.2) {$C_{1}$};
		\node[] at (5.2,.2) {$C_{2}$};
		\node[] at (8.9,.2) {$C_{-2}$};
		\node[] at (10,.2) {$C_{3}$};
		\node[] at (13.7,.2) {$C_{-3}$};
	\end{tikzpicture}
	\includegraphics[width=.28\linewidth]{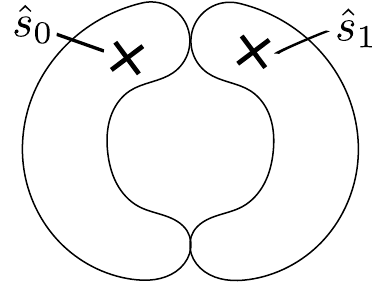}\hspace{.4cm}
	\includegraphics[width=.28\linewidth]{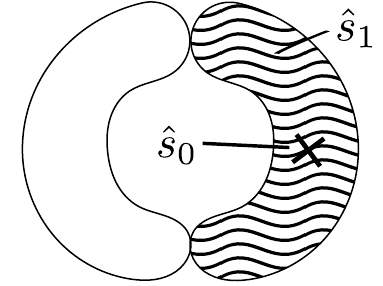}\hspace{.4cm}
	\includegraphics[width=.28\linewidth]{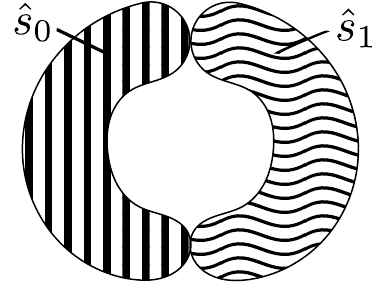}
	\caption{Intersections between the two rational sections and the components of the three types of isolated fibral curves in fibrations constructed from $dP_1$ (polytope $F_3$).
	Shaded components are wrapped by the rational section while crosses indicate a transverse intersection.
	The figures are taken from \cite{Klevers:2014bqa}.}
	\label{fig:f3codim2curves}
\end{figure}
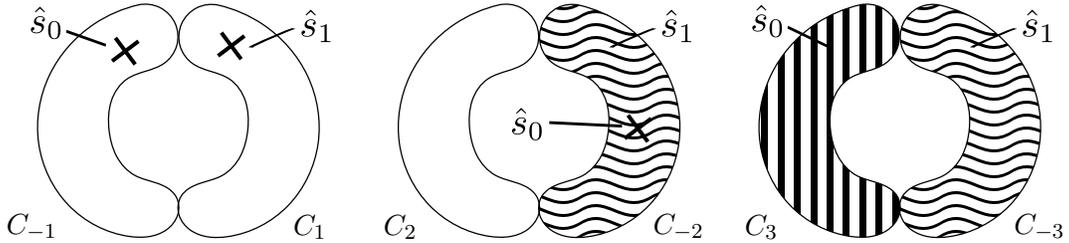

To be concrete let us consider a generic fibration $X$ that can be obtained from fibers in $dP_1$.
There are two independent rational sections $s_0$ and $s_1$ and we can choose $s_0$ as the zero-section~\cite{Klevers:2014bqa}.
The generic gauge group is $G=U(1)$ which reflects the fact that there are no fibral divisors.
There are three types of isolated fibral curves and the intersections with $s_0$ and $s_1$ are indicated in figure \ref{fig:f3codim2curves}.
Note that shaded components are wrapped by the rational section while crosses indicate a transverse intersection.
Recall that up to vertical divisors the image of $s_1$ under the Shioda map is given by
\begin{align}
	\sigma(s_1)=E_1-E_0+\text{(vertical divisors)}\,,
\end{align}
where $E_0$ and $E_1$ are the divisor classes of $s_0$ and $s_1$.
For a curve $C$ that is wrapped by $s_i$ the intersection is $E_i\cdot C=-1$.
Together with the fact that an M2-brane that wraps an isolated fibral curve $C$ has $U(1)$ charge
\begin{align}
	q=\sigma(s_1)\cdot C\,,
\end{align}
this leads to the corresponding labels $C_q$ in figure \ref{fig:f3codim2curves}.

If we now try to Higgs the $U(1)$ with a scalar of charge $q=\pm3$ we have to let the volume of $C_3$ or $C_{-3}$ shrink to zero.
We parametrize the K\"ahler form as in~\eqref{eqn:kaehlerexpansion}.
Then $\tau$ is the volume of the generic fiber while $m$ is the volume of $C_1$ while $t_i,\,i=1,...,h^{1,1}(B)$ are parametrizing the size of curves in the base.
In particular, the sizes of $C_3$ and $C_{-3}$ are
\begin{align}
	\text{vol}(C_3)=3m-\tau\,,\quad\text{vol}(C_{-3})=2\tau-3m\,.
\end{align}
It turns out that if we choose $\tau$ and $m\ne0$ such that $2\tau-3m=0$, the K\"ahler form $\omega$ is always outside the closure of the K\"ahler cone.
However, if $m>0$ the limit $\tau\rightarrow 3m$ is on the boundary of the K\"ahler cone.
It is then possible to deform the complex structure such that the sections $s_i$ merge with two-sections into two three-sections.
The corresponding divisor classes are equivalent up to vertical divisors.

In other words, the topological string partition function on the three-section geometry $X'$ is obtained from the partition function on the geometry with two sections $X$ by substituting
\begin{align}
	\tau\mapsto3\tau\,,\quad m\mapsto\tau\,.
	\label{U1higgsruleexample}
\end{align}
Using the ansatz~\eqref{eqn:jacAns1} and the relations~\eqref{eqn:modhiggs} we find that at base degree $\beta$ it takes the form
\begin{align}
	Z^{X'}_\beta(\tau,\lambda)=\frac{1}{\eta(3\tau)^{12c_1(B)\cdot \beta}\Delta_6(\tau)^{\frac{r_\beta}{3}}}\frac{\phi_\beta'(\tau,\lambda)}{\prod_{l=1}^{b_2(B)}\prod_{s=1}^{\beta_l}\phi_{-2,1}(3\tau,s\lambda)}\,,
\end{align}
where $\phi_\beta'(\tau,\lambda)$ is a weak Jacobi form for $\Gamma_1(3)$ of weight and index such that $Z_\beta^{X'}(\tau,\lambda)$ is a meromorphic Jacobi form of weight zero and index $\frac{1}{6}\beta\cdot(\beta-c_1(B))$.
Recall that $r_{\beta}=\frac{1}{2}b\cdot\beta$ is the index of $Z^X$ with respect to $m$ where $b$ is the height pairing of the section $E_1$ with itself.

Let us denote the class of the three-section in the resulting smooth genus-one fibration $X'$ by $E'$.
From a geometric perspective the result from the previous analysis is that there exists a fibral curve $C$ such that $E'\cdot C=1$.
Therefore, in order to calculate enumerative invariants, we should parametrize the K\"ahler form as
\begin{align}
	\omega=\tau\cdot(E'+D)+\dots\,,
	\label{eqn:omega3section}
\end{align}
where $\tau$ is now \textit{one third} of the volume of the generic fiber and $D$ is vertical divisor such that $\tilde{E}=E'+D$ is orthogonal to a basis of curves in the base.
In general, a curve that intersects the three-section once could be absent but the ansatz remains valid.

An analogous ansatz holds for two- and four-section geometries.
But before we give the expressions, let us note that in every example we find that the factor of $\Delta_{2N}^{r_\beta/N}$ in the denominator is compensated by a factor $\Delta_{2N}^{\lceil r_\beta/N\rceil}$ in the numerator.
Cancellation of poles then implies the congruence relation
\begin{align}
	1-\frac{r_\beta}{N}\equiv\frac12\left[Nc_1(B)-\frac{\tilde{a}}{N}\right]\cdot \beta\,\text{ mod }\,1\,.
	\label{eqn:rcongruence}
\end{align}
In particular, for a genus one fibration with $N$-sections where $N=2,3,4$ that does not have additional fibral divisors or independent multi-sections we find the simplified ansatz
\begin{align}
	Z_\beta^{(N)}(\tau,\lambda)=\frac{1}{\eta(N\tau)^{12c_1(B)\cdot\beta}}\frac{\phi_\beta'(\tau,\lambda)}{\prod_{l=1}^{b_2(B)}\prod_{s=1}^{\beta_l}\phi_{-2,1}(N\tau,s\lambda)}\,,
\end{align}
where the numerator $\phi_\beta'(\tau,\lambda)$ is an element
\begin{align}
	\phi_\beta'(\tau,\lambda)\in M_*(N)[\phi_{-2,1}(N\tau,\lambda),\phi_{0,1}(N\tau,\lambda)]\cdot\Delta_{2N}(\tau)^{1-\frac{r_\beta}{N}\text{ mod }\,1}\,,
\end{align}
such that the overall weight of $Z_\beta$ is zero and the index is $\frac{1}{2N}\beta\cdot(\beta-c_1(B))$.
The rings $M_*(N)$ are defined in~\eqref{eqn:modgens} and the exponent of $\Delta_{2N}(\tau)$ can be obtained from~\eqref{eqn:rcongruence}.

\subsection{Summary of the modular bootstrap ansatz}
\label{ssec:summary}
We will now summarize the ansatz for a genus one fibered Calabi-Yau threefold $M$ with $N$-sections for $N\in\{2,3,4\}$.
The K\"ahler form should be expanded as
\begin{align}
	\omega=\tau\cdot (E_0+D)+\sum_{i=1}^{r}m_i\cdot\sigma(E_i)+\sum_{i=r+1}^{\text{rk}(G)}m_i\cdot D_{f,i-r}+\sum_{i=1}^{b_2(B)}\tilde{t}_i\cdot D_i'\,,
	\label{eqn:kaehlerformN}
\end{align}
where a basis of fibral divisors is denoted by $D_{f,i},\,i=1,...,\text{rk}(G)$ and $E_i,\,i=0,...,r$ are independent $N$-sections.
The vertical divisors $D_i',\,i=1,...,b_2(B)$ are dual to the curves $C_i=E_0\cdot D_i,\,i=1,...,b_2(B)$ with $D_i=\pi^{-1}\tilde{D}_i$ in the sense that
\begin{align}
	D_i'\cdot C_j=N\cdot\delta_{ij}\,.
\end{align}
The ``zero $N$-section'' $E_0$ is shifted by the unique vertical divisor $D$ such that $\tilde{E}_0=E_0+D$ is orthogonal to all curves $C_i,\,i=1,...,b_2(B)$.
Moreover, following the discussion in Section~\ref{ssec:relativeconi}, the shifted K\"ahler parameters $t_i,\,i=1,...,h^{1,1}(B)$ are defined as
\begin{align}
	t_i=\tilde{t}_i+\frac{\tilde{a}_i}{2N}\tau\,,\quad\text{with}\quad \tilde{a}_i=\int_M\tilde{E}_0^2\cdot D_i\,.
	\label{eqn:tshift}
\end{align}
We will also assume that there are no fibral divisors at a generic point of the complex structure moduli space of $M$.

Then, if we expand the topological string partition function as in~\eqref{eqn:topexp}, the coefficients take the form
\begin{align}
	Z_\beta(\tau,\underline{m},\lambda)=\frac{1}{\eta(N\tau)^{12\cdot c_1(B)\cdot\beta}}\frac{\phi_\beta(\tau,\underline{m},\lambda)}{\prod_{l=1}^{b_2(B)}\prod_{s=1}^{\beta_l}\phi_{-2,1}(N\tau,s\lambda)}\,,
	\label{eqn:jacAns2}
\end{align}
where the numerator $\phi_\beta(\tau,\lambda)$ is an element
\begin{align}
	\phi_\beta(\tau,\underline{m},\lambda)\in M_*(N)[\phi_{-2,1}(N\tau,\bullet),\phi_{0,1}(N\tau,\bullet)]\cdot\Delta_{2N}(\tau)^{1-\frac{r_\beta}{N}\text{ mod }\,1}\,,
\end{align}
where $\bullet$ stands for any elliptic parameter $z\in\{\lambda,\underline{m}\}$.
The exponent of $\Delta_{2N}(\tau)$ is determined by the congruence relation
\begin{align}
	1-\frac{r_\beta}{N}\equiv\frac12\left[Nc_1(B)-\frac{\tilde{a}}{N}\right]\cdot \beta\,\text{ mod }\,1\,.
	\label{eqn:rcongruence}
\end{align}
The weight of $\phi_\beta$ is given in~\eqref{eqn:jacobiweight}, while the index with respect to the topological string coupling $\lambda$ is
\begin{align}
	r_\beta^\lambda=\frac{1}{2N}\beta\cdot(\beta-c_1(B))\,.
\end{align}
The index matrix with respect to the geometric elliptic parameters $m_i,\,i=1,...,\text{rk}(G)$ is
\begin{align}
	r^\beta_{ij}=\frac{1}{N}\cdot\left\{\begin{array}{cl}
		-\frac12\pi_*\left(\sigma(E_i)\cdot\sigma(E_j)\right)\cdot\beta&\text{ for }1\le i,j\le r\\
		-\frac12\pi_*\left(D_{f,i}\cdot D_{f,j}\right)\cdot \beta&\text{ for }r<i,j\le\text{rk}(G)\\
		0&\text{ otherwise}
	\end{array}\right.\,.
	\label{eqn:jacobiindex2N}
\end{align}

\section{Duality with heterotic strings on CHL orbifolds}
\label{sec:dualityCHL} 
In this section we will argue that heterotic strings on CHL orbifolds $(K3\times T^2)/\mathbb{Z}_N$ are dual to Type IIA strings on genus one fibered Calabi-Yau threefolds with $N$-sections.
We will then systematically construct Calabi-Yaus that are dual to heterotic compactifications on $(K3\times T^2)/\mathbb{Z}_2$.
The modular bootstrap for multi-section geometries that we developed in the previous sections allows us to compare the topological string partition function against
the heterotic one-loop computations that have been performed in~\cite{Chattopadhyaya:2017zul}.
This provides an all-genus check of our proposal.

\subsection{Heterotic/Type IIA duality and topological strings}
\label{ssec:heterotic/typeIIA}
It is well known that heterotic $E_8\times E_8$ string theory on $K3\times T^2$ is dual to Type II strings on Calabi-Yau manifolds~\cite{Kachru:1995wm}.
Let us further assume that the dual heterotic theory is weakly coupled in the geometric regime of the moduli space of the Type II compactification and that the $T^2$ can be decompactified.
Then one can show that the Calabi-Yau has to be a fibration of elliptic $K3$ surfaces over $\mathbb{P}^1$~\cite{Aspinwall:1996mn}.
Of course, in order to specify a heterotic vacuum one also needs to choose a gauge background on the compactification space.
The unbroken gauge symmetry, the spectrum and therefore also the dual Calabi-Yau depend on this choice of background.

An important observable that can be matched for a given dual pair is the gravitational coupling $F_g(t,\bar{t})$, which enters via the terms
\begin{align}
	S=\int \tilde{F}_g(t,\bar{t})F^{2g-2}R^2\,,
	\label{eqn:gravcoupling}
\end{align}
into the $d=4,\,N=2$ effective action.
Here $F,R$ are the self-dual parts of the graviphoton and the Riemann curvature.
On the heterotic side, all of these couplings are perturbatively one-loop exact but receive non-perturbative corrections.
From the Type IIA perspective, $\tilde{F}_g$ is a $g$-loop correction and corresponds to the topological string free energy at genus $g$~\cite{Antoniadis:1993ze,Antoniadis:1995zn}.
More precisely, the holomorphic limit of the topological string free energies corresponds to
\begin{align}
	F_g^{\text{top}}(t)=\frac{1}{2(2\pi i)^{2g-2}}\bar{F}_g^{\text{hol}}(t)\,,\quad\text{with}\quad F_g^{\text{hol}}(\bar{t})=\lim\limits_{t\rightarrow\infty}\tilde{F}_g(t,\bar{t})\,.
\end{align}
The heterotic dilaton can be identified with the complexified volume of the $\mathbb{P}^1$ that is the base of the $K3$ fibration in the dual Type II compactification.
The complex structure and the complexified volume of the $T^2$ on the heterotic side correspond to linear combinations of the complexified volumes of the elliptic fiber and the $\mathbb{P}^1$ base of the $K3$ fiber.
Wilson lines on the heterotic torus can be matched with volumes of fibral curves in the elliptic fibration.

For particular choices of bundles the one-loop calculation on the heterotic side has been carried out explicitly~\cite{Antoniadis:1995zn,Serone:1996bk,Marino:1998pg}.
The calculation involves an integral of a product of modular forms and theta functions over the fundamental domain of $PSL(2,\mathbb{Z})$.
This has been evaluated~\cite{Marino:1998pg} using the lattice reduction theorem by Borcherds~\cite{Borcherds:1996uda}. 
The modular form that appears in the integrand is closely related to the new supersymmetric index~\cite{Cecotti:1992qh}
\begin{align}
	\mathcal{Z}_{\text{new}}=\frac{1}{\eta(\tau)^2}\text{Tr}_R\left((-1)^FFq^{L_0-\frac{c}{24}}\bar{q}^{\bar{L}_0-\frac{\bar{c}}{24}}\right)\,,
\end{align}
which can be expressed in terms of the elliptic genus of $K3$~\cite{Harvey:1995fq}.
A variant of moonshine relates the latter to representations of the Mathieu group $M_{24}$~\cite{Eguchi:2010ej} and the consequences for compactifications on $K3\times T^2$ were studied by~\cite{Cheng:2013kpa}.

The corresponding dual Calabi-Yau manifolds are $K3$-fibered and this makes it possible to obtain all-genus results for the topological string at least for a restricted set of curve classes~\cite{Klemm:2004km}.
This relies on the fact that genus $g$ curves on elliptically fibered $K3$ can be counted by focusing on certain degenerate configurations where $g$-tori are glued to points of the base $\mathbb{P}^1$~\cite{Yau:1995mv}.
A resolution of the moduli space of such curves together with a choice of bundle is then given by the Hilbert scheme of $g$ points on $K3$ and the enumerative invariant is the corresponding Euler characteristic.
The result for $K3$ can be lifted to obtain product formulas that capture all-genus invariants of $K3\times T^2$~\cite{Katz:1999xq} which in turn admit generalizations to regular $K3$ fibrations~\cite{Klemm:2004km}.
The results from both calculations match which provides highly non-trivial evidence for the duality~\cite{Klemm:2004km}.

\subsection{Genus one fibrations and heterotic strings on $(K3\times T^2)/\mathbb{Z}_N$}
\label{sec:genusoneheterotic}
Soon after heterotic/Type II duality had been established for $K3\times T^2$ compactifications the arguments have been extended to heterotic strings on so-called CHL quotients $(K3\times T^2)/\mathbb{Z}_2$~\cite{Kachru:1997bz}.
A dual geometry was constructed by taking an elliptically fibered Calabi-Yau with a torsional section and taking a $\mathbb{Z}_2$ quotient that involves the corresponding shift along the fiber.
The result is genus one fibered Calabi-Yau threefold with $2$-sections.

However, a detailed discussion of heterotic CHL compactifications on $(K3\times Z^2)/\mathbb{Z}_N$ only appeared much later~\cite{Datta:2015hza}.
Non-standard embeddings of the gauge connection for CHL orbifolds have subsequently been considered in~\cite{Chattopadhyaya:2016xpa} and also the one-loop calculation~\cite{Marino:1998pg} has been generalized, which led to all-genus predictions for the enumerative invariants of dual Calabi-Yau geometries~\cite{Chattopadhyaya:2017zul}.
It turns out that the new supersymmetric index and therefore also the integrand in the one-loop calculation can be expressed in terms of modular forms for $\Gamma_1(N)$.

For CHL quotients of order $2$ such that the element $g'\in M_{23}$ corresponds to the conjugacy class $2A$ of $M_{24}$ it was found~\cite{Chattopadhyaya:2016xpa} that
the new supersymmetric index only depends on the number of vector- and hypermultiplets in the spectrum.
In fact, only the conveniently normalized combination
\begin{align}
	\hat{b}=\frac{1}{144}(N_h-N_v+12)\,,
	\label{eqn:b}
\end{align}
appears to be relevant~\cite{Chattopadhyaya:2016xpa}.
Note that the Hodge numbers of the dual Calabi-Yau are related to the number of vector- and hypermultiplets via
\begin{align}
	h^{1,1}=N_v-1\,,\quad h^{2,1}=N_h-1\,.
	\label{eqn:Nvh}
\end{align}
Two candidate Calabi-Yau duals for $N=2$ CHL compactifications with $\hat{b}=\frac23$ and $\hat{b}=\frac89$ have been identified by explicitly comparing the predictions from~\cite{Chattopadhyaya:2017zul} to the genus zero enumerative invariants of Calabi-Yau threefolds with $h^{1,1}=3$~\cite{Banlaki:2018pcc}.
We will show below that those geometries are again genus one fibered with $2$-sections and, using the modular bootstrap, extend the comparison to all genera.

The action of the $\mathbb{Z}_N$ quotient on $K3\times T^2$ is as follows.
It was shown by Mukai~\cite{MukaiK3} that the automorphisms of a $K3$ surface form a subgroup of the Mathieu group $M_{23}$ which is in turn a subgroup $M_{23}\subset M_{24}$.
A smooth CHL orbifold is therefore obtained from an order $N$ element $g'\in M_{23}$ that acts together with a $1/N$ shift along a cycle of $T^2$.
Other quotients are possible at the level of the CFT~\cite{Cheng:2010pq,Gaberdiel:2011fg} but they do not admit a straightforward geometric interpretation.
This is of course closely related to the moonshine phenomenon that we alluded to in the previous section.

At this point we can already argue on general grounds that $(K3\times T^2)/\mathbb{Z}_N$ should be dual to Type IIA on $N$-section geometries.
Following the usual convention we denote the complex structure and the complexified K\"ahler modulus of the $T^2$ respectively by $U$ and $T$.
For $N=1$ it is easy to see that the theory exhibits a T-duality group
\begin{align}
	\Gamma_{het}=SL(2,\mathbb{Z})\times SL(2,\mathbb{Z}) \times \mathbb{Z}_2\,.
\end{align}
One $SL(2,\mathbb{Z})$ acts on the complex structure $U$ and must clearly leave the theory invariant. T-duality along one of the cycles exchanges $U$ with $T$ and therefore leads to the factor $\mathbb{Z}_2$.
Combining both transformations it follows that the theory must also be invariant under the action of $SL(2,\mathbb{Z})$ on $T$.
In the dual Calabi-Yau geometry $U$ is identified with the K\"ahler modulus of an elliptic fiber
while $T$ is the volume of the base of a $K3$ fiber.
We already argued that the $SL(2,\mathbb{Z})$ action is realized by monodromies in the moduli space of elliptic fibrations
and it is also known that another monodromy acts via $U\leftrightarrow T$~\cite{Klemm:1995tj}.

How does this situation change if we consider CHL orbifolds?
The orbifold group $\mathbb{Z}_N$ acts as an order $N$ shift along one of the cycles of $T^2$ and the action of the duality group on the complex structure parameter of the torus
has to be compatible with this action.
But the subgroup of $SL(2,\mathbb{Z})$ that preserves a subgroup $\mathbb{Z}_N$ of the torus together with a choice of generator is $\Gamma_1(N)$.
On the other hand, we can still perform T-duality along the cycle that is not involved in the CHL quotient to exchange $U$ and $T$.
We therefore conclude that the T-duality group of heterotic strings on a CHL orbifold $(K3\times T^2)/\mathbb{Z}_N$ is
\begin{align}
	\Gamma_{CHL}=\Gamma_1(N)\times\Gamma_1(N)\times \mathbb{Z}_2\,.
\end{align}
From the perspective of the dual IIA compactification the action of this group should again be realized by the monodromies in the quantum K\"ahler moduli space of the Calabi-Yau.
Together with our previous discussion in Section~\ref{ssec:relativeconi} this implies that the Calabi-Yau has to be a genus one fibration with $N$-sections.
For explicit examples that are dual to order $2$ CHL compactifications we checked that $\mathbb{Z}_2$ is also contained in the corresponding monodromy group.

It is natural to ask under which conditions a weakly coupled CHL dual exists for a given $N$-section geometry.
To shed some light on this question we construct all $2$-section fibration over $\mathbb{P}^1\times\mathbb{P}^1$ with $h^{1,1}=3$ that can be realized as hypersurfaces in toric ambient spaces.
This leads to two additional values of $\hat{b}$ for which the topological string amplitudes match with the calculation from the heterotic side.
However, the list of gauge backgrounds that has been obtained by~\cite{Chattopadhyaya:2016xpa} is not exhaustive and we expect that a different construction should provide the missing heterotic duals.
The two models found by~\cite{Banlaki:2018pcc} are contained in this list.

\subsection{Constructing Calabi-Yau duals of CHL orbifolds}
\label{ssec:geometricCHLduals}
To keep the calculations tractable we focus on the case where the gauge symmetry on the heterotic side is maximally Higgsable.
This means that only the four universal vector fields that arise from the Kaluza-Klein reduction
of the metric and the $B$-field along the cycles of the $2$-torus remain massless.
The number of K\"ahler moduli for the dual Calabi-Yau manifolds is therefore $h^{1,1}=3$.

It has been argued on general grounds~\cite{Klemm:1995tj,Aspinwall:1995vk} that weakly coupled heterotic strings can only be dual to Type IIA compactifications on Calabi-Yau manifolds that exhibit a $K3$ fibration.
The heterotic string arises in the IIA picture from a $5$-brane that wraps the $K3$~\cite{Cherkis:1997bx}.
Further dualities relate the elliptic genus of this string to the topological string partition function which in turn encodes the couplings~\eqref{eqn:gravcoupling}.
Moreover, heterotic string theory on $K3\times T^2$ with $12+n$ and $12-n$ instantons embedded into the respective $E_8$ factors is dual to a compactification on an elliptic fibration over the Hirzebruch surfaces $\mathbb{F}_n$~\cite{Morrison:1996na}.
Only for $n=0,1,2$ the gauge theory is maximally Higgsed at a generic point of the hypermultiplet moduli space.

All three geometries lead to the same string when a $5$-brane wraps the restriction of the elliptic fibration to the fiber of the Hirzebruch surface.
To search for Calabi-Yau duals of $\mathbb{Z}_2$ CHL orbifolds we are therefore constructing genus one fibrations over $\mathbb{F}_1=\mathbb{P}^1\times\mathbb{P}^1$ with $2$-sections.
We can then study the strings from both of the $\mathbb{P}^1$'s and compare with the predictions from~\cite{Chattopadhyaya:2017zul}.

Note that a genus one fibration with $2$-sections can always be mapped into a fibration of degree $4$ hypersurfaces in $\mathbb{P}_{112}$ using for example the techniques that have been reviewed in~\cite{Braun:2014oya}.
The systematic construction of elliptically and genus one fibered Calabi-Yau manifolds as hypersurfaces in toric ambient spaces is reviewed in Section~\ref{sec:construction}.
Following that discussion we choose the fiber of the ambient space to be $\mathbb{P}_{112}$ and one can write down generic GLSM charge vectors as the rows of the matrix
\begin{align}
	A=\left(\begin{array}{rrrrrrrr}
		1&1&2&0&0&0&0\\
		-q_1&-2&-q_1&1&1&0&0\\
		-q_2&-2&-q_2&0&0&1&1
	\end{array}\right)\,,
\end{align}
where $q_i=q_i(\mathcal{S}_7-2\mathcal{S}_9)$ in the notation introduced in and below~\eqref{eqn:f4coordinateclass}.
The kernel $B=\text{ker}(A)$ of this matrix is generated by the points of the polytope with integral points
\begin{align}
	\begin{blockarray}{rrrrl}
		&&&&\\
		\begin{block}{(rrrr)l}
			-1&1&0&0&X\quad\leftarrow\text{ 2-section }\\
			-1&-1&0&0&Y\quad\leftarrow\text{ 2-section }\\
			1&0&0&0&Z\quad\leftarrow\text{ 4-section }\\
			0&0&1&0&a_1\\
			-2&\tilde{q}_1&-1&0&a_2\\
			0&0&0&1&b_1\\
			-2&\tilde{q}_2&0&-1&b_2\\
	\end{block}
	\end{blockarray}\,,
	\label{eqn:gen2secpts}
\end{align}
with $\tilde{q}_i=q_i-2$.
We have already indicated the names for the homogeneous coordinates of the associated toric variety.
The $\mathbb{P}^1$ in the base with coordinates $a_1,a_2$ will be denoted by $\mathbb{P}^1_A$ and the other will correspondingly be called $\mathbb{P}^1_B$.
The Stanley-Reisner ideal is then always given by
\begin{align}
	\mathcal{SRI}=\langle a_1a_2,\,b_1b_2,\,XYZ\rangle\,,
\end{align}
and we denote the corresponding divisors by $D_1=[a_1]=[a_2],\,D_2=[b_1]=[b_2],\,E_0=[X]$.
It is then easy to calculate
\begin{align}
	E_0^2\cdot D_1=4-q_2\,,\quad E_0^2\cdot D_2=4-q_1\,,
\end{align}
and therefore
\begin{align}
	\tilde{E}_0=E_0+(q_1-2)D_1+(q_2-2)D_2\,.
\end{align}
We can expand the K\"ahler form as
\begin{align}
	\omega=\tau\cdot \tilde{E}_0+\sum\limits_{i=1}^\infty\left(t_i-\frac{\tilde{a}_i}{4}\right)\cdot D_i\,,
\end{align}
with
\begin{align}
	\tilde{a}_1=\tilde{E}_0^2\cdot D_2=2q_1-4\,,\quad\tilde{a}_2=\tilde{E}_0^2\cdot D_1=2q_2-4\,.
	\label{eqn:atilde2section}
\end{align}

Only for a small set of values for $q_1,q_2$ the convex hull of the points is a reflexive polytope such that the corresponding toric variety is a $\mathbb{P}_{112}$ fibration over $\mathbb{P}^1\times\mathbb{P}^1$ and
the generic Calabi-Yau hypersurface exhibits $h^{1,1}=3$.
Moreover, two pairs of values can lead to polytopes that are identified under a lattice automorphism.
We list a complete set of admissible values in table~\ref{tab:q1q2} that lead to inequivalent reflexive polytopes.
\begin{table}[h!]
	\centering
	\begin{tabular}{rrrrrr}
		$q_1$&$q_2$&$h^{1,1}$&$h^{2,1}$&$\hat{b}_1$&$\hat{b}_2$\\\hline
		\\[-1em]
		4&4&4&148&--&--\\\\[-1em] 
		4&3&3&131&$1$&--\\\\[-1em] 
		4&2&3&115&$\frac89$&$\frac89$\\\\[-1em] 
		4&1&3&99&$\frac79$&--\\\\[-1em]  
		4&0&3&83&$\frac23$&$\frac23$    
	\end{tabular}
	\hspace{1cm}
	\begin{tabular}{rrrrrr}
		$q_1$&$q_2$&$h^{1,1}$&$h^{2,1}$&$\hat{b}_1$&$\hat{b}_2$\\\hline\\[-1em]
		3&3&3&123&--&--\\\\[-1em] 
		3&2&3&115&--&$\frac89$\\\\[-1em] 
		3&1&3&107&--&--\\\\[-1em] 
		2&2&3&115&$\frac89$&$\frac89$   
	\end{tabular}
	\caption{Values for $q_1,q_2$ in~\eqref{eqn:gen2secpts} that lead to reflexive polytopes such that the corresponding toric variety is a $\mathbb{P}_{112}$ fibration over $\mathbb{P}^1\times\mathbb{P}^1$ and the generic Calabi-Yau hypersurface has $h^{1,1}=3$. When the string that arises from a 5-brane wrapping the restriction of the genus one fibration to $\mathbb{P}_A$ or $\mathbb{P}_B$ matches with a 2A CHL string, the corresponding value of $\hat{b}$ is also listed.}
	\label{tab:q1q2}
\end{table}
The list also includes the unique special case where $h^{1,1}=4$ but one of the K\"ahler deformations is non-toric.
This corresponds to the situation where the ramification locus of a toric 2-sections is trivial and it can be identified with the union of two independent sections.
Note that a Calabi-Yau with $(h^{1,1},h^{2,1})=(3,131)$ has already been proposed in~\cite{Kachru:1997bz} as the dual geometry for  a heterotic compactification on a particular $(K3\times T^2)/\mathbb{Z}_2$ orbifold.

Let us denote the corresponding geometries by $M_{q_1,q_2}$.
Using the generic formula for the Euler characteristic from~\cite{Klevers:2014bqa}, the number of complex structure moduli can be expressed as
\begin{align}
	h^{2,1}(q_1,q_2)=144+h^{1,1}(q_1,q_2)-16(q_1+q_2)+8q_1q_2\,.
	\label{eqn:h21q}
\end{align}
Together with~\eqref{eqn:b} and~\eqref{eqn:Nvh} this implies that if a $5$-brane that wraps one of the $K3$ fibers in $M_{q_1,q_2}$ can be identified with the string of a CHL orbifold of the previously discussed type, then 
the gauge background is necessarily such that
\begin{align}
	\hat{b}=\frac{1}{144}\left[160-16(q_1+q_2)+8q_1q_2\right]\,.
\end{align}
Indeed, using the normal form~\cite{Kreuzer:2002uu} for reflexive polytopes one can easily check that the two geometries found by~\cite{Banlaki:2018pcc} respectively correspond to $(q_1,q_2)=(2,2)$ and $(q_1,q_2)=(4,0)$.

For any of the nine geometries we can consider the $K3$ fiber with base $\mathbb{P}^1_A$ or $\mathbb{P}^1_B$ which we will sometimes denote by $K3_1$ and $K3_2$.
This leads to eighteen base degree one partition functions and in all of the genuine genus one fibrations with $h^{1,1}=3$ it can either be written as
\begin{align}
	Z_1^{(1)}(\tau,\lambda)=\frac{\Delta_2(\tau)}{\eta(2\tau)^{24}\phi_{-2,1}(2\tau,\lambda)}\cdot E_2^{(2)}\cdot\left[4(6\hat{b}-5)\cdot\left(E_2^{(2)}\right)^2+2(2-3\hat{b})\cdot E_4\right]\,,
	\label{eqn:ztopk3fiberhetdual}
\end{align}
or
\begin{align}
	Z_1^{(2)}(\tau,\lambda)=\frac{\Delta_2(\tau)^{\frac32}}{\eta(2\tau)^{24}\phi_{-2,1}(2\tau,\lambda)}\cdot \left[\lambda_1\cdot \left(E_2^{(2)}\right)^2+\lambda_2\cdot E_4\right]\,.
\end{align}
As our notation already suggests, it is the first case that matches the predictions from a one-loop calculation in $2A$ CHL orbifolds with a corresponding value for $\hat{b}$.

If we multiply $Z_1$ with $q^{\frac{\tilde{a}}{2N}}$, see~\eqref{eqn:kaehlershift}, the poles at $\tau\rightarrow i\infty$ are cancelled.
Indeed we find that $Z_1^{(1)}$ is a valid ansatz for $K3_i$ if
\begin{align}
	\tilde{a}_i\equiv0\,\text{mod}\,4\,,
\end{align}
and otherwise $Z_1^{(2)}$ matches the genus zero and genus one invariants for an appropriate choice of $\lambda_1,\lambda_2$. 
Expressions for $\tilde{a}_i,\,i=1,2$ in terms of $q_1,q_2$ have been provided in~\eqref{eqn:atilde2section}.

In the special case where $h^{1,1}=4$ the geometry is actually elliptically fibered with two independent sections that descend from one toric $2$-section.
We then find that $K3_1$ is equivalent to $K3_2$ and the base degree one partition function reads
\begin{align}
	Z_1(\tau,\lambda)=\frac23\frac{\Delta_2(\tau)}{\eta(2\tau)^{24}\phi_{-2,1}(2\tau,\lambda)}\cdot E_2^{(2)}\cdot\left[\left(E_2^{(2)}\right)^2-4E_4\right]\,.
\end{align}
\subsection{Comparison with the heterotic one-loop computation}
\label{ssec:comparison}
The one-loop calculation of the couplings~\eqref{eqn:gravcoupling} has been carried out for many CHL orbifolds with standard-embedding of the gauge connection in~\cite{Chattopadhyaya:2017zul}.
However, all of the bundle dependent data that enters the result is contained in the new supersymmetric index.
For many non-standard embeddings on $2A$ orbifolds the index has been obtained in~\cite{Chattopadhyaya:2016xpa} and takes the form
\begin{align}
	\mathcal{Z}_{\text{new}}=-4\sum\limits_{r,s=0}^1\Gamma_{2,2}^{(r,s)}f^{(r,s)}_{2A}\,,
\end{align}
with $\Gamma^{(r,s)}_{2,2}$ a lattice sum that will not be relevant to us and
{\tiny
\begin{align}
	\begin{split}
		f^{(0,0)}_{2A}=&\frac{1}{\eta(\tau)^{24}}\frac{E_4E_6}{2}\,,\\
		f^{(0,1)}_{2A}=&\frac{1}{\eta(\tau)^{24}}\frac14\left[\left(E_6+2E_2^{(2)}\left(\tau\right)E_4\right)\left(\hat{b}\left(E_2^{(2)}\left(\tau\right)\right)^2+\left(\frac23-\hat{b}\right)E_4\right)\right]\,,\\
		f^{(1,0)}_{2A}=&\frac{1}{\eta(\tau)^{24}}\frac14\left[\left(E_6-E_2^{(2)}\left(\frac{\tau}{2}\right)E_4\right)\left(\frac{\hat{b}}{4}\left(E_2^{(2)}\left(\frac{\tau}{2}\right)\right)^2+\left(\frac23-\hat{b}\right)E_4\right)\right]\,,\\
		f^{(1,1)}_{2A}=&\frac{1}{\eta(\tau)^{24}}\frac14\left[\left(E_6-E_2^{(2)}\left(\frac{\tau+1}{2}\right)E_4\right)\left(\frac{\hat{b}}{4}\left(E_2^{(2)}\left(\frac{\tau+1}{2}\right)\right)^2+\left(\frac23-\hat{b}\right)E_4\right)\right]\,,
	\end{split}
\end{align}
}
Recall that the choice of gauge background only enters via the combination $\hat{b}$~\eqref{eqn:b} of the numbers of vector- and hypermultiplets.

A direct evaluation of the one-loop amplitude leads to~\cite{Chattopadhyaya:2017zul}
\begin{align}
	\bar{F}_g^{\text{hol}}=\frac{(-1)^{g-1}}{\pi^2}\sum\limits_{s=0}^{N-1}{\sum\limits_{m}}'e^{-2\pi i n_2s/N}c_{g-1}^{(n_1,s)}\left(\frac{n_1n_2}{2},0\right)\text{Li}_{3-2g}\left(e^{2\pi i m\cdot y}\right)+\text{const.}\,,
	\label{eqn:Fghol}
\end{align}
where $\sum'$ is a sum over the points
\begin{align}
	n_1,\,n_2\ge0\,,\text{ but }(n_1,n_2)\ne(0,0)\,,\text{ and }(n_1,-n_2)\,,\text{ with }n_2>0\text{ and }n_1n_2\le N\,.
\end{align}
and $y=(T,U)$ contains the K\"ahler and complex structure of the torus.
The coefficients $c_{g-1}^{(r,s)}(l,t)$ are defined via 
\begin{align}
	f^{(r,s)}(\tau)\mathcal{P}_{2g+2}(\tau)=\sum\limits_{l\in\frac{\mathbb{Z}}{N},t=0}^{t=g}c_{g-1}^{(r,s)}(l,t)\tau_2^{-t}q^l\,,
\end{align}
where
\begin{align}
	\mathcal{P}_{2k}(\tau)\equiv\mathcal{P}_{2k}(\hat{G}_2,\dots,G_{2k})\,,\quad G_{2k}=2\zeta(2k)E_{2k}\,,\quad \hat{E}_2(\tau)=E_2-\frac{3}{\pi\tau_2}\,,
\end{align}
and $\mathcal{P}_{2k}(x_1,\dots,x_{k})$ is related to the Schur polynomial $\mathcal{S}_k(x_1,\dots,x_k)$ by
\begin{align}
	\mathcal{P}_{2k}(x_1,\dots,x_{k})=-\mathcal{S}_k(x_1,\frac12 x_2,\dots,\frac1k x_k)\,.
\end{align}
The Schur polynomials are in turn defined by
\begin{align}
	\exp\left(\sum\limits_{k=1}^\infty x_kz^k\right)=\sum\limits_{k=0}^\infty\mathcal{S}_k(x_1,...,x_k)z^k\,.
\end{align}
The indices $(r,s)$ label the twisted sectors and are considered modulo $2$.
From this we can extract all Gopakumar-Vafa invariants that have base degree zero with respect to the $K3$ fibration of the dual Calabi-Yau up to arbitrary genus.
The invariants of base degree $1$ with respect to the genus one fibration of the $K3$ are listed in the table~\ref{eqn:k3gv1}.
\begin{table}[h!]
	\centering\tiny
	\begin{tabular}{c|ccccc}
		$d\backslash g$&0&1&2&3&4\\\hline
 0 & 512-288 $\hat{b}$ & 0 & 0 & 0 & 0 \\
 1 & 2376 $\hat{b}$+8128 & 36 $\hat{b}$-32 & 0 & 0 & 0 \\
 2 & 151552-11520 $\hat{b}$ & 576 $\hat{b}$-1024 & 0 & 0 & 0 \\
 3 & 45900 $\hat{b}$+1212576 & -4644 $\hat{b}$-16352 & 48-54 $\hat{b}$ & 0 & 0 \\
 4 & 8671232-158976 $\hat{b}$ & 24768 $\hat{b}$-306176 & 1536-864 $\hat{b}$ & 0 & 0 \\
 5 & 493488 $\hat{b}$+47890048 & -105912 $\hat{b}$-2474048 & 6840 $\hat{b}$+24640 & 72 $\hat{b}$-64 & 0 \\
 6 & 240009216-1410048 $\hat{b}$ & 389376 $\hat{b}$-18255872 & 462848-39168 $\hat{b}$ & 1152 $\hat{b}$-2048 & 0 \\
 7 & 3777570 $\hat{b}$+1055720304 & -1281132 $\hat{b}$-103120800 & 174906 $\hat{b}$+3768496 & -8964 $\hat{b}$-32992 & 80-90 $\hat{b}$ \\
 	\end{tabular}
	\caption{Gopakumar-Vafa invariants of the $K3$ at base degree $1$.}
	\label{eqn:k3gv1}
\end{table}
We find perfect agreement with the predictions from~\ref{eqn:ztopk3fiberhetdual}.

The higher base degrees match with the results from the genus zero and genus one partition functions and can be used to fix the coefficients in the modular ansatz. Evaluation of (\ref{eqn:Fghol}) provide the prediction of Gopakumar-Vafa invariants listed in table~\ref{eqn:k3gv2}.
\begin{table}[h!]
	\centering\tiny
	\begin{tabular}{c|cccc}
		$d\backslash g$&0&1&2&3\\\hline
 0 &  288 $\hat{b}$-32 & 4 & 0 & 0 \\
 1 &151552-11520 $\hat{b}$ & 576 $\hat{b}$-1024 & 0 & 0  \\
 2 & 158976 $\hat{b}$+8387328 & -24768 $\hat{b}$-262464 & 864 $\hat{b}$-128 & 8  \\
 3 & 240009216-1410048 $\hat{b}$ & 389376 $\hat{b}$-18255872 & 462848-39168 $\hat{b}$ &
   1152 $\hat{b}$-2048  \\
 4 & 9596160 $\hat{b}$+4294949632 & -3870144 $\hat{b}$-526389312 & 674208
   $\hat{b}$+27261728 & -54720 $\hat{b}$-524144  \\
 5 & 57704112128-54369792 $\hat{b}$ & 29089152 $\hat{b}$-10135607296 & 865636352-7319808
   $\hat{b}$ & 1016064 $\hat{b}$-39292928  \\
 6 & 268600320 $\hat{b}$+620790389760 & -179967744 $\hat{b}$-143003775232 & 59660928
   $\hat{b}$+17075001600 & -11892672 $\hat{b}$-1217397536 
   \\
 7 & 5647463645184-1191462912 $\hat{b}$ & 962118144 $\hat{b}$-1626724835328 &
   253765050368-397518336 $\hat{b}$ & 104037120 $\hat{b}$-25542406144  \\
 	\end{tabular}
	\caption{Gopakumar-Vafa invariants of the $K3$ at base degree $2$.}
	\label{eqn:k3gv2}
\end{table}

The Gopakumar-Vafa invariants in table~\ref{eqn:k3gv2} can be used to fix the numerator in our Ansatz for the base degree 2 partition function
\begin{align} 
Z_2 = \Bigg(\frac{\Delta_4}{\eta^{24}(2\tau)}\Bigg)^2  \frac{ \phi_2(\tau,\lambda)}{\phi_{-2,1}(2\tau,\lambda)\phi_{-2,1}(4\tau,\lambda)}\,,
\end{align}
and we find that
\begin{footnotesize}
\begin{align}
\begin{split}
\phi_2 =  & \frac{1}{82944} \Bigg[384 A^3 \Bigg(2 (6 \hat{b}-5) g^3+(2-3 \hat{b}) g h\Bigg)^2+8 A^2 B g
   \Bigg(24 (62 \hat{b}-55) g^6   +10 (95-108 \hat{b}) g^4 h\\
   &+(261 \hat{b}-220) g^2
   h^2+7 (2-3 \hat{b}) h^3\Bigg)+A B^2 \Bigg(-48 \Big(4320 \hat{b}^2  -6644 
   \hat{b}+2505\Big) g^8+8 \Big(36 \hat{b} (504 \hat{b}\\
   &-703)
   +8321\Big) g^6 h+\Big(36
   (1159-936 \hat{b}) \hat{b}-11231\Big) g^4 h^2+2 \Big(12 \hat{b} (108
   \hat{b}-109)+179\Big) g^2 h^3\\
   &+3 (11-12 \hat{b}) h^4\Bigg) +4 B^3 g \left(11 g^2-3
   h\right) \Big(12 (8 \hat{b}-7) (36 \hat{b}-25) g^6+\Big(216 (11-8 \hat{b})
   \hat{b}-763\Big) g^4 h\\
   &+\Big(27 \hat{b} (8 \hat{b}-9)
   +50\Big) g^2 h^2+(5-3 \hat{b})
   h^3\Big)\Bigg]\,,
   \end{split}
\end{align}
\end{footnotesize}
where we introduced
\begin{align}
	A=\phi_{0,1}(2\tau,\lambda)\,,\quad B=\phi_{-2,1}(2\tau,\lambda)\,,\quad g=E_2^{(2)}(\tau)\,,\quad h=E_4(\tau)\,.
\end{align}

We can thus check the duality between the proposed pairs of Type IIA compactifications on genus one fibrations with 2-sections
and heterotic compactifications on $(K3\times T^2)/\mathbb{Z}_2$, at least perturbatively, to arbitrary orders.
The contributions to the topological string partition function that arise from curves that also wrap the base of the $K3$-fibration correspond to a highly non-trivial prediction of the non-perturbative corrections on the heterotic side.
\section{Decoupling gravity on genus one fibrations over $(-1)$-curves}
\label{sec:decoupling}
During the last few years there has been a considerable amount of interest in the study of topological strings on non-compact Calabi-Yau manifolds that are elliptically fibered over curves of negative self-intersection~\cite{Klemm:1996hh,Haghighat:2014vxa,Gu:2017ccq,DelZotto:2017mee,Duan:2018sqe,Gu:2018gmy,Gu:2019dan}.
The reason is the relation to elliptic genera of strings in so-called minimal six-dimensional $(1,0)$-superconformal field theories that serve as building blocks in a classification of all such theories~\cite{Morrison:2012np,Heckman:2013pva,Heckman:2015bfa}.
Those strings arise in F-theory from D3-branes that wrap the curve in the base of the fibration.
A natural question is, what happens if we consider a genus one fibration with multi-sections over a curve of negative self-intersection?

Arguably the simplest but most important building block arises from a non-compact fibration without reducible fibers over a curve of self-intersection $-1$.
The corresponding topological string partition function at base degree $d_B$ encodes the elliptic genus of an ensemble of $d_B$ E-strings and
the SCFT can be used to glue theories in the atomic classification.
In this section we will initiate the study of genus one-fibrations over $(-n)$-curves by discussing the analogues to the E-string geometry with $N$-sections for $N=2,3,4$.

\subsection{The refined topological string partition function}
\label{ssec:refined}
The enumerative invariants of an elliptic or genus one fibered Calabi-Yau encode the BPS-spectrum of particles that arise from the strings wrapped along the circle after compactifying from six to five dimensions.
More precisely, the Gopakumar-Vafa invariants~\eqref{eqn:gvexpansion} correspond to a weighted sum of multiplicities of BPS states.
The multiplicities themselves are not invariant and the number of BPS states with a particular mass and spin does, in general, jump across lines of marginal stability in the complex structure moduli space.
However, non-compact Calabi-Yau manifolds are rigid and the topological string partition function can be refined such that it encodes the actual number of BPS states for a given set of quantum numbers.

Physically, the situation is as follows.
M-theory compactified on a Calabi-Yau threefold $M$ leads to a 5d supergravity with eight supercharges.
The five-dimensional little group is $SO(4)=SU(2)_L \times SU(2)_R$ and when gravity is decoupled an additional $SU(2)_I$ R-symmetry emerges.
The decoupling limit corresponds to a decompactification of the Calabi-Yau.
Let us denote with $J_*$ the Cartan generator of $SU(2)_*$.
Then $J_I$ enables a twisting of $J_R$ such that the degeneracies of the BPS states $N_{j_L,j_R'}^\kappa$ are protected~\cite{Choi:2012jz}.
In the following, $j_R$ will denote the spin with respect to the twisted $SU(2)_R$ with Cartan generator $J_R+J_I$.
The refined BPS states with multiplicities $N_{j_L,j_R}^\kappa$ are then labeled by the mass, which is specified by $\kappa \in H_2(X,\mathbb{Z})$,  and the twisted spin representation $[j_L,j_R]$.
They are encoded in the index
\begin{align}
	\mathcal{Z}(\epsilon_L,\epsilon_R,\bm{t})=\text{Tr}_{\mathcal{BPS}}(-1)^{2(J_L+J_R)}e^{-2\epsilon_-J_L}e^{-2\epsilon_+(J_R+J_I)}e^{\beta H}\,.
\end{align}
which is called the refined topological string partition function.

From the perspective of the effective theory, $\mathcal{Z}$ can be obtained from a refinement of the Schwinger-Loop calculation in~\cite{Gopakumar:1998ii}, which was performed by~\cite{Choi:2012jz}.
The result is
\begin{equation}
\log\Big(\mathcal{Z}(\epsilon_1,\epsilon_2,\bm{t})\Big)= \sum_{\kappa \in H_2(X,\mathbb{Z})} \sum_{m>0} (-1)^{g_L +g_R+1} \frac{n_{g_L,g_R}^\kappa}{m}\frac{\big(\sin \frac{m\epsilon_-}{2}\big)^{2g_L}\big(\sin \frac{m\epsilon_+}{2}\big)^{2g_R}}{\sin\frac{m \epsilon_1}{2}\sin\frac{m \epsilon_2}{2}} \bm{Q}^{m \kappa}\,,
\label{eqn:refGV}
\end{equation}
where $n_{g_L,g_R}^\kappa $ are called refined Gopakumar-Vafa invariants and we introduced $\epsilon_i,\,i=1,2$ with $\epsilon_\pm=\frac12(\epsilon_1\pm\epsilon_2)$.
They are related to the refined BPS invariants $N_{j_L,j_R}^\kappa$ via 
\begin{align}
\sum_{j_L,j_R\in\frac{ \mathbb{N}}{2}}  N_{j_L,j_R}^\kappa \chi_{j_L} (x_L) \chi_{j_R} (x_R) = \sum_{g_L,g_R \in \mathbb{N}} n_{g_L,g_R}^\kappa \Bigg(x_L^{\frac{1}{2}} - x_L^{-\frac{1}{2}}\Bigg)^{2g_L} \Bigg(x_R^{\frac{1}{2}} - x_R^{-\frac{1}{2}}\Bigg)^{2g_R} \,,
\end{align}
where
\begin{equation}
\chi_{j_*}(x_*) = \frac{x_*^{2j_*+1}-x_*^{-2j_*-1}}{x_* -x_*^{-1}}\,,
\end{equation}
and $x_*$ is a formal variable.
One can also define refined free energies $\mathcal{F}^{(n,g)}(\bm{t})$ via
\begin{equation}
\log\Big(\mathcal{Z}(\epsilon_1,\epsilon_2,\bm{t})\Big)= \sum_{n,g = 0}^\infty (\epsilon_1 + \epsilon_2)^{2n} (\epsilon_1 \epsilon_2)^{g-1} \mathcal{F}^{(n,g)} (\bm{t})\,.
\label{eqn:refGV}
\end{equation}
In the limit $\epsilon_+ \rightarrow 0$, (\ref{eqn:refGV}) reproduces the instantons part of the unrefined topological string partition function in (\ref{eqn:gvexpansion}) and the unrefined Gopakumar-Vafa invariants are
\begin{align}
n_g^\beta=(-1)^{g} \sum_{g_R}n_{g,g_R}^\kappa\,. 
\end{align}

The refined topological string partition function can also be expanded in terms of base parameters $Q^\beta,\,\beta\in H_2(B)$
\begin{align}
	\mathcal{Z}(\epsilon_1,\epsilon_2,\bm{t},\tau,\bm{m})=\mathcal{Z}_0\left(1+\sum\limits_{\beta\in H_2(B)}\mathcal{Z}_\beta(\epsilon_1,\epsilon_2,\tau,\bm{m})Q^\beta\right)\,.
\end{align}
The coefficient $\mathcal{Z}_\beta(\epsilon_1,\epsilon_2,\tau,\bm{m})$ corresponds to the elliptic genus of a string that wraps the class $\beta$ in an F-theory compactification on $M$~\cite{Klemm:1996hh,Haghighat:2013gba,Haghighat:2014vxa}.
Moreover, after compactifying on an additional circle the theory is dual to M-theory on $M$ and the wrapped strings lead to the BPS particles that are counted by the refined invariants.

\subsection{Genus one fibrations and E-strings with Wilson lines}
\label{ssec:wilsonline}
We construct the non-compact fibration as local geometries in compact genus one fibered Calabi-Yau threefolds over the base $B=\mathbb{F}_1$.
One can then show that the ramification divisor of the $N$-section intersects the $(-1)$-curve and therefore we expect the
local geometry to be genus one fibered as well.
We only consider maximally Higgsed geometries with only one linearly independent $N$-section and no fibral divisors.
It seems to be the case that the results are then independent of the global structure of the fibration.
We will therefore only consider one Calabi-Yau for each value $N=2,3,4$.

More precisely, we will consider three genus one fibrations with $N$-sections
\begin{equation}
M_N = \begin{cases}\label{eqn:Nsecfibs}
(F_4 \rightarrow \mathbb{F}_1 )[\mathbb{Z}_2]^{-224}_3 \text{ for } N=2 \,,  \\
(F_1 \rightarrow \mathbb{F}_1 )[\mathbb{Z}_3]^{-144}_3 \text{ for } N=3\,,  \\
(\mathbb{P}^3 \rightarrow \mathbb{F}_1 )[\mathbb{Z}_4]^{-104}_3 \text{ for } N=4 \,.  \\
\end{cases}.
\end{equation}
We also consider three elliptic fibrations with ``pseudo-$N$-section''. By this we mean elliptically fibered Calabi-Yau 3-folds that are realized as hypersurfaces or complete intersections in toric ambient spaces but exhibit non-toric divisors
such that from the toric perspective multiple sections or merged into an $N$-section with trivial monodromy.
The corresponding geometries are
\begin{equation}
M'_N = \begin{cases}\label{eqn:pseudoNsecfibs}
(F_4 \rightarrow \mathbb{F}_1 )[\mathbb{Z}_2]^{-288}_{4(1)} \text{ for } N=2 \,,  \\
(F_1 \rightarrow \mathbb{F}_1 )[\mathbb{Z}_3]^{-192}_{5(2)} \text{ for } N=3\,,  \\
(\mathbb{P}^3 \rightarrow \mathbb{F}_1 )[\mathbb{Z}_4]^{-128}_{6(3)} \text{ for } N=4 \,. \\
\end{cases}
\end{equation}
All of the geometries are hypersurfaces in toric ambient spaces except for $M_4$ and $M'_4$ which are complete intersections.
We will not discuss the geometries in detail but provide the toric data in~\ref{app:localgeometries}.
In each case we want to consider the local limit of the base $\hat{B} = \mathcal{O}(-1) \rightarrow \mathbb{P}^1$~\cite{Haghighat:2014vxa}.
This is achieved when $\text{Vol}(C_F) \rightarrow \infty$, where $C_F$ is the class of the fiber in $\mathbb{F}_1$.
We denote the $(-1)$-curve in the base of $\mathbb{F}_1$ by $C_B$ and let $\beta = b\cdot C_B  \in H_2(\hat{B},\mathbb{Z})$.

From our discussion of the modular ansatz for genus one fibrations in Section~\ref{ssec:modgenusone}, we expect that one can obtain the
ansatz for genus one fibrations over $(-1)$-curves from the E-string ansatz~\cite{DelZotto:2016pvm,Gu:2017ccq} by substituting $\tau\rightarrow N\tau$.
The result is
\begin{align}\label{eqn:jacRefAns1}
	\mathcal{Z}_\beta^{(N)} (\tau, \epsilon_+,\epsilon_-) = \frac{1}{\eta(N\tau)^{12c_1(B)\cdot\beta}}  \frac{\varphi_\beta (\tau,\epsilon_+,\epsilon_-)}{\prod_{\ell=1}^{b_2(B)} \prod_{s=1}^{\beta_\ell}\left[ \phi_{-1,\frac{1}{2}}(N \tau,s \epsilon_1 )\phi_{-1,\frac{1}{2}}(N \tau,s \epsilon_2 )\right]}\,.
\end{align}
where we have set the mass parameters of the E-string to zero.
The numerator is now an element
\begin{equation}
	\varphi_\beta \in M_{*}(N)[ \phi_{-2,1}(N\tau,\bullet),\phi_{0,1}(N\tau,\bullet)] \cdot \Delta_{2N}^{1-\frac{r_\beta}{N} \text{mod} \phantom{\,,} 1}\,,
\end{equation}
where $\bullet\in\{\epsilon_+,\epsilon_-\}$, such that the overall indices of $\mathcal{Z}_{\beta}^{(N)}$ with respect to $\epsilon_-,\epsilon_+$ are
\begin{equation}\label{eqn:refind}
r_\beta^{-} = \frac{1}{2} \beta\cdot\left(\beta-c_1(B)\right)\,, \quad r_\beta^+ = \frac{1}{2} \beta\cdot\left(\beta+c_1(B)\right) -2(\bm{1}\cdot\beta)\,.
\end{equation}
Recall that the exponent of $\Delta_{2N}$ was defined via the congruence relation~\eqref{eqn:rcongruence}.
The indices of the numerator with respect to $\epsilon_+$ and $\epsilon_-$~\eqref{eqn:refind} are
\begin{align}\label{eqn:refindex2}
\begin{split}
r_\beta^+ =\frac{b}{3}(b^2+3b-4) \,,\quad r_\beta^- = \frac{b}{3}(b^2-1) \,. 
\end{split}
\end{align} 

Let us denote the numerators in $\mathcal{Z}_{b\cdot C_B}(\tau,\epsilon_+,\epsilon_-)$ for $M_N$ and $M_N'$ by $\phi_b^{(N)}$ and $\phi_b^{(N')}$.
We then find that
\begin{align}
	\begin{split}
		\phi_1^{(1)}(\tau)=&-\frac{E_4(\tau)}{\eta(\tau)^{12}}\,,\\
	\phi_1^{(2)}(\tau)=&-16\frac{\Delta_4(\tau)}{\eta(2\tau)^{12}}\,,\quad \phi_1^{(2')}(\tau)=-2\frac{\sqrt{\Delta_4(\tau)}E_2^{(2)}(\tau)}{\eta(2\tau)^{12}}\,,\\
		\phi_1^{(3)}(\tau)=&-9\frac{\Delta_6(\tau)^{\frac23}}{\eta(3\tau)^{12}}\,,\quad \phi_1^{(3')}(\tau)=-\frac{3}{216}\frac{E_4(\tau)-E_2^{(3)}(\tau)^2}{\eta(3\tau)^{12}}\,,\\
		\phi_1^{(4)}(\tau)=&-\frac12\frac{\Delta_8(\tau)^{\frac14}\left(E_2^{(2)}(\tau)-E_2^{(4)}(\tau)\right)}{\eta(4\tau)^{12}}\,,\quad\phi_1^{(4')}(\tau)=-4\frac{\sqrt{\Delta_8(\tau)}}{\eta(4\tau)^{12}}\,,
	\end{split}
\end{align}
where $N=1$ corresponds to the ordinary E-string.
It turns out that the elliptic genus $\mathcal{Z}_1^{(2)}$ already appeared in~\cite{Kim:2014dza} where it was obtained from the E-string on a circle with a non-zero Wilson loop for the affine $E_8$-flavor symmetry turned on.
More generally, up to overall factors of $q$ we obtain the expression
\begin{align}
	\phi_1^{(N)}(\tau)=-\frac{2}{\eta(N\tau)^{12}}\sum\limits_{i=2}^4\prod\limits_{j=1}^8 \theta_i\left(N\tau,v^{(N)}_j\cdot\tau\right)\,,
\end{align}
with Wilson loop parameters $\vec{v}^{\,(N)}$ given by
\begin{align}
	\begin{split}
		\vec{v}^{\,(1)}=&(0,0,0,0,0,0,0,0)\,,\\
		\vec{v}^{\,(2)}=&(0,0,0,0,0,0,0,2)\,,\quad\vec{v}^{\,(2')}=(0,0,0,0,0,0,1,1)\,,\\
		\vec{v}^{\,(3)}=&(0,0,0,1,1,1,1,2)\,,\quad\vec{v}^{\,(3')}=(0,0,0,0,0,1,1,2)\,,\\
		\vec{v}^{\,(4)}=&(0,0,0,1,1,2,2,2)\,,\quad\vec{v}^{\,(4')}=(0,0,0,0,1,1,1,3).
	\end{split}
\end{align}
The definition of the Jacobi theta functions $\theta_i(\tau,z),\,i=1,...,4$ was given in~\eqref{eqn:jacobitheta}.
E-strings with wilson lines have also been discussed in~\cite{Eguchi:2002nx}.

Moreover, we provide the numerator $\phi_b^{(N)}$ of $M_2$ for base degree $b=2$ : 
\begin{tiny}
\begin{align}
\begin{split}
\phi_2^{(2)}=&\frac{\Delta_4^2}{\eta^{24}(2\tau)}\Bigg(-\frac{497 \mathcal{A}_{+,2}^4 \mathcal{A}_{-,2}^2 E_{2,2}^6}{373248}+\frac{29 \mathcal{A}_{+,2}^4
   \mathcal{A}_{-,2}^2 E_{2,2}^4 E_4}{124416}+\frac{13 \mathcal{A}_{+,2}^4 \mathcal{A}_{-,2}^2
   E_{2,2}^2 E_4^2}{41472} -\frac{\mathcal{A}_{+,2}^4 \mathcal{A}_{-,2}^2
   E_4^3}{13824} \\
   &+\frac{127 \mathcal{A}_{+,2}^4 \mathcal{A}_{-,2} \mathcal{B}_{-,2}
   E_{2,2}^5}{93312} -\frac{35 \mathcal{A}_{+,2}^4 \mathcal{A}_{-,2} \mathcal{B}_{-,2} E_{2,2}^3
   E_4}{31104} +\frac{\mathcal{A}_{+,2}^4 \mathcal{A}_{-,2} \mathcal{B}_{-,2} E_{2,2}
   E_4^2}{5184}-\frac{\mathcal{A}_{+,2}^4 \mathcal{B}_{-,2}^2 E_{2,2}^4}{2916}\\
   &+\frac{5 \mathcal{A}_{+,2}^4 \mathcal{B}_{-,2}^2 E_{2,2}^2 E_4}{13824} -\frac{\mathcal{A}_{+,2}^4
   \mathcal{B}_{-,2}^2 E_4^2}{13824}+\frac{533 \mathcal{A}_{+,2}^3 \mathcal{A}_{-,2}^2
   \mathcal{B}_{+,2} E_{2,2}^5}{93312} -\frac{23 \mathcal{A}_{+,2}^3 \mathcal{A}_{-,2}^2 \mathcal{B}_{+,2}
   E_{2,2}^3 E_4}{10368}+\frac{\mathcal{A}_{-,2}^2 \mathcal{B}_{+,2}^4
   E_4}{10368}\\
   &+\frac{\mathcal{A}_{+,2}^3 \mathcal{A}_{-,2}^2 \mathcal{B}_{+,2}
   E_{2,2} E_4^2}{5184}
   -\frac{257 \mathcal{A}_{+,2}^3 \mathcal{A}_{-,2} \mathcal{B}_{+,2}
   \mathcal{B}_{-,2} E_{2,2}^4}{46656}+\frac{55 \mathcal{A}_{+,2}^3 \mathcal{A}_{-,2} \mathcal{B}_{+,2}
   \mathcal{B}_{-,2} E_{2,2}^2 E_4}{20736} +\frac{\mathcal{A}_{-,2} \mathcal{B}_{+,2}^4 \mathcal{B}_{-,2}
   E_{2,2}}{5832}\\
   & -\frac{7 \mathcal{A}_{+,2}^3 \mathcal{A}_{-,2}
   \mathcal{B}_{+,2} \mathcal{B}_{-,2} E_4^2}{20736}
   +\frac{29 \mathcal{A}_{+,2}^3 \mathcal{B}_{+,2}
   \mathcal{B}_{-,2}^2 E_{2,2}^3}{23328}-\frac{\mathcal{A}_{+,2}^3 \mathcal{B}_{+,2} \mathcal{B}_{-,2}^2
   E_{2,2} E_4}{2592}-\frac{193 \mathcal{A}_{+,2}^2 \mathcal{A}_{-,2}^2 \mathcal{B}_{+,2}^2
   E_{2,2}^4}{31104}\\
   &+\frac{25 \mathcal{A}_{+,2}^2 \mathcal{A}_{-,2}^2 \mathcal{B}_{+,2}^2 E_{2,2}^2
   E_4}{13824}-\frac{\mathcal{A}_{+,2}^2 \mathcal{A}_{-,2}^2 \mathcal{B}_{+,2}^2
   E_4^2}{13824}+\frac{11 \mathcal{A}_{+,2}^2 \mathcal{A}_{-,2} \mathcal{B}_{+,2}^2 \mathcal{B}_{-,2}
   E_{2,2}^3}{1944}-\frac{1}{648} \mathcal{A}_{+,2}^2 \mathcal{A}_{-,2} \mathcal{B}_{+,2}^2
   \mathcal{B}_{-,2} E_{2,2} E_4\\
   &-\frac{7 \mathcal{A}_{+,2}^2 \mathcal{B}_{+,2}^2 \mathcal{B}_{-,2}^2
   E_{2,2}^2}{7776}
   +\frac{\mathcal{A}_{+,2}^2 \mathcal{B}_{+,2}^2 \mathcal{B}_{-,2}^2
   E_4}{10368}+\frac{59 \mathcal{A}_{+,2} \mathcal{A}_{-,2}^2 \mathcal{B}_{+,2}^3
   E_{2,2}^3}{23328}-\frac{5 \mathcal{A}_{+,2} \mathcal{A}_{-,2}^2 \mathcal{B}_{+,2}^3 E_{2,2}
   E_4}{7776}+\frac{\mathcal{B}_{+,2}^4 \mathcal{B}_{-,2}^2}{5832}\\
   &-\frac{23 \mathcal{A}_{+,2} \mathcal{A}_{-,2} \mathcal{B}_{+,2}^3 \mathcal{B}_{-,2}
   E_{2,2}^2}{11664}+\frac{7 \mathcal{A}_{+,2} \mathcal{A}_{-,2} \mathcal{B}_{+,2}^3 \mathcal{B}_{-,2}
   E_4}{15552}-\frac{\mathcal{A}_{+,2} \mathcal{B}_{+,2}^3 \mathcal{B}_{-,2}^2
   E_{2,2}}{5832}-\frac{\mathcal{A}_{-,2}^2 \mathcal{B}_{+,2}^4
   E_{2,2}^2}{2916}\Bigg)\,.
      \end{split}
   \end{align}
\end{tiny}
Here we introduced the compact notation $\mathcal{A}_{\pm,N} \equiv \phi_{-2,1}(N\tau,\epsilon^\pm)\,,  \mathcal{B}_{\pm,N} \equiv \phi_{0,1}(N\tau,\epsilon^\pm)$, and $E_2^{(N)}\equiv E_{2,N}$.
Analogous expressions for the case $M_3$ and $M_4$ can be found in Appendix~\ref{app:modexp}.
Expressions $\phi_b^{(N)}$ for $M_N'$ are quite similar to the $M_N$ cases and hence we omit them.
Instead, we provide some refined BPS invariants in Appendix \ref{app:refBPS} up to $b=2$.
We provide a notebook online~\cite{DataLink} that contains higher degree invariants up to $b=3$ for all cases that we consider.

Note that the E-string has a dual interpretation in heterotic $E_8\times E_8$ theory.
From the Horava-Witten perspective it arises from M2-branes that are stretched between an M5 and M9 brane on the interval that is used to compactify M-theory.
If we consider a generic elliptic fibration over the Hirzebruch surface $\mathbb{F}_1$, the E-string arises from D3-branes wrapping the $(-1)$-curve while the corresponding heterotic string arises
from D3-branes that wrap the $\mathbb{P}^1$ fiber of the base.
In the previous section we have matched the invariants from the restriction of the genus-one fibration to the fiber $\mathbb{P}^1$ for $2$-section geometries with the one-loop calculation in 
a heterotic compactification on $(K3\times T^2)/\mathbb{Z}_N$.
It is therefore natural to expect an interpretation of the flavour group for the strings from the $(-1)$-curve in terms of small instantons in the corresponding CHL-model.
We leave a detailed investigation of this question to future work.

\section{Examples}
\label{sec:examples}
In this section we will illustrate the general discussion of the previous sections with several examples of genus one fibered Calabi-Yau threefolds.
One set of geometries is related via a chain of birational transformations that do not change the base of the fibration.
In the effective action of the corresponding F-theory compactification this is manifested as Higgsing and un-Higgsing the gauge group.
We will therefore use the name \textit{Higgs chain} for a set of genus one fibered Calabi-Yau manifolds that are related via extremal transitions which do not involve blow ups or blow downs of the base.
Moreover, when we refer to the gauge group of a geometry this is to be understood as the gauge group of the corresponding F-theory effective action.

All of our examples are obtained using the fiber based construction that we review in the following Section~\ref{sec:construction}.
We use the notation
\begin{align}
	(F_{i}\rightarrow B)[G]_{h^{1,1}}^{\chi}\,,
\end{align}
for a genus one fibered Calabi-Yau with base $B$ such that the generic fiber is constructed as a hypersurface in $\mathbb{P}_{F_i}$
The gauge group of the corresponding F-theory vacuum is $G$ and the hodge numbers are determined by $h^{1,1}$ and the Euler characteristic $\chi$.
If some of the divisors are not torically realized we indicate this by writing e.g. $h^{1,1}=3(1)$ if one out of three generators is non-toric.

In general this data does not determine the geometry uniquely.
The reason is that for a given fiber polytope the geometry is determined by fixing two line bundles on the base.
Since the number of divisors and the gauge group are the same for all sufficiently generic choices of the bundles there is only one constraint imposed by the Euler characteristic.
However, in all the cases that we discuss in this section the bundles are non-generic and there is no such ambiguity.

This being said, we will discuss the Higgs chain that includes geometries with the data listed in \eqref{eqn:higgschains}:
{
\begin{align}
	\begin{split}
		\begin{array}{ccccc}
			(F_{10}\rightarrow \mathbb{P}^2)[SU(2)]_{3}^{-216}&\rightarrow&(F_6\rightarrow\mathbb{P}^2)[U(1)]_{3}^{-216}&\rightarrow&(F_{4}\rightarrow \mathbb{P}^2)[\mathbb{Z}_2]_{2}^{-252}
		\end{array}
	\end{split}
	\label{eqn:higgschains}
\end{align}
}
But first we are going to review the general strategy to construct elliptic and genus one fibered Calabi-Yau manifolds with the tools of toric geometry.

\subsection{Eliptically fibered Calabi-Yau as toric hypersurfaces}
\label{sec:construction}
Still the most abundant source of Calabi-Yau threefolds are hypersurfaces and complete intersections in toric ambient spaces.
The Batyrev construction associates a mirror pair of $d$-dimensional Calabi-Yau hypersurfaces to every $d+1$-dimensional reflexive polyotope~\cite{1993alg.geom.10003B}.
As a slight abuse of terminology we will call such varieties toric Calabi-Yau hypersurfaces.
Four dimensional reflexive polytopes have been fully classified and the complete Kreuzer-Skarke list with 473,800,776 entries is available online~\cite{Kreuzer:2000xy}.
On the other hand, complete intersection Calabi-Yau $d$-folds of codimension $m$ are related to nef-partitions of $d+m$-dimensional reflexive polytopes~\cite{Batyrev:1994pg}.
For recent progress on the classification of five-dimensional reflexive polytopes see~\cite{Scholler:2018apc}.

A given Calabi-Yau $X$ is said to be genus one fibered over a base $B$ if there exists a surjective map $\pi:\,X\rightarrow B$ such that the generic fiber over $B$ is a torus.
If the projection $\pi$ admits a section then the fibration is called elliptic.
It is well known that the majority of toric Calabi-Yau hypersurfaces is genus one fibered in at least one and often multiple ways.
For a recent discussion and pointers to the literature see e.g. \cite{Huang:2018esr}.
To obtain explicit examples of genus one fibered Calabi-Yau hypersurfaces there are three general strategies:
\begin{figure}[t!]
	\centering
	\includegraphics[width=.6\linewidth]{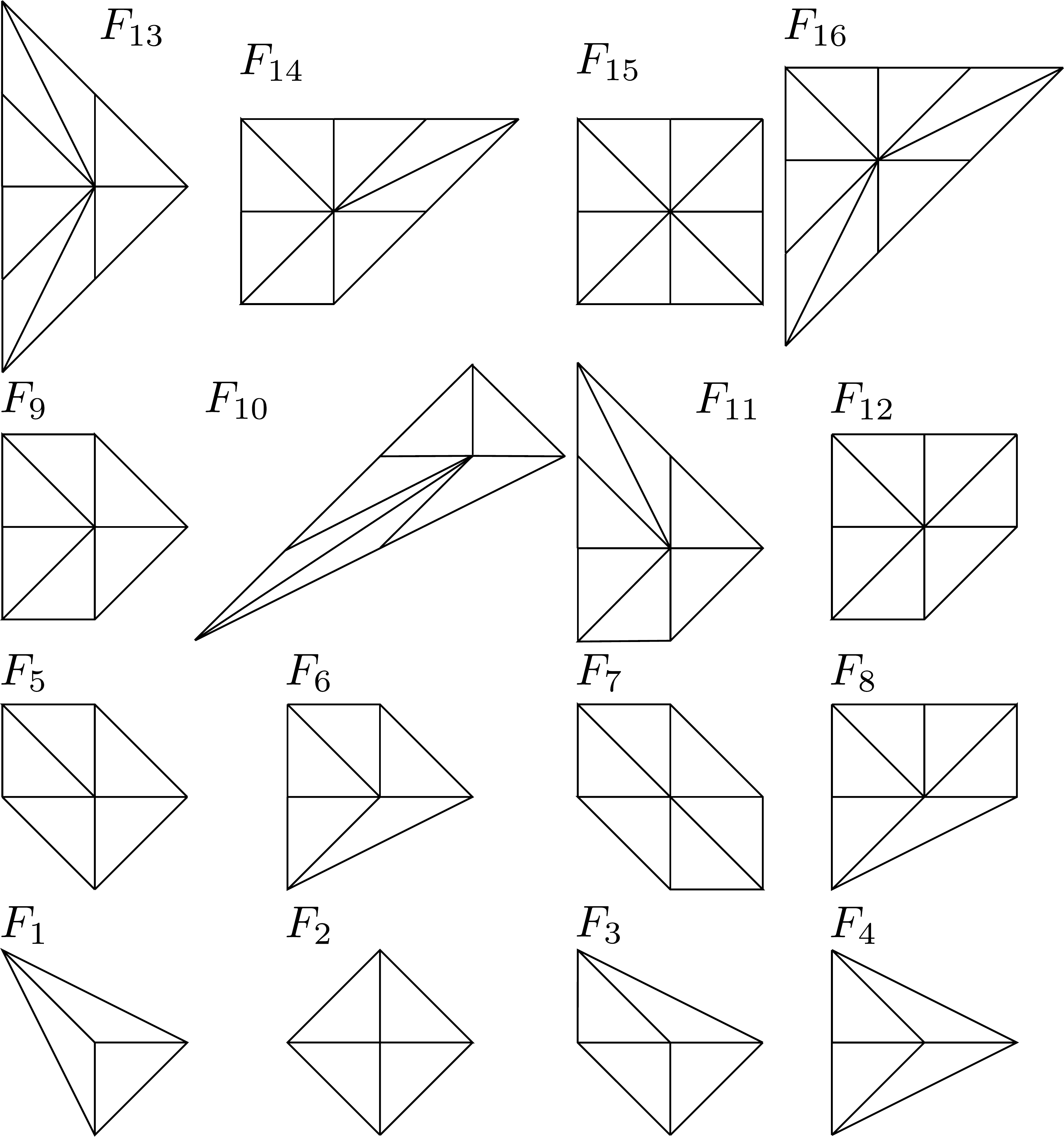}
	\caption{The 16 two-dimensional reflexive polytopes. The image is taken from~\cite{Klevers:2014bqa}.}
	\label{fig:2dreflexive}
\end{figure}

\begin{enumerate}
	\item It is possible to manually engineer polytopes that correspond to toric fibrations such that the generic hypersurface cuts out a genus one curve from the generic fiber.
	\item For a given reflexive polytope one can systematically search for the toric fibrations of the corresponding toric variety.
		This has been used by \cite{Braun:2011ux} and \cite{Scholler:2018apc} to scan for fibrations in the complete Kreuzer-Skarke list.
	\item One can first construct an elliptic curve as a hypersurface in a toric variety. The coefficients of the hypersurface equation can then be lifted to sections of line bundles over some base that we will also assume to be toric.
		For every  pair of fiber and base this leads to a finite number of ``reasonable'' choices. 
		This is what we refer to as the ``fiber based approach''.
\end{enumerate}

The fiber based construction has the advantage that it is quite general and fibrations with very specific properties can be engineered.
A comphrehensive study of the properties of the fibers that can be obtained from the 16 two-dimensional reflexive polytopes that are shown in figure \ref{fig:2dreflexive} has been performed in~\cite{Klevers:2014bqa} and we will frequently refer to these results.
Furthermore, a reflexive polytope that corresponds to the total space can often be recovered from the choice of fiber, base and bundles.

We will now review the general construction at the hand of a particular example.
Note that although we only work with hypersurfaces the discussion can be easily generalized to complete intersections.
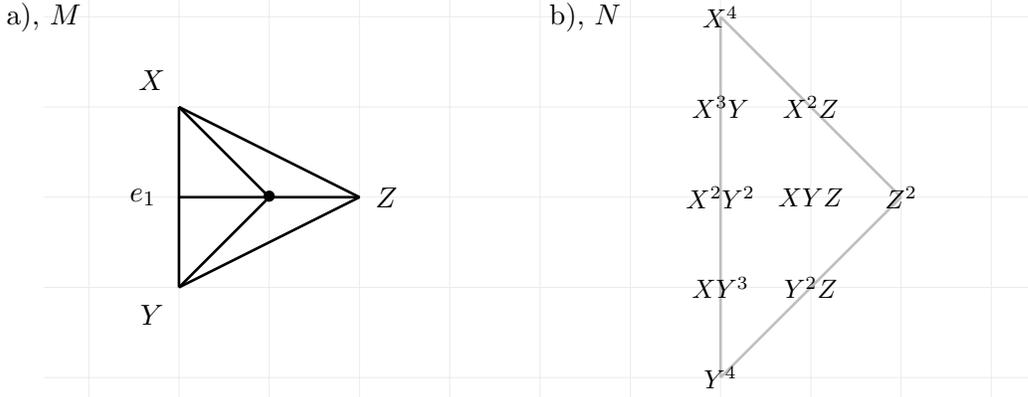
\begin{figure*}[t]
\centering
\vspace{3cm}
\begin{tikzpicture}[remember picture,overlay,node distance=4mm, >=latex',block/.style = {draw, rectangle, minimum height=65mm, minimum width=83mm,align=center},]
\begin{scope}[shift={(3,0)},scale=1.2]
\draw[help lines, overlay, lightgray!30] (-8.5,-0.5) grid (-3.5,4.5);
\node[overlay] at (-8.5,4) {a), $M$};
\draw [line width=1] (-7,3) -- (-5,2);
\draw [line width=1] (-7,3) -- (-7,1);
\draw [line width=1] (-7,1) -- (-5,2);
\draw [line width=1] (-7,1) -- (-6,2);
\draw [line width=1] (-7,3) -- (-6,2);
\draw [line width=1] (-5,2) -- (-6,2);
\draw [line width=1] (-7,2) -- (-6,2);

\node[overlay] at (-7.3, 3.3) {$X$};
\node[overlay] at (-7.4, 2) {$e_1$};
\node[overlay] at (-7.3, .7) {$Y$};
\node[overlay] at (-4.7, 2) {$Z$};

\node at (-6,2) {\textbullet};
\end{scope}

\begin{scope}[shift={(3,0)},scale=1.2]
\draw[help lines, overlay, lightgray!30] (-3.5,-0.5) grid (2.5,4.5);
\node[overlay] at (-2.5,4) {b), $N$};
\draw [line width=1, lightgray!100] (-1,4) -- (-1,0);
\draw [line width=1, lightgray!100] (-1,4) -- ( 1,2);
\draw [line width=1, lightgray!100] (-1,0) -- ( 1,2);

\node[overlay] at (-1,4) {\small$X^4$};
\node[overlay] at (-1,0) {\small$Y^4$};
\node[overlay] at ( 1,2) {\small$Z^2$};
\node[overlay] at ( 0,2) {\small$XYZ$};
\node[overlay] at (-1,3) {\small$X^3Y$};
\node[overlay] at (-1,2) {\small$X^2Y^2$};
\node[overlay] at (-1,1) {\small$XY^3$};
\node[overlay] at ( 0,3) {\small$X^2Z$};
\node[overlay] at ( 0,1) {\small$Y^2Z$};
\end{scope}
\end{tikzpicture}
\caption{
The dual pair of polytopes $F_4$ and $F_{13}$ is shown in a) and b).
We also indicate the toric fan obtained from a complete star triangulation of $F_4$ and labelled the homogeneous coordinates.
In the monomials that correspond to points of $F_{13}$ we have set $e_1=1$. The dependence can be easily restored.
}
\label{fig:f4f13}
\end{figure*}
Figure \ref{fig:f4f13} shows the dual pair of reflexive polytopes $F_4$ and $F_{13}$ that are subsets of the respective lattices $M$ and $N$.
We construct the generic fiber as a hypersurface in the toric variety that corresponds to the face fan of $F_4$.
In terms of homogeneous coordinates the ambient space consists of the points
\begin{align}
	[X:Y:Z:e_1]\in\mathbb{C}^4\backslash\left(\{X=Y=0\}\cup\{e_1=Z=0\}\right)\,,
\end{align}
that are identified under the equivalence relation
\begin{align}
	[X:Y:Z:e_1]\sim[\lambda_1 X:\lambda_1^{-1}\lambda_2^{-2}Y:\lambda_2Z:\lambda_2^{-1}e_1]\,,
	\label{eqn:equivalence}
\end{align}
for all $\lambda_1,\lambda_2\in\mathbb{C}^*$.
We will denote this space as $\mathbb{P}_{F_4}$ which is in line with the conventions of \cite{Klevers:2014bqa}~\footnote{Note that this is at odds with e.g. \cite{Cox:2000vi} where the name $\mathbb{P}_{F_{13}}$ would have been used.}.
A generic section of the anti-canonical line bundle takes the form
\begin{align}
	\begin{split}
	p_{F_4}=&e_1^2(d_1X^4+d_2X^3Y+d_3X^2Y^2+d_4XY^3+d_5Y^4)\\
	&+e_1(d_6X^2+d_7XY+d_8Y^2)Z+d_9Z^2\,.
	\end{split}
	\label{eqn:f4polynomial}
\end{align}

At this level the coefficients $d_i,\,i=1,...,9$ are complex numbers and redundantly parametrize the complex structure of the elliptic curve $\{p_{F_4}=0\}\subset\mathbb{P}_{F_4}$.
We can construct a genus one fibration over a base $B$ by choosing four line bundles of which $X,Y,Z,e_1$ are taken to be sections.
The requirement that $p_{F_4}$ is also a section of a well-defined bundle and that $X=\{p_{F_4}=0\}$ is a Calabi-Yau manifold then also fixes the bundles of which the coefficients have to be sections.
In fact, the choice of four bundles is redundant since the equivalence relation \eqref{eqn:equivalence} can be used to let e.g. $X^{-1}Z$ and $e_1$ be sections of the trivial bundle.
Again following the conventions laid out in \cite{Klevers:2014bqa} we pick two line bundles $\mathcal{S}_7,\mathcal{S}_9$ on $B$ and fix the classes
\begin{align}
	\begin{split}
		&[X]=H-E_1+\mathcal{S}_9-c_1(B)\,,\quad [Y]=H-E_1-\mathcal{S}_7+\mathcal{S}_9\,,\\
		&[Z]=2H-E_1+\mathcal{S}_9-c_1(B)\,,\quad [e_1]=E_1\,,
	\end{split}
	\label{eqn:f4coordinateclass}
\end{align}
for the toric divisors of $\mathbb{P}_{F_4}$.

\paragraph{The class of coefficients} The Calabi-Yau requirement imposes
\begin{align}
	[p_{F_4}]=c_1(B)+c_1(\mathbb{P}_{F_4})=c_1(B)+[X]+[Y]+[Z]+[e_1]\,,
\end{align}
and therefore fixes the class of the coefficient $[d_7]=c_1(B)$ that corresponds to the unique inner point of $F_{13}$.
It is now easy to see that the charge of the coefficient that multiplies a monomial corresponding to any $m\in F_{13}$ is given by
\begin{align}
	[d_m]=c_1(B)+\sum\limits_{\rho\in\Sigma(1)}\langle m,\rho\rangle [D_\rho]_B\,.
	\label{eqn:coefficientclass}
\end{align}
Here we introduced the notation $D_\rho$ for the divisor that corresponds to the generator $\rho\in\Sigma(1)$ and $[D]_B$ for the base part of the class of a divisor.
Inserting the classes of $[X],[Y],[Z]$ and $[e_1]$ this determines
\begin{align}
	\begin{split}
		&[d_1]=3c_1(B)-\mathcal{S}_7-\mathcal{S}_9\,,\quad [d_2]=2c_1(B)-\mathcal{S}_9\,,\quad [d_3]=c_1(B)+\mathcal{S}_7-\mathcal{S}_9\,,\\
		&[d_4]=2\mathcal{S}_7-\mathcal{S}_9\,,\quad [d_5]=-c_1(B)+3\mathcal{S}_7-\mathcal{S}_9\,,\quad [d_6]=2c_1(B)-\mathcal{S}_7\,,\\
		&[d_7]=c_1(B)\,,\quad [d_8]=\mathcal{S}_7\,,\quad [d_9]=c_1(B)-\mathcal{S}_7+\mathcal{S}_9\,.
	\end{split}
	\label{eqn:f4coefficientClass}
\end{align}

From now on we will assume that the base $B$ is itself a smooth toric variety.
The line bundles on $B$ are then determined by charge vectors in terms of a basis of effective divisors.
Let us assume that the cone of effective divisors is generated by $k$ independent classes and therefore the charges are elements $\vec{q}\in\mathbb{Z}^k$.
We will also assume that the cone is simplicial i.e. $k$-dimensional although this assumption can be easily dropped. 
In any case, \eqref{eqn:coefficientclass} can now be read component wise.
If we demand that every coefficient $d_i,\,i=1,...,9$ appears in $p_{F_4}$ this leads to $k$ inequalities
\begin{align}
	q_j([d_m])=q_j\left(c_1(B)+\sum\limits_{\rho\in\Sigma(1)}\langle m,\rho\rangle [D_\rho]_B\right)\ge 0\,,\quad j=1,...,k\,,
	\label{eqn:effectiveconstraint}
\end{align}
in terms of the $k$-th charges of $\mathcal{S}_7$ and $\mathcal{S}_9$ for every monomial $m\in F_{12}$.
The inequalities define $k$ polytopes in $k$ $\left(q_i(\mathcal{S}_7),q_i(\mathcal{S}_9)\right)$-planes and each one is related via a lattice automorphism to $q_i(c_1(B))\cdot F_4$.

Note that if we allow some of the coefficients to generically vanish we might as well work with the toric ambient space that is associated to the dual of the Newton polytope of the non-vanishing monomials.
Due to \eqref{eqn:coefficientclass} it is not possible that the hull of the set of points that correspond to coefficients in an effective class is non-convex.
We can therefore impose \eqref{eqn:effectiveconstraint} wihout loss of generality.

\paragraph{Example $B=\mathbb{F}_1$} To make this somewhat abstract discussion more concrete we illustrate it for $B=\mathbb{F}_1=\mathbb{P}_{F_3}$.
This can be constructed as the quotient
\begin{align}
	[u:v:w:e_1]\in\frac{\mathbb{C}^4\backslash\left(\{v=u=0\}\cup\{w=e_1=0\}\right)}{\sim}\,,
\end{align}
where the equivalence relations are
\begin{align}
	[u:v:w:e_1]\sim [\lambda_1\lambda_2^{-1}u:\lambda_2v:\lambda_1^{-1}w:\lambda_1e_1]\,,
	\label{eqn:relationsF1}
\end{align}
for all $\lambda_1,\lambda_2\in\mathbb{C}^*$.
The cone of effective divisors on $B$ is generated by $u$ and $w$ and the charge of the first Chern class with respect to this basis is $\vec{q}(c_1(B))=(1,2)$. 
The solutions to the inequalities \eqref{eqn:effectiveconstraint} are the points in the shaded regions shown in figure \ref{fig:f4overf1charges}.

\begin{figure}[h]
\centering
\begin{tikzpicture}
\begin{scope}[shift={(4,0)},scale=1]
	\draw[help lines, overlay, lightgray!30] (0,-2) grid[step={(1,1)}] (4,4);
\draw[fill=cyan, ultra thick] (0,-2) -- (4,2) -- (2,4) -- cycle; 
\draw [->,line width=1] (0,-.3) -- (0,3.7);
\draw [->,line width=1] (-.3,0) -- (3.7,0);
	\node at (0,4.3) {$q_2(\mathcal{S}_9)$};
	\node at (4.7,0) {$q_2(\mathcal{S}_7)$};
	\node at (2,4.4) {$(2,4)$};
	\node at (4.8,2) {$(4,2)$};
	\node at (-.2,-2.5) {$(0,-2)$};
\end{scope}
\begin{scope}[shift={(-4,0)},scale=1]
	\draw[help lines, overlay, lightgray!30] (0,-2) grid[step={(2,2)}] (4,4);
\draw[fill=magenta, ultra thick] (0,-2) -- (4,2) -- (2,4) -- cycle; 
\draw [->,line width=1] (0,-.3) -- (0,3.7);
\draw [->,line width=1] (-.3,0) -- (3.7,0);
	\node at (0,4.3) {$q_1(\mathcal{S}_9)$};
	\node at (4.7,0) {$q_1(\mathcal{S}_7)$};
	\node at (2,4.4) {$(1,2)$};
	\node at (4.8,2) {$(2,1)$};
	\node at (-.2,-2.5) {$(0,-1)$};
\end{scope}
\end{tikzpicture}
\label{fig:f4overf1charges}
\end{figure}
The properties of the resulting fibrations for a generic choice of $\mathcal{S}_7,\mathcal{S}_9$ have been summarized in \cite{Klevers:2014bqa}.
If one chooses $\mathcal{S}_7,\mathcal{S}_9$ such that the charge is on the boundary of both polytopes the class of some of the coefficients $d_i$ will be trivial.
It is still easy to deduce the resulting properties from the generic expressions that have been provided in \cite{Klevers:2014bqa}.
In particular the sections, multi-sections and singularities of the fibrations in various codimensions have been determined for all hypersurfaces that correspond to the 16 two-dimensional reflexive polytopes.
A powerful strategy is therefore to select a fiber-polytope that leads to the desired gauge group via F-theory of which the non-Abelian part can be further broken down with a particular choice of $\mathcal{S}_7,\mathcal{S}_9$.
We will now review how for any such choice of fiber, toric base and bundles one can recover the reflexive polytope of the total ambient space.

\paragraph{Recovering the polytope}
Again we provide an example that can easily be generalized.
Let us stick to $\mathbb{P}_{F_4}$ as the ambient space of the fiber and $B=\mathbb{F}_1=\mathbb{P}_{F_3}$ as the base.
A generic choice of bundles would be $\mathcal{S}_7=(1,3)$ and $\mathcal{S}_9=(2,2)$.
From \eqref{eqn:equivalence} and \eqref{eqn:relationsF1} can deduce that this corresponds to the relations
\begin{align}
	\begin{split}
		&[X:Y:Z:e_1:u:v:w:e_1']\\\sim &[\lambda^{l^{(i)}_1}X:\lambda^{l^{(i)}_2}Y:\lambda^{l^{(i)}_3}Z:\lambda^{l^{(i)}_4}e_5:\lambda^{l^{(i)}_6}u:\lambda^{l^{(i)}_7}v:\lambda^{l^{(i)}_8}w:\lambda^{l^{(i)}_9}e_1']\,,
	\end{split}
\end{align}
for all $\lambda\in\mathbb{C}^*$ and $i=1,...,4$, where
		\begin{equation}
			\begin{tabu}{rrrrrrrrr}
		l^{(1)}=&( 1,& 1,& 1,& 0,& 1,& 1,& 0,&-1)\,,\\
		l^{(2)}=&( 0,&-1,& 0,& 0,& 0,& 0,& 1,& 1)\,,\\
		l^{(3)}=&( 1,& 1,& 0,&-2,& 0,& 0,& 0,& 0)\,,\\
		l^{(4)}=&( 0,& 0,& 1,& 1,& 0,& 0,& 0,& 0)\,.
			\end{tabu}
		\end{equation}
The last four entries of $l^{(1)},l^{(2)}$ form a basis of charges for the equivalence relations that define $\mathbb{F}_1$ such that $u$ has charge $1$ under $l^{(1)}$ and charge $0$ under the relation $l^{(1)}$ while the opposite holds for $w$.
This implies that the first four entries of $l^{(i)},\,i=1,2$ consist of the charges $q_i(X),q_i(Y),q_i(Z),q_i(e_1)$ that are listed in table \ref{eqn:f4overf1charges}.
\begin{table}[h!]
	\centering
	\begin{tabular}{c|cccc}
		&$X$&$Y$&$Z$&$e_1$\\
		$q_1$&$1$&$1$&$1$&$0$\\
		$q_2$&$0$&$-1$&$0$&$0$
	\end{tabular}
	\caption{Charges of the homogeneous coordinates on $\mathbb{P}_{F_4}$ with respect to the basis of effective divisors on $B=\mathbb{F}_1$ given by $u$,$w$ for the choice $\mathcal{S}_7=(1,3)$ and $\mathcal{S}_9=(2,2)$.}
	\label{eqn:f4overf1charges}
\end{table}
Those can in turn be obtained from $q_i(\mathcal{S}_7)$ and $q_i(\mathcal{S}_9)$ via equation \eqref{eqn:f4coordinateclass}.
The relations $l^{(3)}$ and $l^{(4)}$ directly correspond to relations among the fiber coordinates \eqref{eqn:equivalence}.

Let us denote the $8\times 4$ matrix of relations as $Q$, i.e. $Q_{ij}=l^{(j)}_i$.
Note that the exponents of the scaling relations among the homogeneous coordinates of a toric variety directly correspond to coefficients in linear relations among the points that generate the one-dimensional cones $\Sigma(1)$ of the fan.
One can often obtain a reflexive polytope from the set of relations by considering the kernel
\begin{align}
	V=\ker(Q^t)\,,
\end{align}
and then finding a basis of the sublatttice $V\cap\mathbb{Z}^8$.
We now take the elements of such a basis as the columns of a matrix
\begin{align}
	A=\left(\begin{array}{rrrr}
		1&0&0&0\\
		-1&1&0&0\\
	       -1&-1&0&0\\
		0&0&1&0\\
		0&0&0&1\\
		0&1&-1&-1\\
		-1&1&-1&0
	\end{array}\right)\,.
	\label{eqn:pointsf4overf3}
\end{align}
The rows of $A$ determine a set of points in $\mathbb{Z}^4$ and the convex hull is the desired reflexive polytope $\Delta$.
This algorithm to obtain the polytope from a basis of relations has been reviewed e.g. in \cite{Aspinwall:2015zia}.

Looking at the points in \eqref{eqn:pointsf4overf3} we see that there is a 2-dimensional sublattice $L$ that intersects $\Delta$ as $L\cap \Delta=F_4$.
On the other hand, projection on the last two coordinates recovers the polytope $F_3$ that correpsonds to the base of the fibration.
In fact this structure is necessary for the corresponding toric variety to be an $\mathbb{P}_{F_4}$ fibration over $\mathbb{P}_{F_3}$.

\subsection{The Higgs chain $SU(2)\rightarrow U(1)\rightarrow \mathbb{Z}_2$}
\begin{figure}[t]
	\centering
\includegraphics[width=.7\linewidth]{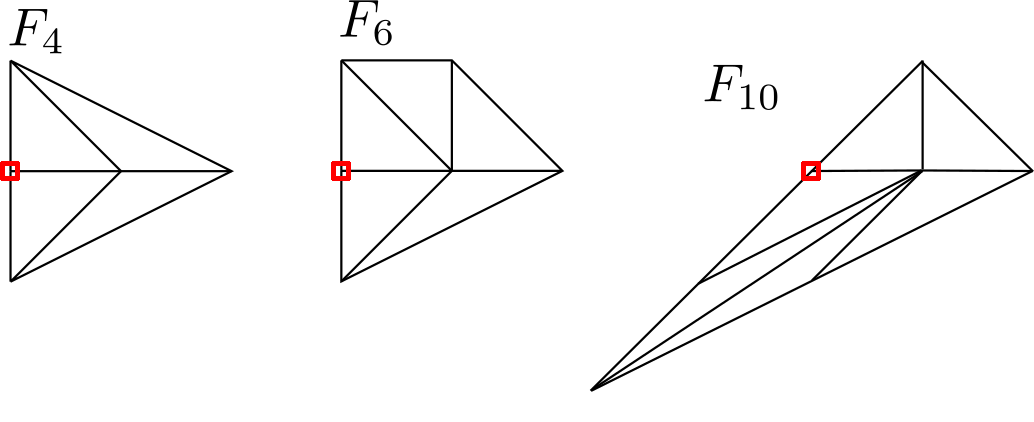}
	\caption{Fibers of the Higgs chain $\mathbb{Z}_2\leftarrow U(1)\leftarrow SU(2)$.}
	\label{fig:chain2}
\end{figure}
As our first set of examples we study the Higgs chain that is indicated in figure \ref{fig:chain2} where the base for all of the fibrations is $B=\mathbb{P}^2$.
At the bottom of the chain we start with a genus one fibered Calabi-Yau threefold $M^h_1=(F_{4}\rightarrow\mathbb{P}^2)[\mathbb{Z}_2]_{2}^{-252}$ that only admits a two-section and no fibral divisors.

This is related via an extremal transition to the elliptic fibration $M^h_2=(F_{6}\rightarrow\mathbb{P}^2)[U(1)]_{3}^{-216}$ that has two independent sections.
In particular, the corresponding F-theory effective action exhibits hypermultiplets of charge one and two.
In geometric terms the charge two matter arises from fibral curves that are wrapped by the section that generates the Mordell-Weil group and are intersected transversely by the zero section.

On a subslice of the complex structure moduli space of $M^h_2$ some of the isolated fibral curves deform into families of $I_2$ fibers over a genus $g=10$ Riemann surface in the base. 
This enhances the Abelian gauge group of the corresponding F-theory vacuum into $SU(2)$.
Other fibral curves remain isolated and lead to $n=72$ hypermultiplets in the fundamental representation.
Note that the number of the charge two loci in $M^h_2$ is $2g-2=18$ while the number of charge one loci is $2n=144$.

We construct this subslice as a hypersurface in a toric ambient space $M^h_3=(F_{10}\rightarrow\mathbb{P}^2)[SU(2)]_{3}^{-216}$ where $g=10$ complex structure deformations are non-polynomial.
The discriminant becomes reducible and on one component the fibral divisor collapses to a curve in the base.

\subsubsection{$M^h_1=(F_{4}\rightarrow \mathbb{P}^2)[\mathbb{Z}_2]_{2}^{-252}$}
The fiber of $M^h_1$ is a hypersurface in $\mathbb{P}_{F_{4}}$ and the base remains $\mathbb{P}^2$.
Choosing the line bundles $\mathcal{S}_7=3H,\,\mathcal{S}_9=0$ with $H$ being the class of the hyperplane in $\mathbb{P}^2$ leads to the toric data given in \eqref{eqn:f4overf1toricData},
\begin{align}
\begin{blockarray}{crrrrrrl}
	&&&&&C_1&C_2\\
\begin{block}{c(rrrr|rr)l}
	x_1& 1& 0&-1&0& 0& 1&\leftarrow\text{vertical divisor }D_b\\
	x_2& 0& 1&-1&0& 0& 1&\phantom{x}\hspace{1.5cm}\text{\ditto}\\
	x_3&-1&-1&-1&0& 0& 1&\phantom{x}\hspace{1.5cm}\text{\ditto}\\
	x& 0& 0&-1& 1& 1& 0&\leftarrow\text{two-section }E_0\\
	y& 0& 0&-1&-1& 1&0&\phantom{x}\hspace{1.5cm}\text{\ditto}\\
	z& 0& 0& 1& 0& 2& 3&\leftarrow\text{four-section}\\
	& 0& 0& 0& 0& -4&-6&\\
\end{block}
\end{blockarray}\,.
\label{eqn:f4overf1toricData}
\end{align}
The relevant points of the polytope admit two different regular fine star triangulations.
Only one of the triangulations is compatible with the toric morphism that induces the fibration.
But the curve that is flopped when moving from one geometric phase to the other is not contained in the generic Calabi-Yau hypersurface.
Therefore the flop that connects the two phases is ineffective and we should work with the intersection of the Mori cones.
The latter is generated by the curves $C_1$ and $C_2$ that are intersecting the toric divisors as listed in \eqref{eqn:f4overf1toricData}.

The geometry of the image of the toric divisors on the Calabi-Yau is as follows.
As usual, $D_b$ is the class of the pullback of the hyperplane in $\mathbb{P}^2$.
In addition there are two equivalent two-sections $\{x=0\}$ and $\{y=0\}$ with $E_0=[x]=[y]$ and a four section $\{z=0\}$.
The class of the generic fiber is $C_1'=D_b^2$ while $C_2'=E_0\cdot D_b$ is the restriction of the two-section to the generator of the Mori cone of the base.
The latter is a double cover of a $\mathbb{P}^1$.
It is clear that $E\cdot C_1'=D_b \cdot C_2'=2$.

But the generic fiber is in fact not a generator of the Mori cone of $M^h_1$.
There are isolated $I_2$ singular fibers over $144$ points of $B$ and the two-sections intersect each of the two components once~\cite{Klevers:2014bqa}.
Physically, this is a consequence of charge minimality and the fact that the discrete symmetry that corresponds to the multi-section has to arise from a Higgsed $U(1)$ gauge symmetry.
After this Higgsing, the $144$ hypermultiplets with discrete charge $q=1$ are the remnants of the same number of hypermultiplets that are minimally charged under the $U(1)$.
We will study the un-Higgsed geometry below.

To fix the numerator of the topological string partition function for the lowest base degree, let us calculate the Gopakumar-Vafa invariants at genus zero and genus one.
To this end we choose a basis $J_1=E_0,\,J_2=D_b$ of the K\"ahler cone such that
\begin{align}
	\int_{C_i'}J_j=2\delta_{i,j}\,,
\end{align}
and use it to parametrize the complexified K\"ahler form as
\begin{align}
	\omega=t^1J_1+t^2J_2\,.
\end{align}
Note that $t^1$ is the volume of each component of the isolated $I_2$ fibers while $t^2$ is half of the volume of $C_2'$.
It turns out that this is the correct parametrization to obtain integer Gopakumar-Vafa invariants.
The topological invariants~\eqref{eqn:topdef} are
\begin{align}
	h^{1,1}=2\,,\quad\chi=-252\,,\quad \vec{b}=\int\limits_{M^h_1} c_2\left(M^h_1\right)\cdot \vec{J}=(24,36)^T\,,
\end{align}
and the only non-vanishing triple intersection number is $c_{122}=2$.
With this data the genus zero free energy can then be obtained e.g. using the techniques reviewed in~\cite{Hosono:1994av}.

To obtain the genus one free energy we will briefly discuss the mirror geometry $W$.
The polynomial that determines $W$ in the dual ambient space is of the form
\begin{align}
	\begin{split}
	P=&x_1^2 + (x_3x_4x_5)^4 + z_1(x_2x_6x_7)^4+ z_2(x_5x_6)^6 \\
	&+ (x_2x_3)^6  + (x_4x_7)^6+ x_1x_2x_3x_4x_5x_6x_7 \,.
	\end{split}
\end{align}
Here the coordinates $z_1,z_2$ on the complex structure moduli space are identified via mirror symmetry with $q_i+\mathcal{O}(q^2)$, where $q_i=e^{2\pi i t^i}$.
The Picard-Fuchs system is generated by the operators
\begin{align}
	\begin{split}
	\mathcal{D}_1=&\Theta _1^2-4z_1 \left(4 \Theta _1+6 \Theta _2+1\right) \left(4 \Theta _1+6 \Theta _2+3\right)\,,\\
	\mathcal{D}_2=&\Theta _2^3-8z_2 \left(4 \Theta _1+6 \Theta _2+1\right) \left(4 \Theta _1+6 \Theta _2+3\right) \left(4 \Theta _1+6 \Theta _2+5\right)\,.
	\end{split}
\end{align}
The discriminant is irreducible and corresponds to the vanishing locus of the polynomial
\begin{align}
	\Delta=1 - 192z_1 + 12288z_1^2 - 262144z_1^3 - 3456z_2 - 663552z_1z_2 + 2985984z_2^2\,.
\end{align}
We then find that the ansatz
\begin{align}
	\begin{split}
	F_1=&-\frac12\left(3+h^{1,1}-\frac{\chi}{12}\right)K-\frac12\log\,\det\,G^{-1}\\
	&-\frac{1}{24}\sum\limits_{i=1}^2(b_i+12)\log\,z^i-\frac{1}{12}\log\,\Delta\,,\\
	\end{split}
\end{align}
for the free energy at genus one leads to integer Gopakumar-Vafa invariants.

Let us now discuss the ``modular parametrization'' of the K\"ahler form.
Following our discussion in~\ref{ssec:relativeconi}, we calculate
\begin{align}
	E_0\cdot C_2'=0
\end{align}
and therefore $\tilde{E}_0=E_0$. This also implies that
\begin{align}
	\tilde{a}_b=\tilde{E}_0^2\cdot D_b=0\,.
\end{align}
We therefore parametrize the K\"ahler form as
\begin{align}
	\omega=\tau\cdot E_0+t\cdot D_b\,.
\end{align}

If we assume that poles in $q$ are always cancelled we can write down the general ansatz
\begin{align}
	Z_d(\tau,\lambda)=\frac{\Delta_4^{3d}}{\eta(2\tau)^{36d}}\frac{\phi_d(\tau,\lambda)}{\prod_{k=1}^d\phi_{-2,1}(2\tau,k\lambda)}\,,
\end{align}
where $\phi_d$ is a Jacobi form of weight $4d$ and index
\begin{align}
	\tilde{r}_\lambda^d=\frac14d(d-3)+\frac12\sum\limits_{k=1}^dk^2=\frac16d(d-1)(d+4)\,.
\end{align}
Together with the dimensions of the spaces of modular forms for $\Gamma_1(2)$,
\begin{align}
	\text{dim}\,M_w(2)=\left\lfloor\frac{w}{4}\right\rfloor+1\,,
\end{align}
one find that there are
\begin{align}
	\alpha_d=\tilde{r}_\lambda^d+(\tilde{r}_\lambda^d+1)d+\left\lfloor\frac14 \tilde{r}_\lambda^d(\tilde{r}_\lambda^d+1)\right\rfloor\,,
\end{align}
coefficients that determine $\phi_d$.
Using the Gopakumar-Vafa invariants at genus zero and genus one we can fix
\begin{align}
	\phi_1(\tau,\lambda)=192\left[12\left(E_2^{(2)}\right)^2+E_4\right]\,.
	\label{eqn:f4overf1z1}
\end{align}
To obtain the partition function at base degree two we have to make some assumption about the vanishing of Gopakumar-Vafa invariants.
Some invariants at base degree one that can be extracted from~\eqref{eqn:f4overf1z1} are shown in table~\ref{tab:f4overf1gv1} and one can observe that $n^g_{d_F,1}$ vanishes for $d_F<2g$.
\begin{table}[h!]
	\centering
	{\tiny
	\begin{align*}
\begin{array}{|c|cccc|}
\hline
d_F\backslash g & 0 & 1 & 2 & 3\\\hline
 0 & 2496 & 0 & 0 & 0\\ 
 1 & 216576 & 0 & 0 & 0\\
 2 & 6391296 & -4992 & 0 & 0\\
 3 & 104994816 & -433152 & 0 & 0\\
 4 & 1209337344 & -12797568 & 7488 & 0\\
 5 & 10917983232 & -211289088 & 649728 & 0\\
 6 & 82279299072 & -2457042432 & 19213824 & -9984\\
 7 & 538501165056 & -22467667968 & 318449664 & -866304 \\\hline
\end{array}
\end{align*}
	}
	\caption{Gopakumar-Vafa invariants for $(F_{4}\rightarrow \mathbb{P}^2)[\mathbb{Z}_2]_{2}^{-252}$ at base degree $1$.}
	\label{tab:f4overf1gv1}
\end{table}
It turns out that this does not hold for $d_B>1$ but at $d_B=2$ we can impose that $n^g_{d_F,1}$ vanishes for $d_F<2\lfloor g/2\rfloor$.
This fixes
{\small
\begin{align}
	\begin{split}
	\phi_2(\tau,\lambda)=&\frac{32}{9}A^4\cdot \left(12 g^2+h\right)^2+A^3B\frac{4}{27} g \left(1072 g^4-7832 g^2 h-797 h^2\right)\\
		&-\frac{1}{54}A^2B^2\cdot \left(4 g^2-h\right) \left(25504 g^4+6924 g^2 h+227 h^2\right)\\
		&+AB^3\cdot \frac{g \left(1425683 g^6+7311527 g^4 h-733303 g^2 h^2-154563 h^3\right)}{1728}\\
		&+B^4\cdot\frac{2550099 g^8-20848992 g^6 h+2131870 g^4 h^2+885304 g^2 h^3+8887 h^4}{6912}\,,
	\end{split}
	\label{eqn:f4overf1z2}
\end{align}
}
where we introduce
\begin{align}
	A=\phi_{0,1}(2\tau,\lambda)\,,\quad B=\phi_{-2,1}(2\tau,\lambda)\,,\quad g=E_2^{(2)}(\tau)\,,\quad h=E_4(\tau)\,.
\end{align}
\begin{table}[h!]
{\tiny
\begin{align*}
\begin{array}{|c|ccccc|}
\hline
 d_F\backslash g & 0 & 1 & 2 & 3 & 4 \\\hline
 0 & 223752 & -492 & 0 & 0 & 0 \\
 1 & 152031744 & 69120 & 0 & 0 & 0 \\
 2 & 19638646848 & 104982288 & -304464 & -1476 & 0 \\
 3 & 1180450842624 & 11531535360 & -214941696 & 207360 & 0 \\
 4 & 43199009739072 & 582562932240 & -23399572104 & 308769960 & -601056 \\
 5 & 1107266933984256 & 18197544339456 & -1223655651840 & 35175416832 & -430989312 \\
 6 & 21665294606886144 & 403387081306944 & -40593035175168 & 1888636322256 & -48423847008 \\
 7 & 342620943505772544 & 6879702812129280 & -975207871309824 & 64188987386880 & -2630615021568 \\\hline
\end{array}
\end{align*}
}
\caption{Gopakumar-Vafa invariants for $(F_{4}\rightarrow \mathbb{P}^2)[\mathbb{Z}_2]_{2}^{-252}$ at base degree $2$.}
\label{tab:f4overf1gv2}
\end{table}
\subsubsection{$M^h_2=(F_{6}\rightarrow \mathbb{P}^2)[U(1)]_{3}^{-216}$}
\label{sec:m22}
The geometry $M^h_{1}$ is related via an extremal transition to what we call $M^h_2$.
The choice of line bundles in the conventions of~\cite{Klevers:2014bqa} is still $\mathcal{S}_7=3H,\,\mathcal{S}_9=0$ and the toric data is given in \eqref{eqn:f6overf1toricData},
\begin{align}
\begin{blockarray}{crrrrrrrl}
	&&&&&C_1&C_2&C_3\\
\begin{block}{c(rrrr|rrr)l}
	x_1& 1& 0&-1&0&0& 0& 1&\leftarrow\text{vertical divisor }D_b\\
	x_2& 0& 1&-1&0& 0& 0& 1&\phantom{x}\hspace{1.5cm}\text{\ditto}\\
	x_3&-1&-1&-1&0&0& 0& 1&\phantom{x}\hspace{1.5cm}\text{\ditto}\\
	v& 0& 0&-1&-1& 1&0&0&\leftarrow\text{two-section}\\
	e_2& 0& 0& -1& 1&-1& 1& -3&\leftarrow\text{holomorphic section }E_0\\
	w& 0& 0& 1& 0&0& 1& 0&\leftarrow\text{three-section}\\
	u& 0& 0& 0& 1& 2&-1& 3&\leftarrow\text{rational section }E_1\\
	& 0& 0& 0& 0& -2&-1& -3&\\
\end{block}
\end{blockarray}\,.
\label{eqn:f6overf1toricData}
\end{align}
Again we encounter the situtation that the relevant points admit two different regular fine star triangulations.
Only one is compatible with the fibration of the ambient space but the phase boundary is lifted by the hypersurface equation.
The curves $C_i,\,i=1,...,3$ in \eqref{eqn:f6overf1toricData} generate the intersection of the two Mori cones.

The geometry of the images of the toric divisors on $M^h_2$ is as follows.
The divisor $D_b=[x_1]=[x_2]=[x_3]$ is again the pullback of the hyperplane class of the base.
Futhermore, ${e_2=0}$ is a holomorphic section while ${u=0}$ is a linearly independent rational section.
We denote the corresponding divisors by $E_0=[e_2]$ and $E_1=[u]$.
That ${u=0}$ is not holomorphic manifests itself in the fact that it wraps a curve in the fiber over 18 points of the base~\footnote{This can easily be determined by studying the hypersurface equation
of $M$ or, even easier, using the general results from \cite{Klevers:2014bqa}.}.
Each of these curves is transversely intersected by ${e_2=0}$.
On the other hand $[v]=E_1+E_0$ intersects the generic fiber twice and $[w]=[v]+E_0+3D_b$ intersects it thrice.

Let us now study the Mori cone of $M^h_2$ directly.
It is generated by four curves.
The first, $C_b^1=E_0\cdot D_b$, is the restriction of the holomorphic section to the generator of the Mori cone of the base.
The second, $C_b^2=E_1\cdot D_b$, is the restriction of the rational section.
The other two curves arise from resolutions of singularities in the fiber.
Over 18 points in the base the fiber develops an $I_2$ singularity that leads to matter with charge $q=2$.
Another 144 $I_2$ singularities lead to matter with charge $q=1$.
In the resolved geometry there are two spheres $C_A^q,\,C_B^q$ over each point and the respective numerical equivalence class only depends on the charge of the matter from that locus.
The intersections are as follows:
	\begin{equation}
	\begin{array}{c|rrrrrr}
		&C_A^1&C_B^1&C_A^2&C_B^2&C_b^1&C_b^2\\\hline
		E_0& 0& 1& 1& 0& 3&-3\\
		E_1& 1& 0&-1& 2&-3& 3\\
		D_b& 0& 0& 0& 0& 1& 1
	\end{array}
	\end{equation}
From this we see that $C_b^1,C_b^2,C_A^1,C_A^2$ generate the Mori cone on $M^h_2$.

This being said we will introduce $J_1=[v],\,J_2=[w],\,J_3=D_b$ and expand the complexified K\"ahler form as
\begin{align}
	\omega=t^1\cdot J_1+t^2\cdot J_2+t^3\cdot J_3\,,
\end{align}
where we have replaced the basis element $E_0+3D_b$ with the element $[w]$ from the interior of the K\"ahler cone.
This enables us to apply the usual machinery to calculate Gopakumar-Vafa invariants of toric hypersurfaces.

The topological invariants~\eqref{eqn:topdef} are
\begin{align}
	h^{1,1}=3\,,\quad\chi = -216\,,\quad  \vec{b}=\int\limits_{M^h_2} c_2\left(M^h_2\right)\cdot \vec{J}=(24,\, 126,\, 36)^T\,,
\end{align}
and the non-vanishing triple intersection numbers are encoded in the polynomial
\begin{align}
	\mathcal{J}=63 J_2^3+18 J_1 J_2^2+15 J_3 J_2^2+3 J_3^2 J_2+6 J_1 J_3 J_2+2 J_1 J_3^2\,.
\end{align}
The section that determines the mirror $W$ in the dual ambient space is given by
\begin{align}
	\begin{split}
	P=&-z_3\cdot  x_{2}^6 x_{3}^6 x_{4}^3 - z_2\cdot  x_{1}^2 x_{4} x_{5} x_{6} + z_1\cdot x_{2}^3 x_{4}^2 x_{5}^2 x_{6}^2 x_{7}^3 x_{8}^3 + x_{10}^6 x_{5}^3 x_{8}^6\\
	&+ x_{1} x_{2} x_{3} x_{4} x_{5} x_{6} x_{7} x_{8} x_{9}x_{10}  + x_{1} x_{10}^3 x_{3}^3 x_{9}^3 + x_{10}^4 x_{2} x_{3}^4 x_{7} x_{8} x_{9}^4 + x_{6}^3 x_{7}^6 x_{9}^6\,,
	\end{split}
\end{align}
where $z_i,\,i=1,...,3$ are the Batyrev coordinates on the complex structure moduli space.
The Picard-Fuchs system is comparatively simple and given in the Appendix~\ref{app:discpff6}.
When expressed in terms of the Batyrev variables the discriminant $\Delta$ is irreducible and also given in~\ref{app:discpff6}.
The logarithm of $\Delta$ contributes to the genus one free energy with the usual factor of $-1/12$.

We will now discuss the Shioda map and the height pairing.
Let us fix $\{e_2=0\}$ as the holomorphic zero section and $\{u=0\}$ as the generator of the Mordell-Weil group.
Then the image of the generator under the Shioda map is given by
\begin{align}
	\sigma(\{u=0\})=E_1-E_0-2\cdot\pi^*c_1(B)=3J_1-2J_2\,,
\end{align}
where we have used that $\pi^*\pi_*(E_1\cdot E_0)=\pi^*c_1(B)$.
The corresponding height pairing is
\begin{align}
	b_{11}=-\pi(D_b)\cdot \pi\left(\sigma(\{u=0\})\cdot\sigma(\{u=0\})\right)=12\,.
\end{align}
We also calculate
\begin{align}
	E_0^2\cdot D_b=-3\,,
\end{align}
and therefore introduce
\begin{align}
	\tilde{E}_0=E_0+3D_b\,,
\end{align}
with
\begin{align}
	\tilde{a}_b=\tilde{E}_0^2\cdot D_b=3\,.
\end{align}
The correct modular parametrization of the K\"ahler form is therefore
\begin{align}
	\begin{split}
		\omega=&\tau\cdot\tilde{E}_0+m\cdot\sigma(\{u=0\})+\tilde{t}\cdot D_b\\
		=&\tau\cdot (-J_1+J_2)+m\cdot(3J_1-2J_2)+\tilde{t}\cdot J_3\,,
	\end{split}
\end{align}
where $\tau$ is the volume of the generic fiber, $m$ is the volume of the isolated fibral curve $C_A^1$ and $\tilde{t}$ is the volume of $C_b^1$ and $C_b^2$.
The new parameters are related to $t_i,\,i=1,...,3$ via
\begin{align}
	\tau=2t^1+3t^2\,,\quad m=t^1+t^2\,,\quad \tilde{t}=t^3\,.
\end{align}
The shifted K\"ahler parameter is the base is $t=\tilde{t}+\frac32\tau$.
We can then use the Gopakumar-Vafa invariants at genus zero and genus one to obtain
\begin{align}
	Z_1(\tau,m,\lambda)=\frac{1}{\eta(\tau)^{36}}\frac{\phi_1(\tau,m)}{\phi_{-2,1}(\tau,\lambda)}\,,
\end{align}
with
\begin{align}
	\begin{split}
	\phi_1=&\frac{1}{2^{16}2^7}\left[-A^6\cdot h(31h^3+113g^2)-6A^5B\cdot g(115h^3+29g^2)\right.\\
	&-3A^4B^2\cdot h^2(203h^3+517g^2)-4A^3B^3\cdot hg(479h^3+241g^2)\\
	&-3A^2B^4(51h^6+581h^3g^2+88g^4)-6AB^5\cdot h^2g(19h^3+125g^2)\\
	&\left.+B^6\cdot(9h^6-49h^3g^2-104g^4)\right]\,,
	\end{split}
\end{align}
where
\begin{align}
	A=\phi_{0,1}(\tau,m)\,,\quad B=\phi_{-2,1}(\tau,m)\,,\quad h=E_4(\tau)\,,\quad g=E_6(\tau)\,.
\end{align}
One can easily check that $Q\cdot Z_1(2\tau,\tau,\lambda)$ matches the result for $Q\cdot Z_1(\tau,\lambda)$ from $M^h_1$.
\subsubsection{$M^h_3=(F_{10}\rightarrow \mathbb{P}^2)[SU(2)]_{3}^{-216}$}
The final geometry in the second Higgs chain describes a 101 dimensional subslice inside the 111 dimensional complex structure moduli space of $M^h_2$.
On this subslice the $2g-2=18$ isolated $I_2$ fibers of $M^h_2$ that lead to matter with charge $q=2$ under the $U(1)$ deform into a genus $g=10$ curve of $I_2$ singularities
and therefore the $U(1)$ un-Higgses into an $SU(2)$ gauge symmetry.
The $144$ charge $q=1$ multiplets from $M^h_2$ arrange into $72$ hyper multiplets in the fundamental representation of $SU(2)$.
Geometrically these arise from $I_3$ enhancements of the $I_2$ singular fibers over $72$ points in the base of $M^h_3$.

We realize this subslice as a hypersurface in a toric ambient space where $g=10$ complex structure deformations are non-polynomial.
The toric data is given in~\eqref{eqn:f10toricdata},
\begin{align}
\begin{blockarray}{crrrrrrrl}
	&&&&&C_1&C_2&C_3\\
\begin{block}{c(rrrr|rrr)l}
	x_1&-1 & -1 & -1 & 0 & 0 & 0 & 1&\leftarrow\text{vertical divisor }D_b\\
	x_2&0 & 1 & -1 & 0 & 0 & 0 & 1&\phantom{x}\hspace{1.5cm}\text{\ditto}\\
	x_3&1 & 0 & -1 & 0 & 0 & 0 & 1&\phantom{x}\hspace{1.5cm}\text{\ditto}\\
	v&0 & 0 & 0 & 1 & 1 & 0 & 0&\leftarrow\text{two-section} \\
	w&0 & 0 & 1 & 0 & 0 & 1 & 0&\leftarrow\text{three-section} \\
	u&0 & 0 & -1 & -1 & 3 & -2 & 6&\leftarrow\text{fibral divisor }D_f \\
	e_3&0 & 0 & -3 & -2 & -1 & 1 & -3&\leftarrow\text{holomorphic section }E_1 \\
	&0 & 0 & 0 & 0 & -3 & 0 & -6 \\
\end{block}
\end{blockarray}\,.
	\label{eqn:f10toricdata}
\end{align}
In the conventions of~\cite{Klevers:2014bqa} this geometry corresponds to the choice of line bundles $\mathcal{S}_7=c_1(B),\,\mathcal{S}_9=0$ on $B=\mathbb{P}^2$.
The curve of $I_2$ fibers is in the class $S_{SU(2)}=2\mathcal{S}_7-\mathcal{S}_9=2c_1(B)$.
Note that the fibral divisor $D_f=[u]$ corresponds to the {affine} node that is intersected by the holomorphic zero section $E_1=[e_3]$.
We therefore introduce $D_f'=S_{SU(2)}-D_f$ to denote the other fibral divisor such that $D_f'\cdot E_1=0$.
If we introduce $J_1=[v],\,J_2=[w]$ and $J_3=[x_1]$ the topological data of $M^h_3$ is identical to that of $M^h_2$.

In Batyrev coordinates $z_i,\,i=1,...,3$ the hypersurface equation that defines the mirror $W$ in the dual ambient space reads
\begin{align}
	\begin{split}
	P=&- z_1\cdot x_3^3x_4^3x_5^3 +z_2\cdot x_1^2 + z_3\cdot x_2^6x_3^6 + x_5^6x_6^6+ x_4^6x_7^6  \\
	&+ x_1x_2x_3x_4x_5x_6x_7 + x_1x_2^3x_6^3x_7^3 + x_2^6x_6^6x_7^6\,.
	\end{split}
\end{align}
Identifying the mirror maps we find that the Batyrev coordinates of $M^h_3$ can be mapped to those of $M^h_2$ via
\begin{align}
	z_1\mapsto z_1'\frac{(1+2z_2')^3}{1+z_2'}\,,\quad z_2\mapsto z_2'\frac{1+z_2'}{(1+2z_2')^2}\,,\quad z_3\mapsto z_3'\frac{(1+2z_2')^6}{(1+z_2')^3}\,,
\end{align}
where $z_i',\,i=1,...,3$ parametrize the complex structure moduli space of $M^h_2$.

The discriminant of $W$ is reducible.
The principle component $\Delta_1$ is again provided in the Appendix~\eqref{eqn:f10disc} while the other component reads
\begin{align}
	\Delta_2=1-4z_2\,.
\end{align}
The generators of the Picard-Fuchs system are also provided in~\eqref{eqn:f10pc}.
In this case we did not try to simplify the operators and expect that a more economical choice can be made.

If we expand the K\"ahler form as
\begin{align}
	\begin{split}
	\omega=&\tau\cdot (E_1+c_1(B))+m\cdot D_f'+\tilde{t}\cdot J_3\\
	=&\tau\cdot (-J_1+J_2)-m\cdot(3J_1-2J_2)+\tilde{t}\cdot J_3\,,
	\end{split}
\end{align}
where $\tau$ is the volume of the generic fiber, $m$ is twice the volume of the fiber of $D_f$ and $C_B$ is the volume of $C_b=E_1\cdot D_b$.
The new parameters are related to $t_i,\,i=1,...,3$ via
\begin{align}
	\tau=2t_1+3t_2\,,\quad m=-(t_1+t_2)\,,\quad \tilde{t}=t_3\,.
\end{align}
The shifted K\"ahler parameter is the base is again $t=\tilde{t}+\frac32\tau$.
Since the family $M^h_3$ corresponds to a subslice in the complex structure moduli space of $M^h_2$, the topological string partition functions are identical.

\subsection{A genus-one fibrations over $\mathbb{F}_1$: $(F_4\rightarrow\mathbb{F}_1)[SU(2)\times \mathbb{Z}_2]_4^{-144}$}
\label{sec:genusOneOverHF1}
Now we will demonstrate the modular bootstrap for genus one fibrations at the hand of a 2-section geometry over the Hirzebruch surface $\mathbb{F}_1$ that also exhibits a fibral divisor.
To engineer this geometry we can again use the fiber polytope $F_4$.
Following the discussion in~\label{sec:construction} we choose the bundles
\begin{align}
	\mathcal{S}_7=2B+4F\,,\quad\mathcal{S}_9=2B+3F\,,
\end{align}
where $B$ and $F$ are the base and fiber of the Hirzebruch surface.
This leads to the following toric data:
\begin{align}
\begin{blockarray}{crrrrrrrrl}
	&&&&&C_1&C_2&C_3&C_4\\
\begin{block}{c(rrrr|rrrr)l}
	x_1& 1& 0& 0& 0& 0& 1& 0& 0&\leftarrow\text{vertical divisor }\pi^{-1}(F+B)\\
	x_2& 0& 1& 0& 0& 0& 0& 1& 0&\leftarrow\text{vertical divisor }\pi^{-1}F\\
	x_3&-1&-1&-1&-1& 0& 0& 1& 0&\phantom{x}\hspace{1.5cm}\text{\ditto}\\
	x_4&-1& 0& 0& 0& 0& 1&-1& 0&\leftarrow\text{vertical divisor }\pi^{-1}B\\
	x  & 0& 0&-1& 1& 1& 0& 0& 0&\leftarrow\text{two-section }E_0\\
	e_1& 0& 0&-1& 0&-2& 0& 0& 1&\leftarrow\text{fibral divisor }D_f'\\
	y  & 0& 0&-1&-1& 1& 0&-1& 0&\leftarrow\text{two-section }\\
	z  & 0& 0& 1& 0& 0& 0& 0& 1&\leftarrow\text{four-section}\\
	   & 0& 0& 0& 0& 0&-2& 0&-2&\\
\end{block}
\end{blockarray}
\label{eqn:f4overf3su2toricData}
\end{align}
There are three regular fine star triangulations of the relevant points of the polytope and we provided the data for the Mori cone
that corresponds to the unique triangulation that is compatible with the fibration.

A convenient basis of divisors to obtain the free energies at genus $0$ and genus $1$ is
\begin{align}
	J_1=E_0\,,\quad J_2=D_1'\,,\quad J_3=D_2'\,,\quad J_4=[z]=D_f'+2\cdot E_0\,.
\end{align}
The relevant topological invariants are
\begin{align}
	h^{1,1}=4\,,\quad\chi=-144\,,\quad \vec{b}=\int\limits_Mc_2(M)\cdot\vec{J}=(36,\,36,\,\,24,\,68)^T\,,
\end{align}
and the triple intersections are encoded in the polynomial
\begin{align}
	\begin{split}
		\mathcal{J}=&8 J_4^3+4 J_1 J_4^2+12 J_2 J_4^2+4 J_3 J_4^2+2 J_1^2 J_4+4 J_2^2 J_4+6 J_1 J_2 J_4\\
		&+2 J_1 J_3 J_4+4 J_2 J_3 J_4+2 J_1 J_2^2+2 J_1^2 J_2+2 J_1 J_2 J_3\,.
	\end{split}
\end{align}
The principal component of the discriminant contains 427 terms and will be provided as supplementary data online~\cite{DataLink}.

Let us now construct the appropriate parametrization of the K\"ahler form to perform the modular bootstrap.
To this end we introduce
\begin{align}
	D_1'=\pi^{-1}(F+B)\,,\quad D_2'=\pi^{-1}(F)\,,\quad D_1=\pi^{-1}(F)\,,\quad D_2=\pi^{-1}(B)\,,
\end{align}
and $E_0=[x]$ as well as the curves $C_i=E_0\cdot D_i,\,i=1,2$ such that
\begin{align}
	D_i'\cdot C_j=2\cdot \delta_{ij}\,.
\end{align}
Moreover, $\tilde{E}_0=E_0-D_1$ is orthogonal to those curves and we can calculate
\begin{align}
	\tilde{a}_i=\int_M\tilde{E}_0^2\cdot D_i=\left\{\begin{array}{rl}0&\text{ for }i=1\\-2&\text{ for }i=2\end{array}\right.\,.
\end{align}
From~\eqref{eqn:f4polynomial} and~\eqref{eqn:f4coefficientClass} we can see that the fibral divisor $D_f'=[e_1]$ is fibered over a divisor in the class 
\begin{align}
	[d_9]=c_1(B)-\mathcal{S}_7+\mathcal{S}_9=2B+2F\,.
\end{align}
The fibral divisor $D_f'=[e_1]$ is not orthogonal to $E_0$ but we can construct the linear combination of fibral divisors
\begin{align}
	D_f=\frac12\left([d_9]-2\cdot D_f'\right)\,,
\end{align}
such that $\tilde{E}_0\cdot D_f\cdot D_i=0$ for $i=1,2$.
We will then expand the K\"ahler form as
\begin{align}
	\omega=\tau\cdot\tilde{E}_0+m\cdot D_f+\sum\limits_{i=1}^2\left(t_i-¸\frac{\tilde{a}_i}{4}\tau\right)\cdot D_i'\,.
\end{align}
The index of $Z_\beta(\lambda,\tau,m)$ with respect to the geometric elliptic parameter $m$ is
\begin{align}
	r^\beta_{11}=(B+F)\cdot \beta\,.
\end{align}

Using the Ansatz~\eqref{eqn:jacAns2} and the genus zero free energy we can immediately fix the numerators
\begin{align}
	\begin{split}
		\phi_F=&\frac{2}{9}\left(\Delta_4\right)^2\left[-8A^2g +AB\left(4 g^2+h\right) +B^2g \left(18 g^2-5 h\right)\right]\,,\quad \phi_B=-2\sqrt{\Delta_4}g\,,\\
		\phi_{2B}=&\frac{\Delta_4}{288}\left[16A^2g^2 +8A Bg\left(h-2g^2\right) +B^2h\left(3h-11g^2\right)\right]\,,\\
		\phi_{B+F}=&\frac{\left(\Delta_4\right)^{\frac52}}{216}\left[8A\left(8 C^2 g^2- C D g \left(4 g^2+h\right)- D^2 g^2 \left(18 g^2-5 h\right)\right)\right.\\
		&B\left(-4 C^2 g \left(4 g^2+33 h\right)+ 8C D \left(-4 g^4+14 g^2 h+h^2\right)\right.\\
		&\left.\left.- D^2 g \left(4 g^4-331 g^2 h+91 h^2\right)\right)\right]\,,
		\label{eqn:f4su2mod}
	\end{split}
\end{align}
where we have introduced
\begin{align}
	\begin{split}
	A=&\phi_{0,1}(2\tau,\lambda)\,,\quad B=\phi_{-2,1}(2\tau,\lambda)\,,\quad C=\phi_{0,1}(2\tau,m)\,,\quad D=\phi_{-2,1}(2\tau,m)\,,\\
	g=&E_2^{(2)}(\tau)\,,\quad h=E_4(\tau)\,.
		\end{split}
\end{align}
To obtain expressions for other base degrees we have to use additional data.
This could, for example, be knowledge about the vanishing of certain Gopakumar-Vafa invariants as was demonstrated in~\cite{Huang:2015sta}.

Some of the Gopakumar-Vafa invariants corresponding to $\beta=B$ and $\beta=B+F$ are listed in the tables~\ref{tab:f4su2gvB} and~\ref{tab:f4su2gvBF}.
We label the class of a curve $C$ by the degrees
\begin{align}
	d_F=C\cdot \tilde{E}_0\,,\quad d_E=C\cdot D_f\,,
\end{align}
and a class $\beta\in H_2(B)$.
The reflection symmetry along the vertical axis among the invariants~\ref{tab:f4su2gvBF} with $\beta=B+F$ is a consequence of the invariance under $m\rightarrow-1$.
However, there is also a curious periodicity which appears to be present for all genera.
This is not at all manifest in the modular expression for $\phi_{B+F}$~\eqref{eqn:f4su2mod}.
It should severely constrain the number of free parameters that have to be fixed in the ansatz and it would be very interesting to get a better understand of the
origin of this pattern.
\begin{table}[h!]
\begin{align*}
\begin{array}{|c|ccccccc|}
	\hline
 g\backslash d_F & 1 & 2 & 3 & 4 & 5 & 6 & 7 \\\hline
 1 & 56 & 0 & 0 & 0 & 0 & 0 & 0 \\
 2 & 276 & -4 & 0 & 0 & 0 & 0 & 0 \\
 3 & 1360 & -112 & 0 & 0 & 0 & 0 & 0 \\
 4 & 4718 & -564 & 6 & 0 & 0 & 0 & 0 \\
 5 & 15960 & -3056 & 168 & 0 & 0 & 0 & 0 \\
 6 & 46284 & -11108 & 860 & -8 & 0 & 0 & 0 \\
 7 & 130064 & -40528 & 4976 & -224 & 0 & 0 & 0 \\
 8 & 334950 & -123112 & 18660 & -1164 & 10 & 0 & 0 \\
 9 & 837872 & -367552 & 72160 & -7120 & 280 & 0 & 0 \\
 10 & 1980756 & -989236 & 226952 & -27392 & 1476 & -12 & 0 \\
 11 & 4564224 & -2603520 & 712128 & -111360 & 9488 & -336 & 0 \\\hline
\end{array}
\end{align*}
\caption{GV invariants $n^{(g)}_{d_F,d_E=0}$ for $(F_4\rightarrow\mathbb{F}_1)[SU(2)\times \mathbb{Z}_2]_4^{-144}$ with base class $\beta=B$.}
\label{tab:f4su2gvB}
\end{table}
\begin{table}[h!]
{\tiny
\begin{align*}
\begin{array}{|c|ccccccccc|}
	\hline d_F\backslash d_E & -4&-3&-2&-1&0&1&2&3&4 \\\hline
 2 & 0 & 0 & 0 & 0 & 0 & 0 & 0 & 0 & 0 \\
 3 & 0 & 0 & 0 & -412 & -1056 & -412 & 0 & 0 & 0 \\
 4 & 0 & 0 & -1056 & -16432 & -35072 & -16432 & -1056 & 0 & 0 \\
 5 & 0 & -412 & -35072 & -72444 & -73408 & -72444 & -35072 & -412 & 0 \\
 6 & 0 & -16432 & -73408 & 4727056 & 9905152 & 4727056 & -73408 & -16432 & 0 \\
 7 & -1056 & -72444 & 9905152 & 108929428 & 202048512 & 108929428 & 9905152 & -72444 & -1056 \\
 8 & -35072 & 4727056 & 202048512 & 1400552368 & 2439058688 & 1400552368 & 202048512 & 4727056 & -35072 \\\hline
\end{array}
\end{align*}
}
\caption{GV invariants $n^{(g=2)}_{d_F,d_E}$ for $(F_4\rightarrow\mathbb{F}_1)[SU(2)\times \mathbb{Z}_2]_4^{-144}$ with base class $\beta=B+F$.}
\label{tab:f4su2gvBF}
\end{table}
\section{Conclusion}
\label{sec:conclusion}

Using homological mirror symmetry we analysed in depth the action of the integral symplectic transformations  on the central charges of  Type II A and B strings compactified on genus one fibered  Calabi-Yau 3-folds $M$ and 
their  mirrors $W$. We considered  the case that $M$ had multiple $N$-sections as well as fibral divisors, which respectively lead to Abelian and non-Abelian gauge symmetry enhancements in the Type II -- and F-theory vacua. 
We established that certain auto-equivalences of the category of branes act as $\Gamma_1(N)$ on the stringy K\"ahler moduli space and can be expressed in terms of generic Conifold and large volume monodromies.
Together with monodromies that generate the Weyl group this restricts in crucial ways the correlation function of the physical theories.
In particular it follows that  the topological string partition function can be expressed in terms of Weyl invariant meromorphic Jacobi forms under $\Gamma_1(N)$, $N=1,2,3,4$ for each base degree.

We got further  insights in the properties of these topological amplitudes by considering 
their behaviour under those geometric transitions that correspond physically to Higgsing 
the gauge symmetries. We found that this implies  very non-trivial identities  among  
the rings of Jacobi-Foms that occur at the different stages of the Higgsing tree, since  the effect of 
the Higgsing  is that the Coulomb branch parameters of the gauge symmetries, which are 
elliptic parameters of the Jacobi-Forms, are identified with the elliptic arguments which 
becomes an $N$-th multiple of itself as for example in~\eqref{U1higgsruleexample}.
This allows us to generalize the Ansatz for the modular bootstrap on elliptic fibrations to genus one fibrations.
At least for low base degrees the partition function can then be fixed by additional boundary conditions. 

This partition function is geometrically the most detailed  information that is available for the BPS spectrum  of the  effective physical theories  that arise in six, 
five and four dimensions from F-theory, M-theory and Type II compactifications respectively.
For example in 6d F-theory compatifications already the BPS invariants that correspond to the  
rational curves with base degree zero give the multiplicities of the matter multiplets that arise in co-dimension two in the base.    

In the future it would be very interesting to understand the modular properties for genus one fibrations with $N$-sections where $N>4$.
Already for $N=5$ the groups $\Gamma_0(N)$ and $\Gamma_1(N)$ require more than $2$ generators and it is not clear how the additional generator manifests as a monodromy in the stringy K\"ahler moduli space.
Also the derivation of the modular properties of the base degree zero partition function cannot easily be generalized.
This is particularly exciting because, as we stressed above, the base degree zero part already contains the full information about the spectrum of the corresponding F-theory vacuum.
Moreover, it appears that only genus one fibrations with $N\le 4$ can be realized as complete intersections in toric ambient spaces.
If modularity persists for geometries with $N>4$, this should imply an even more intriguing relation involving the monodromies in the stringy K\"ahler moduli space, the spectra of exotic F-theory vacua, non-toric realizations
of genus one fibered Calabi-Yau manifolds and the theory of Jacobi forms.
A potential starting point to study this question are the genus one fibrations with $5$- and $6$-sections that have respectively been constructed in~\cite{Kimura:2019bzv} and~\cite{Anderson:2019kmx}.

In the context of heterotic - Type II duality we used the general discussion of the monodromies to argue 
that heterotic compactifications on CHL orbifolds $(K3\times T^2)/\mathbb{Z}_N$ should be dual to 
Type IIA compactifications on genus one fibrations with $N$-sections.
We then constructed novel duals of heterotic compactifications on $(K3\times T^2)/\mathbb{Z}_2$ where $\mathbb{Z}_2$ acts as an automorphism of class $2A$ on the $K3$
and used the modular bootstrap as well as known results for the corresponding one-loop amplitude on the heterotic side~\cite{Chattopadhyaya:2017zul} to perform all order checks of the duality.
The restriction to $\mathbb{Z}_2$, and therefore automorphisms of class $2A$, is due to the fact that the one-loop amplitudes that can be matched with the topological string partition function has only been calculated for non-standard embeddings on these geometries~\cite{Chattopadhyaya:2016xpa,Chattopadhyaya:2017zul}.
More precisely, the result of the one-loop amplitude can be expressed in terms of the new supersymmetric index and the new supersymmetric index has only been calculated for non-standard embedding on those geometries.
Extending the calculation of the new supersymmetric index to compactifications with non-standard embeddings on $(K3\times T^2)/\mathbb{Z}_N$ with $N>2$ should be relatively straight forward.
Matching the results for the corresponding one-loop amplitudes with the topological string partition function on genus one fibrations with $3$- and $4$-sections would be
a highly non-trivial check of our proposal that we plan to perform in the future.

From the perspective of the $F$-theory effective theory, vacua that arise from $N$-section geometries are obtained by Higgsing $U(1)$ factors with matter of charge $N$.
These are six dimensional, i.e. their effective theory is a chiral six-dimensional supergravity theory with discrete gauge group $\mathbb{Z}_N$~\cite{deBoer:2001wca,Morrison:2014era}.      
$F$-theory duals to the CHL orbifolds of the hererotic string in $8$ dimension have been constructed in~\cite{Berglund:1998va} for $N=2,3,4$.
On the type II side they are elliptically fibered $K3$ with reduced fiber monodromy and reduced rank of the gauge group, which matches the one of the CHL orbifold on $T^2$ with an $N$ shift on one of the circles~\footnote{A different 
construction using a $2$-torsional section and half-integral  $B$-field  on the $\mathbb{P}^1$ basis has been proposed in \cite{Bershadsky:1998vn}.}.
A candidate  construction for an F-theory/CHL duality in six dimensions could be a fiberwise extension of these eight dimensional CHL orbifolds, similar as the one in the heterotic bundle construction of~\cite{Friedman:1997yq} on elliptic $K3$.
However to maintain  the CHL $\mathbb{Z}_N$ shift symmetry the elliptically fibered $K3$ would have to have an $N$-torsional section.               

In the recent literature a major amount of effort is dedicated towards a geometric classification of 
5d SCFTs~\cite{Jefferson:2017ahm,Jefferson:2018irk,Bhardwaj:2018yhy,Bhardwaj:2018vuu,DelZotto:2017pti,Apruzzi:2018nre,Apruzzi:2019vpe,Apruzzi:2019opn,Apruzzi:2019enx,Bhardwaj:2019jtr}.
In particular, the work~\cite{Bhardwaj:2019fzv} discusses a class of 5d SCFTs that is obtained by deformation from KK reductions of 6d SCFTs with twists along the compactification circle (see also~\cite{Anderson:2019kmx}).
The latter class of theories can be described by M-theory compactifications on local genus one fibered Calabi-Yau threefolds without sections.
These rigid theories that come from the $N$-section discussed in Section~\ref{sec:decoupling} can be also solved by the elliptic blow-up equations similar to the theories discussed in~\cite{Gu:2017ccq,Gu:2018gmy,Gu:2019dan}\footnote{We thank Xin Wang for pointing this out to us.}. 
In five dimensions, the blow up equations are even more powerful in the sense that the classical topological data is in many cases enough to reconstruct all BPS 
states~\cite{GKSWprogress}, i.e. data similar to that which is used to determine the monodromies and anomaly polynomials.
It would be very interesting to apply the modular bootstrap for genus one fibrations that we developed in this paper, to study the BPS spectrum of these theories and in particular to use the 5d blow-up equations and a refined   
modular bootstrap approach to the ones that involve a twisted circle compactification~\cite{Bhardwaj:2019fzv}.
    
\newpage
\appendix
\label{Appendix}  
\section{A brief review of the F-theory dictionary}
\label{app:ftheory}
In this section we review the origin of gauge symmetries in F-theory and the corresponding anomaly coefficients in six dimensions.
Most readers will be familiar with the material and can safely skip it.
For a thorough review of F-theory we refer to \cite{Weigand:2018rez} and for a review that focuses on Abelian and discrete gauge symmetries see \cite{Cvetic:2018bni}.
We mostly follow the notation of \cite{Klevers:2014bqa}.
\subsection{Gauge symmetries and matter in F-theory}
Up to a choice of flux and possibly T-brane data the effective physics of a given F-theory vacuum is entirely encoded in the geometry of an elliptically or genus one fibered Calabi-Yau variety $\pi: X\rightarrow B$.
We will only consider smooth threefolds $X$ without multiple or non-flat fibers such that the effective theory is a six-dimensional $(1,0)$-supergravity.
Recall that if the theory is further compactified on a circle there is a dual interpretation via M-theory compactified on $X$.

A non-Abelian gauge group $G_I$ arises for every irreducible divisor $S^b_{G_I}:=\{\Delta_I=0\}\subset B$ in the base such that 
the generic fiber over $S^b_{G_I}$ is a union of rational curves that intersect like the negative Cartan matrix $-C_{ij}^{G_I}$ of the affine Lie algebra associated to $G_I$~\footnote{Here we assume that
there are no non-simply laced gauge groups.}.
All but one component of the reducible fiber correspond to simple roots $\alpha_i,\,i=1,...,\text{rk}(G_I)$ of $G_I$ and will be denoted by $c^{G_I}_{-\alpha_i}$.
The so-called \textit{fibral divisors} in $X$ that are obtained by fibering $c^{G_I}_{-\alpha_i}$ over $S^b_{G_I}$ will be denoted by $D_i^{G_I},\,i=1,...,\text{rk}(G_I)$.

In the M-theory interpretation the 3-form $C_3$ can be expanded along the harmonic forms that correspond to the divisors $D_i^{G_I}$.
This leads to $\text{rk}(G_I)$ massless vector bosons that gauge the Abelian subgroup $U(1)^{\text{rk}(G_I)}\subset G_I$.
When the volume of some $c^{G_I}_{-\alpha_i}$ vanishes, additional massless gauge bosons arise from wrapped M2 branes.
In the F-theory limit the volume of the whole fiber is set to zero and one recovers the full unbroken gauge group $G_I$.

An elliptically fibered manifold can admit a section that is a rational map $\hat{s}:B\rightarrow X$ such that $\pi\circ \hat{s}=\text{id}$.
If $X$ admits at least one section, we can choose a section $\hat{s}_0$ and the affine node of the affine Dynkin diagrams associated to the reducible fibers over all $S^b_{G_I}$ such that $\hat{s}_0$ does
not intersect any of the $c^{G_I}_{-\alpha_i}$.
Then $\hat{s}_0$ is called the zero-section and the name reflects the fact that once a zero-section is chosen the sections form a group with the zero-section as the identity.
This group is called the Mordell-Weil group MW$(X)$ of $X$.

According to the M-theory interpretation the harmonic form that corresponds to the divisor of the zero-section again leads to a massless gauge boson.
This can be identified with the Kaluza-Klein gauge boson that arises from reducing the six-dimensional metric along the circle.
It disappears in the limit where the circle is decompactified and therefore a single section does not lead to any gauge symmetry in six dimensions.

Additional sections lead to Abelian gauge symmetry if they do not correspond to elements of finite order in the Mordell-Weil group.
Given such a non-torsional section $\hat{s}$ we have to ``orthogonalize'' it such that M2 branes wrapping the generic fiber or the curves $c^{G_I}_{-\alpha_i}$ are not charged under the corresponding
gauge gauge symmetry.
In addition it has to be compatible with the F-theory limit which requires that the dual cohomology class has ``one leg along the fiber''.
Both is achieved by applying the Shioda map $\sigma:\text{MW}(X)\rightarrow \text{NS}(X,\mathbb{Q})$,
\begin{align}
	\sigma(\hat{s})=S-S_0+\pi^{-1}([K_B]-\pi(S\cdot S_0))+\sum\limits_{I}(S\cdot c_{-\alpha_i}^{G_I})(C^{-1}_{G_I})^{ij}D_j^{G_I}\,,
	\label{eqn:shioda}
\end{align}
where $\text{NS}(X,\mathbb{Q})$ is the Néron-Severi group of divisors modulo numerical equivalence, $S,S_0$ are the divisors that correspond to $\hat{s},\hat{s}_0$ respectively and $K_B$ is the canonical class of the base.
Note that $\text{MW}(X)\rightarrow \text{NS}(X,\mathbb{Q})$ is a group homomorphism and $\text{NS}(X,\mathbb{Q})$ is torsionless.
This implies that the image of any torsional section will be trivial.
On the other hand, the harmonic form that corresponds to the image of a non-torsional generator of MW$(X)$ leads to an Abelian gauge group that survives decompactifying the circle.

However, the presence of a torsional section $\hat{s}^t$ still has a physical effect.
To understand this we first recall the origin of matter in F-theory.
Let us consider loci $C$ that are of codimension two in the base where the fiber over $C$ is reducible with $m$ irreducible components.
Furthermore let $C$ be contained in $n\ge 0$ divisors $S^b_{G_{J_k}},\,k=1,...,n$ with $J_a\ne J_b$ for $a\ne b$ and assume that $m>1+\sum_{k=1}^n\text{rk}(G_{J_k})$.
The components of the generic fiber over $C$ will intersect like the affine Dynkin diagram of a group $G_C$ with
\begin{align}
	\bigoplus\limits_{k=1}^n\mathfrak{g}_{J_k}\subset \mathfrak{g}_C\,.
\end{align}
The Dynkin label of the representation with respect to $G_{J_k}$ of M2-branes that wrap components of the fiber over $C$ is given by
\begin{align}
	\lambda^c_{J_k,i}=D_i^{G_{J_k}}\cdot c\,,
\end{align}
for $k=1,...,n$.
Considering all possible ways in which an M2-brane can wrap components of the fiber (while not wrapping the affine node)
this leads to one adjoint representation for every $G_{J_k}$ and additional representations that correspond to hypermultiplets from M2-branes wrapping e.g. a curve $c$ and some combination of $-c^{G_{J_k}}_{-\alpha}$ for $k=1,...,n$ and $\alpha$ roots of the corresponding gauge algebra. 
The $U(1)$ charges of the latter can be calculated via
\begin{align}
	q_j=\sigma(\hat{s}_j)\cdot c\,,\quad j=1,...,\text{rk}(MW)\,.
\end{align}

Now given a torsional section $\hat{s}^t$ we already noted that $\sigma(\hat{s}^t)$ is a trivial divisor.
This implies that
\begin{align}
	c\cdot \sum\limits_{I}(S^t\cdot c_{-\alpha_i}^{G_I})(C^{-1}_{G_I})^{ij}D_j^{G_I}=\sum\limits_{I}(S^t\cdot c_{-\alpha_i}^{G_I})(C^{-1}_{G_I})^{ij}\lambda^c_{I,j}\in\mathbb{Z}\,,
\end{align}
for every curve $c$. This is a non-trivial constraint on the Dynkin labels and therefore a torsional section restricts the representations of matter that can occur.
In fact, an analogous condition arises also from non-torsional sections.

Finally we have to discuss what happens when the fibration does not admit a section but only $n$-sections with $n>1$.
An $n$-section is locally the union of $n$ sections that are permuted when moving around some divisor in the base.
There are various dual pictures to describe this situation but we will only discuss it from the perspective of Higgsing an Abelian gauge group.
Consider an elliptic fibration with non-zero Mordell-Weil rank and a generating section $\hat{s}$ as well as isolated fibral curves $C_{\pm q}$ that
intersect with the Shioda map as $\sigma(\hat{s})\cdot C_{\pm q}=\pm q$ and therefore lead to hypermultiplets of charge $q$.
One can then tune the K\"ahler moduli such that the volume of one of these curves goes to zero and the corresponding hypermultiplet becomes massless.
Turning on a vacuum expectation value for the scalar field amounts to a complex structure deformation that merges $q$ sections and the resulting geometry is a genus one fibration with $q$-sections.
The $U(1)$ factor of the gauge symmetry that corresponded to $\hat{s}$ is broken to $\mathbb{Z}_q$.
A ``Shioda map'' for the multi-sections in the genus one fibration can be obtained by orthogonalization and ensuring the condition that it has ``one leg along the fiber''.

\subsection{The geometric origin of the anomaly polynomial}
\label{app:anpol}
In a six-dimensional F-theory vacuum the gauge symmetries itself are anomalous and the anomalies have to be cancelled by a generalized Green-Schwarz mechanism.
To this end we note that the six-dimensional spectrum also includes $T=h^{1,1}(B)-1$ tensor multiplets that contain anti-self-dual two-forms $B^\alpha,\,\alpha=2,...,T+1$ and another self-dual two-form $B^{1}$ belongs to the gravity multiplet.
Let us introduce a basis $D_\alpha,\,\alpha=1,...,T+1$ of divisors on the base $B$ and write
\begin{align}
	\Omega_{\alpha\beta}=D_\alpha\cdot D_\beta\,.
\end{align}
The Green-Schwarz counterterm takes the form
\begin{align}
	S_{GS}=-\frac12\int_{M_6}\Omega_{\alpha\beta}B^\alpha\wedge X_4^\beta\,,
	\label{eqn:gscounterterm}
\end{align}
with
\begin{align}
	X_4^\alpha=\frac12a^\alpha\text{tr}R\wedge R+2\sum\limits_{I}\frac{b_I^\alpha}{\lambda_I}\text{tr}F^I\wedge F^I+2\sum\limits_{a,b}b_{ab}^\alpha F^a\wedge F^b\,.
\end{align}
Here $R$ is the gravitational field strength, $F^I$ is the field strength associated to the factor $G_I$ of the non-Abelian gauge group and $F^a$ is the Abelian field strength associated to the section $\hat{s}_a$.
The anomaly coefficients $a^\alpha,\,b_I^\alpha$ and $b^\alpha_{ab}$ are given by
\begin{align}
	a^\alpha=D^\alpha\cdot [K_B]\,,\quad b_I^\alpha=D^\alpha\cdot \pi(S^b_{G_I})\,,\quad b_{ab}^\alpha=-D^\alpha\cdot\pi(\sigma(\hat{s}_a)\cdot\sigma(\hat{s}_b))\,,
\end{align}
where $D^\alpha=(\Omega^{-1})^{\alpha\beta}D_\beta$ and $b_{ab}=-\pi(\sigma(\hat{s}_a)\cdot\sigma(\hat{s}_b))$ is also called the height pairing.
Moreover, $\lambda_I$ is a group theoretical normalization constant and for $G_I=SU(N)$ we have $\lambda_I=1$.

The anomalies can only be cancelled via the counterterm \eqref{eqn:gscounterterm} if the one-loop anomaly polynomial $I_8$ factorizes as
\begin{align}
	I_8=-\frac12\Omega_{\alpha\beta}X_4^\alpha\wedge X_4^\beta\,.
\end{align}
This requires non-trivial relations among the multiplicites of representations and the anomaly coefficients.
It is striking that the geometrical properties of genus one fibered Calabi-Yau manifold seems to guarantee that these conditions are always satisfied.

\section{An identity of characteristic classes}
\label{app:characteristic}  
In this Appendix we show that
\begin{align}
	\int_Mc_2(M)D=12\int_Bc_1(B)\tilde{D}\,,\quad\text{for }D=\pi^*\tilde{D},\,\tilde{D}\in H^{1,1}(B)\,,
	\label{eqn:eqnx}
\end{align}
where $M$ is an elliptically fibered threefold that has been constructed from one of the sixteen toric hypersurfaces that have been discussed in~\cite{Klevers:2014bqa}.
This implies that the relation holds for a large class of elliptic and genus-one fibrations that exhibit $I_n$ fibers, multiple sections and multi-sections.
A generic proof of this relation for elliptic fibrations has been given in~\cite{Esole:2018bmf}.

Let us denote the rays of the two-dimensional reflexive fiber polytope $\Delta$ by $\rho_i,\,i=1,...,k$ and the corresponding divisors by $D_i$
such that they are in clockwise order.
The Chern class of the total space of the fibration is given by the restriction of
\begin{align}
	c(M)=\frac{\left(1+c_1(B)\right)\prod\limits_{i=1}^k(1+D_i)}{1+c_1(B)+\sum\limits_{i=1}^kD_i}\,,
\end{align}
and therefore the second Chern class $c_2(M)$ is the restriction of
\begin{align}
	c_2(M)=c_1(B)\sum\limits_{i=1}^k+\sum\limits_{i<j}^kD_iD_j\,.
\end{align}
For a vertical divisor $D_b=\pi^*\tilde{D}$ we can therefore replace
\begin{align}
	\int_Mc_2(M)D_b=\int\limits_{\mathbb{P}}D_b\left(c_1(B)\sum\limits_{i=1}^kD_i+\sum\limits_{i<j}^kD_iD_j\right)\left(c_1(B)+\sum\limits_{i=1}^kD_i\right)\,,
\end{align}
where $\mathbb{P}$ is the total ambient space.
Further manipulation leads to
\begin{align}
	\int_Mc_2(M)D_b=\underbrace{\int\limits_{\mathbb{P}}D_bc_1(B)\left[\left(\sum\limits_{i=1}^kD_i\right)^2+\sum\limits_{i<j}^kD_iD_j\right]}_{A}+\underbrace{\int\limits_{\mathbb{P}}D_b\sum\limits_{i\ne j}^k D_i^2D_j}_{B}\,.
\end{align}
Since $D_bc_1(B)$ is the class of a multiple of the generic fiber we can rewrite
\begin{align}
	\begin{split}
		A=&\left(\int\limits_B \tilde{D}c_1(B)\right)\left(\int\limits_{\mathbb{P}_\Delta}\left[\left(\sum\limits_{i=1}^kD_i\right)^2+\sum\limits_{i<j}^kD_iD_j\right]\right)\\
		=&\left(\int\limits_B \tilde{D}c_1(B)\right)\int\limits_{\mathbb{P}_\Delta}\left(c_1(F)^2+c_2(F)\right)\,,
	\end{split}
\end{align}
where $\mathbb{P}_\Delta$ is the ambient space of the fiber and $c_i(F)$ are the corresponding Chern classes.
Now the Hirzebruch-Riemann-Roch theorem tells us that
\begin{align}
	\int\limits_{\mathbb{P}_\Delta}\text{Td}(F)=\frac{1}{12}\int\limits_{\mathbb{P}_\Delta}\left(c_1(F)^2+c_2(F)\right)=1\,,
\end{align}
where the second equality reflects that every toric surface is rational.
Therefore
\begin{align}
	A=12\int\limits_Bc_1(B)\tilde{D}\,,
\end{align}
and our remaining task is to show that $B=0$.

Linear equivalences among toric divisors are of the form
\begin{align}
	\sum\limits_{i=1}^k\langle m,\rho_i\rangle D_i\sim 0\,,
\end{align}
for any point $m\in\mathbb{Z}^2$.
If we consider a bundle where the fibers are given by the toric variety, the right hand side is replaced by $D_m=\pi^*\tilde{D}_m$.
Note that the map $m\mapsto D_m$ has to be linear.
Furthermore note that only the divisors that correspond to neighboring rays in the two-dimensional fan have non-zero intersection.

Let us extend our notation such that $D_{k+1}=D_1$ and $D_0=D_k$.
Then
\begin{align}
	B=\int\limits_{\mathbb{P}}D_b\sum_{i=1}^k\left[D_i^2D_{i+1}+D_i^2D_{i-1}\right]\,.
\end{align}
Let us denote by $m_i$ the ray that is dual to the edge that contains $D_i$ and $D_{i+1}$, i.e. $\langle m_i,\rho_i\rangle=\langle m_i,\rho_{i+1}\rangle=-1$.
Then we can use the linear equivalence corresponding to $m_i$ and replace $D_i^2D_{i+1}$ with $-D_i D_{i+1}^2+D_i D_{i+1}D_{m_i}$.
But this implies
\begin{align}
	\begin{split}
		B=&\int\limits_{\mathbb{P}}D_b\sum\limits_{i=1}^kD_i D_{i+1}D_{m_i} =\int\limits_{B}\tilde{D}\sum\limits_{i=1}^k\left(\int\limits_{\mathbb{P}_\Delta}D_i D_{i+1}\right)\tilde{D}_{m_i}\\
		=&\int\limits_B \tilde{D}\sum\limits_{i=1}^k\tilde{D}_{m_i}\,.
	\end{split}
\end{align}
By linearity of the map $m\mapsto \tilde{D}_m$ we only have to show that
\begin{align}
	\sum\limits_{i=1}^k m_i=0\,.
	\label{eqn:sum}
\end{align}

One can easily verify this for the sixteen reflexive two-dimensional polytopes but let us give an easy argument why this has to be true.
As was explained in~\cite{Oehlmann:2016wsb} the intersection of a toric divisor that corresponds to the vertex of the polytope with the generic anti-canonical hypersurface is equal to the number of points on the dual edge minus one. 
Therefore \eqref{eqn:sum} is nothing but the sum over the vertices of $\Delta^\circ$ where each vertex is weighted with the number of intersections of the corresponding divisor with that curve.
Note that the divisors on $\mathbb{P}_{\Delta^\circ}$ that correspond to points in the interior of facets do \textit{not} intersect that curve.
Therefore
\begin{align}
	\sum\limits_{i=1}^k m_i=\sum_{i=1}^{k'}(D_i'\cdot C) \rho_i'\,,
\end{align}
where $\rho_i'$ are the points on the boundary of $\Delta^\circ$, $D_i'$ are the corresponding divisors and $C$ is the class of the anti-canonical hypersurface.
But the intersections of toric divisors with a curve correspond to the coefficients in a linear equivalence among the corresponding rays~\footnote{A nice explanation of this fact can be found in~\cite{DelaOssa:2001blj}.}.
Therefore the sum has to vanish and~\eqref{eqn:eqnx} follows.
\newpage

\section{Discriminants and Picard-Fuchs systems}
\label{app:Pfdisc}
\subsection{$(F_{6}\rightarrow \mathbb{P}_2)[U(1)]_{3}^{-216}$}
\label{app:discpff6}
\paragraph{Picard-Fuchs operators}
{\small
\begin{align}
	\begin{split}
		\mathcal{D}_1=&\Theta _1^2+z_1\left(\Theta _1+1\right) \left(\Theta _1-\Theta _2+3 \Theta _3\right)\\
		&-2z_1z_2 \left(2 \Theta _1+\Theta _2+3 \Theta _3+1\right) \left(3 \Theta _1+\Theta _2+3 \Theta _3+3\right)-2z_2 \left(2 \Theta _1-\Theta _2+3 \Theta _3\right) \Theta _1 \,,\\
		\mathcal{D}_2=&\Theta _1 \left(\Theta _2-3 \Theta _3\right)-z_1\left(\Theta _1-\Theta _2+3 \Theta _3\right) \left(\Theta _2+3 \Theta _3\right)\\
		&-2z_1z_2 \left(2 \Theta _1+\Theta _2+3 \Theta _3+1\right) \left(4 \Theta _1+\Theta _2+3 \Theta _3+3\right) -4z_2 \Theta _1 \left(2 \Theta _1-\Theta _2+3 \Theta _3\right)\,,\\
		\mathcal{D}_3=&\Theta _2 \left(\Theta _1-\Theta _2+3z_2 \Theta _3\right)+\left(2 \Theta _1-\Theta _2+3 \Theta _3\right) \left(2 \Theta _1+\Theta _2+3 \Theta _3+1\right) \,,\\
		\mathcal{D}_4=&\Theta _1 \left(2 \Theta _1-\Theta _2+3 \Theta _3-1\right) \left(2 \Theta _1-\Theta _2+3 \Theta _3\right)\\
		&+z_1\left(\Theta _1-\Theta _2+3 \Theta _3\right) \left(2 \Theta _1+\Theta _2+3 \Theta _3+1\right) \left(2 \Theta _1+\Theta _2+3 \Theta _3+2\right) \,,\\
		\mathcal{D}_5=&-\Theta _1 \Theta _2 \left(2 \Theta _1-\Theta _2+3 \Theta _3\right)\\
		&+z_1z_2\left(2 \Theta _1+\Theta _2+3 \Theta _3+1\right) \left(2 \Theta _1+\Theta _2+3 \Theta _3+2\right) \left(2 \Theta _1+\Theta _2+3 \Theta _3+3\right) \,.
	\end{split}
	\label{eqn:f6pfsystem}
\end{align}
}
\paragraph{Discriminant}
{\small
\begin{align}
	\begin{split}
		\Delta=&1 + 3 z_1 + 3 z_1^2 + z_1^3 + 3 z_2 - 84 z_1 z_2 - 258 z_1^2 z_2 - 
 252 z_1^3 z_2 - 81 z_1^4 z_2 \\
		&+ 3 z_2^2 - 465 z_1 z_2^2 + 1782 z_1^2 z_2^2 + 
 7110 z_1^3 z_2^2 + 7047 z_1^4 z_2^2 + 2187 z_1^5 z_2^2 \\
		&+ z_2^3 - 
 858 z_1 z_2^3 + 18939 z_1^2 z_2^3 + 10500 z_1^3 z_2^3 - 55161 z_1^4 z_2^3 - 
 65610 z_1^5 z_2^3 \\
		&- 19683 z_1^6 z_2^3 - 672 z_1 z_2^4 + 56064 z_1^2 z_2^4 - 
 194112 z_1^3 z_2^4 - 456192 z_1^4 z_2^4 \\
		&- 209952 z_1^5 z_2^4 - 
 192 z_1 z_2^5 + 76032 z_1^2 z_2^5 - 955008 z_1^3 z_2^5 - 
 1057536 z_1^4 z_2^5 \\
		&- 139968 z_1^5 z_2^5 + 49152 z_1^2 z_2^6 - 
 1978368 z_1^3 z_2^6 - 995328 z_1^4 z_2^6 + 12288 z_1^2 z_2^7 \\
		&- 
 2138112 z_1^3 z_2^7 - 331776 z_1^4 z_2^7 - 1179648 z_1^3 z_2^8 - 
 262144 z_1^3 z_2^9 \\
		&+ 27 z_3 - 2268 z_1 z_2 z_3 - 5832 z_1 z_2^2 z_3 + 
 65448 z_1^2 z_2^2 z_3 - 3456 z_2^3 z_3 \\
		&- 9072 z_1 z_2^3 z_3 + 
 307152 z_1^2 z_2^3 z_3 - 699840 z_1^3 z_2^3 z_3 - 10368 z_2^4 z_3 \\
		&- 
 355104 z_1 z_2^4 z_3 - 657072 z_1^2 z_2^4 z_3 - 5458752 z_1^3 z_2^4 z_3 + 
 944784 z_1^4 z_2^4 z_3 \\
		&- 10368 z_2^5 z_3 - 1684800 z_1 z_2^5 z_3 - 
 2592000 z_1^2 z_2^5 z_3 - 12177216 z_1^3 z_2^5 z_3 \\
		&- 3456 z_2^6 z_3 - 
 2996352 z_1 z_2^6 z_3 - 2954880 z_1^2 z_2^6 z_3 - 8118144 z_1^3 z_2^6 z_3 \\
		&- 
 2322432 z_1 z_2^7 z_3 - 1990656 z_1^2 z_2^7 z_3 - 663552 z_1 z_2^8 z_3 - 
 663552 z_1^2 z_2^8 z_3 \\
		&- 93312 z_2^3 z_3^2 - 9517824 z_1 z_2^4 z_3^2 - 
 16796160 z_1 z_2^5 z_3^2 - 15116544 z_1^2 z_2^5 z_3^2 \\
		&+ 
 2985984 z_2^6 z_3^2 + 6718464 z_1 z_2^6 z_3^2 + 8957952 z_2^7 z_3^2 + 
 4478976 z_1 z_2^7 z_3^2 \\
		&+ 8957952 z_2^8 z_3^2 + 2985984 z_2^9 z_3^2 + 
 80621568 z_2^6 z_3^3\,.
\end{split}
	\label{eqn:f6discriminants}
\end{align}
}

\subsection{$(F_{10}\rightarrow \mathbb{P}_2)[SU(2)]_{3}^{-216}$}
\paragraph{Picard-Fuchs operators}
{\small
\begin{align}
	\begin{split}
		\mathcal{D}_1=&-\Theta _2 \left(-\Theta _1+\Theta _2-3 \Theta _3\right)+{z_2} \left(3 \Theta _1-2 \Theta _2+6 \Theta _3-1\right) \left(3 \Theta _1-2 \Theta _2+6 \Theta _3\right)\,,\\
		\mathcal{D}_2=&\Theta _1 \left(3 \Theta _1-2 \Theta _2+6 \Theta _3\right)+{z_1} \left(\Theta _1-\Theta _2+3 \Theta _3\right) \left(3 \Theta _1+2 \Theta _2+6 \Theta _3+3\right)\\
		&-2 {z_1} {z_2} \left(9 \Theta _1^2+6 \Theta _2 \Theta _1+36 \Theta _3 \Theta _1+24 \Theta _1-4 \Theta _2^2+36 \Theta _3^2-8 \Theta _2\right.\\
		&\left.+12 \Theta _2 \Theta _3+48 \Theta _3+3\right)\,,\\
		\mathcal{D}_3=&2 \Theta _1^2-\left(\Theta _1-\Theta _2+3 \Theta _3\right) \left(3 \Theta _1+2 \Theta _2+6 \Theta _3+3\right) {z_1}^2\\
		&+2 {z_2} \left(9 \Theta _1^2+6 \Theta _2 \Theta _1+36 \Theta _3 \Theta _1+24 \Theta _1-4 \Theta _2^2\right.\\
		&\left.+36 \Theta _3^2-8 \Theta _2+12 \Theta _2 \Theta _3 +48 \Theta _3+3\right) {z_1}^2 \\
		&-16 {z_2}^2 \left(2 \Theta _2+3\right) \left(3 \Theta _1-2 \Theta _2+6 \Theta _3\right) {z_1} +\left(-\Theta _1^2+2 \Theta _1-2 \Theta _2+6 \Theta _3\right) {z_1} \\
		&-4 {z_2} \left(9 \Theta _1^2+18 \Theta _3 \Theta _1+12 \Theta _1+4 \Theta _2^2+2 \Theta _2+18 \Theta _3+3\right) {z_1}\,,\\
		\mathcal{D}_4=&-\Theta _1 \Theta _2 \left(3 \Theta _1-2 \Theta _2+6 \Theta _3\right)\\
		&+3 {z_1} {z_2} \left(\Theta _1+2 \Theta _3+1\right) \left(3 \Theta _1+6 \Theta _3+1\right) \left(3 \Theta _1+6 \Theta _3+2\right)\,,\\
		\mathcal{D}_5=&8 \Theta _3^3+738 \Theta _3^3 {z_1}^4+3072 {z_2} \Theta _3^3 {z_1}^3-256 \Theta _3^3 {z_1}^3+1728 {z_2}^2 \Theta _3^3 {z_1}^2-1176 {z_2} \Theta _3^3 {z_1}^2\\
		&+3 {z_3} \left(\Theta _1-\Theta _2+3 \Theta _3\right) \left(381 \Theta _1^2+80 \Theta _2 \Theta _1+288 \Theta _3 \Theta _1-48 \Theta _1\right.\\
		&\left.+16 \Theta _2^2+144 \Theta _3^2+48 \Theta _2+48 \Theta _2 \Theta _3+144 \Theta _3+63\right) {z_1}^2\\
		&-6 {z_2} {z_3} \left(531 \Theta _1^3+138 \Theta _2 \Theta _1^2+2178 \Theta _3 \Theta _1^2+1317 \Theta _1^2-64 \Theta _2^2 \Theta _1\right.\\
		&\left.+2988 \Theta _3^2 \Theta _1-68 \Theta _2 \Theta _1+384 \Theta _2 \Theta _3 \Theta _1+3336 \Theta _3 \Theta _1+771 \Theta _1-32 \Theta _2^3+1944 \Theta _3^3\right.\\
		&\left.-144 \Theta _2^2+1512 \Theta _3^2-190 \Theta _2+96 \Theta _2^2 \Theta _3+480 \Theta _2 \Theta _3+1290 \Theta _3+81\right) {z_1}^2\\
		&+{z_3} \left(141 \Theta _1^3+12 \Theta _2^2 \Theta _1-144 \Theta _2 \Theta _1-32 \Theta _1-648 \Theta _3^3+72 \Theta _2^2-648 \Theta _3^2\right.\\
		&\left.+32 \Theta _2-96 \Theta _3\right) {z_1}+8 {z_2} {z_3} \left(537 \Theta _1^3-408 \Theta _2 \Theta _1^2+2241 \Theta _3 \Theta _1^2-180 \Theta _1^2\right.\\
		&\left.-94 \Theta _2^2 \Theta _1+2538 \Theta _3^2 \Theta _1+223 \Theta _2 \Theta _1+57 \Theta _1-96 \Theta _2^3+1944 \Theta _3^3-238 \Theta _2^2\right.\\
		&\left.+1350 \Theta _3^2-191 \Theta _2+279 \Theta _3-30\right) {z_1}+8 {z_2}^2 {z_3} \left(351 \Theta _1^3-96 \Theta _2 \Theta _1^2+3834 \Theta _3 \Theta _1^2\right.\\
		&\left.+738 \Theta _1^2-488 \Theta _2^2 \Theta _1+11124 \Theta _3^2 \Theta _1-1776 \Theta _2 \Theta _1-408 \Theta _2 \Theta _3 \Theta _1+5796 \Theta _3 \Theta _1\right.\\
		&\left.-837 \Theta _1+384 \Theta _2^3+9720 \Theta _3^3+1384 \Theta _2^2-432 \Theta _2 \Theta _3^2+8640 \Theta _3^2+1224 \Theta _2\right.\\
		&\left.-1152 \Theta _2^2 \Theta _3-4296 \Theta _2 \Theta _3-1242 \Theta _3+216\right) {z_1}\\
		&+2592 {z_2}^2 {z_3}^2 \Theta _1 \left(2 \Theta _2+3\right) \left(3 \Theta _1-2 \Theta _2+6 \Theta _3\right)\\
		&-1296 {z_2} {z_3}^2 \Theta _1 \left(\Theta _1-\Theta _2+3 \Theta _3\right) \left(5 \Theta _1+2 \Theta _2+6 \Theta _3\right)\\
		&+64 {z_2}^3 {z_3} \left(1296 \Theta _1^3-1800 \Theta _2 \Theta _1^2+5076 \Theta _3 \Theta _1^2-585 \Theta _1^2+600 \Theta _2^2 \Theta _1+4752 \Theta _3^2 \Theta _1\right.\\
		&\left.+228 \Theta _2 \Theta _1-3600 \Theta _2 \Theta _3 \Theta _1-1440 \Theta _3 \Theta _1-219 \Theta _1+8 \Theta _2^3-432 \Theta _3^3+72 \Theta _2^2\right.\\
		&\left.-540 \Theta _3^2+100 \Theta _2-144 \Theta _2 \Theta _3-438 \Theta _3-15\right)\\
		&+24 {z_2} {z_3} \left(11 \Theta _1^3-12 \Theta _2^2 \Theta _1-25 \Theta _2 \Theta _1-16 \Theta _1+4 \Theta _2^3-108 \Theta _3^3+12 \Theta _2^2-108 \Theta _3^2\right.\\
		&\left.+14 \Theta _2-42 \Theta _3\right)+8 {z_3} \left(-\Theta _2^3+3 \Theta _1 \Theta _2^2-3 \Theta _2^2+6 \Theta _1 \Theta _2-2 \Theta _2+27 \Theta _3^3+27 \Theta _3^2\right.\\
		&\left.+2 \Theta _1+6 \Theta _3\right)+48 {z_2}^2 {z_3} \left(261 \Theta _1^3-136 \Theta _2 \Theta _1^2+1062 \Theta _3 \Theta _1^2-72 \Theta _1^2-100 \Theta _2^2 \Theta _1\right.\\
		&\left.+1116 \Theta _3^2 \Theta _1+86 \Theta _2 \Theta _1+57 \Theta _1-8 \Theta _2^3+216 \Theta _3^3-40 \Theta _2^2+360 \Theta _3^2-48 \Theta _2+144 \Theta _3\right)\,.
	\end{split}
	\label{eqn:f10pc}
\end{align}
}
\paragraph{Discriminant components}
{\small
\begin{align}
	\begin{split}
		\Delta_1=&
1 + 3z_1 + 3z_1^2 + z_1^3 - 108z_1z_2 - 297z_1^2z_2 - 270z_1^3z_2 - 81z_1^4z_2 \\
		&+ 4536z_1^2z_2^2 + 11016z_1^3z_2^2 + 8667z_1^4z_2^2 + 2187z_1^5z_2^2 - 1296z_1^2z_2^3 \\
		&- 95904z_1^3z_2^3 - 187920z_1^4z_2^3 - 113724z_1^5z_2^3 - 19683z_1^6z_2^3 \\
		&+ 93312z_1^3z_2^4 + 1143072z_1^4z_2^4 + 1469664z_1^5z_2^4 + 472392z_1^6z_2^4 \\
		&- 2239488z_1^4z_2^5 - 8118144z_1^5z_2^5 - 4723920z_1^6z_2^5 + 559872z_1^4z_2^6 \\
		&+ 20715264z_1^5z_2^6 + 25194240z_1^6z_2^6 - 20155392z_1^5z_2^7 - 75582720z_1^6z_2^7 \\
		&+ 120932352z_1^6z_2^8 - 80621568z_1^6z_2^9 + 27z_3 - 324z_2z_3 - 2268z_1z_2z_3 \\
		&+ 1296z_2^2z_3 + 25920z_1z_2^2z_3 + 65448z_1^2z_2^2z_3 - 5184z_2^3z_3 - 
 103680z_1z_2^3z_3 \\
		&- 740016z_1^2z_2^3z_3 - 699840z_1^3z_2^3z_3 - 186624z_1z_2^4z_3 + 1632960z_1^2z_2^4z_3 \\
		&+ 7138368z_1^3z_2^4z_3 + 944784z_1^4z_2^4z_3 + 5038848z_1^2z_2^5z_3 \\
		&- 18475776z_1^3z_2^5z_3 - 18895680z_1^4z_2^5z_3 - 15676416z_1^2z_2^6z_3 \\
		&- 15676416z_1^3z_2^6z_3 + 151165440z_1^4z_2^6z_3 + 80621568z_1^3z_2^7z_3 \\
		&- 604661760z_1^4z_2^7z_3 + 1209323520z_1^4z_2^8z_3 - 967458816z_1^4z_2^9z_3 \\
		&- 93312z_2^3z_3^2 + 1119744z_2^4z_3^2 - 9517824z_1z_2^4z_3^2 - 4478976z_2^5z_3^2 \\
		&+ 116453376z_1z_2^5z_3^2 - 15116544z_1^2z_2^5z_3^2 + 8957952z_2^6z_3^2 \\
		&- 474771456z_1z_2^6z_3^2 + 241864704z_1^2z_2^6z_3^2 + 644972544z_1z_2^7z_3^2 \\
		&- 1451188224z_1^2z_2^7z_3^2 + 3869835264z_1^2z_2^8z_3^2 - 3869835264z_1^2z_2^9z_3^2 \\
		&+ 80621568z_2^6z_3^3 - 967458816z_2^7z_3^3 + 3869835264z_2^8z_3^3 - 5159780352z_2^9z_3^3\,,\\
		\Delta_2=&1-4z_2\,.
	\end{split}
	\label{eqn:f10disc}
\end{align}
}

\section{Toric data for geometries with local limits}
\label{app:localgeometries}
Multi-section geometries:\\
$(F_4\rightarrow\mathbb{F}_1)[\mathbb{Z}_2]^{-224}_3$\\
\begin{align}
	\left(\begin{array}{rrrr|rrr}
		 1& 0&-1& 0& 0& 1& 0\\
		-1& 0&-1& 0& 0& 1&-1\\
		-1& 1&-1& 0& 0& 0& 1\\
		 0&-1&-1& 0& 0& 0& 1\\
		 0& 0&-1&-1& 1& 0& 0\\
		 0& 0&-1& 1& 1& 0& 0\\
		 0& 0& 1& 0& 2& 2& 1\\
		 0& 0& 0& 0&-4&-4&-2
	\end{array}\right)
\end{align}
$(F_1\rightarrow\mathbb{F}_1)[\mathbb{Z}_3]^{-144}_3$\\
\begin{align}
	\left(\begin{array}{rrrr|rrr}
		 1& 0& 0& 0& 0& 1& 0\\
		-1& 0& 0& 0& 0& 1&-1\\
		-1& 1& 0& 0& 0& 0& 1\\
		 0&-1& 0& 0& 0& 0& 1\\
		 0& 0& 1& 0& 1& 0& 0\\
		 0& 0& 0& 1& 1& 0& 0\\
		 0& 0&-1&-1& 1& 0& 0\\
		 0& 0& 0& 0&-3& 0& 0
	\end{array}\right)
\end{align}
$(\mathbb{P}^3\rightarrow\mathbb{F}_1)[\mathbb{Z}_4]^{-104}_3$\vspace{.5cm}\\
The label $\Delta_i,\,i\in\{1,2\}$ indicates to which part of the nef-partition the corresponding point belongs:
\begin{align}
	\begin{blockarray}{rrrrrr|rrr}
		\begin{block}{r(rrrrr|rrr)}
			\Delta_2& 1& 0& 0& 0& 0& 0& 1& 0\\
			\Delta_1&-1& 0& 0& 0& 0& 0& 1&-1\\
			\Delta_2&-1& 1& 0& 0& 0& 0& 0& 1\\
			\Delta_1& 0&-1& 0& 0& 0& 0& 0& 1\\
			\Delta_1& 0& 0& 1& 0& 0& 1& 0& 0\\
			\Delta_1& 0& 0& 0& 1& 0& 1& 0& 0\\
			\Delta_2& 0& 0& 0& 0& 1& 1& 0& 0\\
			\Delta_2& 0& 0&-1&-1&-1& 1& 0& 0\\
			& 0& 0& 0& 0& 0&-4& 0& 0\\
		\end{block}
	\end{blockarray}
	\begin{array}{rrrrr|rrr}
	\end{array}
\end{align}
Pseudo multi-section geometries:\\
$(F_4\rightarrow\mathbb{F}_1)[\mathbb{Z}_2]^{-288}_{4(1)}$\\
\begin{align}
	\left(\begin{array}{rrrr|rrr}
		 1& 0&-1&-1& 0& 1& 0\\
		-1& 0&-1&-1& 0& 1&-1\\
		-1& 1&-1&-1& 0& 0& 1\\
		 0&-1&-1&-1& 0& 0& 1\\
		 0& 0&-1&-1& 1&-2&-1\\
		 0& 0&-1& 1& 1& 0& 0\\
		 0& 0& 1& 0& 2& 0& 0\\
		 0& 0& 0& 0&-4& 0& 0
	\end{array}\right)
\end{align}
$(F_1\rightarrow\mathbb{F}_1)[\mathbb{Z}_3]^{-192}_{5(2)}$\\
\begin{align}
	\left(\begin{array}{rrrr|rrr}
		 1& 0& 1& 0& 0& 1& 0\\
		-1& 0& 1& 0& 0& 1&-1\\
		-1& 1& 1& 0& 0& 0& 1\\
		 0&-1& 1& 0& 0& 0& 1\\
		 0& 0& 1& 0& 1&-2&-1\\
		 0& 0& 0& 1& 1& 0& 0\\
		 0& 0&-1&-1& 1& 0& 0\\
		 0& 0& 0& 0&-3& 0& 0
	\end{array}\right)
\end{align}
$(\mathbb{P}^3\rightarrow\mathbb{F}_1)[\mathbb{Z}_4]^{-128}_{6(3)}$\\
The label $\Delta_i,\,i\in\{1,2\}$ indicates to which part of the nef-partition the corresponding point belongs:
\begin{align}
	\begin{blockarray}{rrrrrr|rrr}
		\begin{block}{r(rrrrr|rrr)}
			\Delta_1& 1& 0& 0& 0& 1& 0& 1& 0\\
			\Delta_1&-1& 0& 0& 0& 1& 0& 1&-1\\
			\Delta_2&-1& 1& 0& 0& 1& 0& 0& 1\\
			\Delta_1& 0&-1& 0& 0& 1& 0& 0& 1\\
			\Delta_1& 0& 0& 1& 0& 0& 1& 0& 0\\
			\Delta_1& 0& 0& 0& 1& 0& 1& 0& 0\\
			\Delta_1& 0& 0& 0& 0& 1& 1& 0& 0\\
			\Delta_2& 0& 0&-1&-1&-1& 1& 0& 0\\
			& 0& 0& 0& 0& 0&-4& 0& 0\\
		\end{block}
	\end{blockarray}
	\begin{array}{rrrrr|rrr}
	\end{array}
\end{align}

\section{Modular expressions for refined partition functions}
\label{app:modexp}

\begin{tiny}
\begin{align}
\begin{split}
\phi_2^{(3)} = & \frac{\Delta_6^{\frac{4}{3}}}{\eta^{24}(3\tau)} \Bigg(-\frac{145 \mathcal{A}_{+,3}^4 \mathcal{A}_{-,3}^2 E_{2,3}^3 E_6}{1417176}-\frac{21313
   \mathcal{A}_{+,3}^4 \mathcal{A}_{-,3}^2 E_{2,3}^2 E_4^2}{181398528}-\frac{1549
   \mathcal{A}_{+,3}^4 \mathcal{A}_{-,3}^2 E_{2,3} E_4 E_6}{34012224}-\frac{3847
   \mathcal{A}_{+,3}^4 \mathcal{A}_{-,3}^2 E_4^3}{1088391168}\\
   &-\frac{\mathcal{A}_{+,3}^4
   \mathcal{A}_{-,3}^2 E_6^2}{419904}+\frac{77 \mathcal{A}_{+,3}^4 \mathcal{A}_{-,3} \mathcal{B}_{-,3}
   E_{2,3}^3 E_4}{209952}+\frac{2501 \mathcal{A}_{+,3}^4 \mathcal{A}_{-,3} \mathcal{B}_{-,3}
   E_{2,3}^2 E_6}{7558272}+\frac{2749 \mathcal{A}_{+,3}^4 \mathcal{A}_{-,3} \mathcal{B}_{-,3}
   E_{2,3} E_4^2}{30233088}\\
   &+\frac{\mathcal{A}_{+,3}^4 \mathcal{A}_{-,3} \mathcal{B}_{-,3}
   E_4 E_6}{139968}-\frac{295 \mathcal{A}_{+,3}^4 \mathcal{B}_{-,3}^2 E_{2,3}^2
   E_4}{1119744}-\frac{257 \mathcal{A}_{+,3}^4 \mathcal{B}_{-,3}^2 E_{2,3}
   E_6}{1259712}-\frac{49 \mathcal{A}_{+,3}^4 \mathcal{B}_{-,3}^2
   E_4^2}{1259712}\\
   &+\frac{847 \mathcal{A}_{+,3}^3 \mathcal{A}_{-,3}^2 \mathcal{B}_{+,3} E_{2,3}^3
   E_4}{839808}+\frac{5089 \mathcal{A}_{+,3}^3 \mathcal{A}_{-,3}^2 \mathcal{B}_{+,3} E_{2,3}^2
   E_6}{7558272}+\frac{3719 \mathcal{A}_{+,3}^3 \mathcal{A}_{-,3}^2 \mathcal{B}_{+,3} E_{2,3}
   E_4^2}{30233088}+\frac{\mathcal{A}_{+,3}^3 \mathcal{A}_{-,3}^2 \mathcal{B}_{+,3} E_4
   E_6}{279936}\\
  & -\frac{1555 \mathcal{A}_{+,3}^3 \mathcal{A}_{-,3} \mathcal{B}_{+,3} \mathcal{B}_{-,3}
   E_{2,3}^2 E_4}{559872}-\frac{1157 \mathcal{A}_{+,3}^3 \mathcal{A}_{-,3} \mathcal{B}_{+,3}
   \mathcal{B}_{-,3} E_{2,3} E_6}{629856}-\frac{1535 \mathcal{A}_{+,3}^3 \mathcal{A}_{-,3}
   \mathcal{B}_{+,3} \mathcal{B}_{-,3} E_4^2}{5038848}\\
   &+\frac{55 \mathcal{A}_{+,3}^3 \mathcal{B}_{+,3}
   \mathcal{B}_{-,3}^2 E_{2,3}^3}{15552}-\frac{25 \mathcal{A}_{+,3}^3 \mathcal{B}_{+,3} \mathcal{B}_{-,3}^2
   E_{2,3} E_4}{31104}-\frac{\mathcal{A}_{+,3}^3 \mathcal{B}_{+,3} \mathcal{B}_{-,3}^2
   E_6}{7776}-\frac{965 \mathcal{A}_{+,3}^2 \mathcal{A}_{-,3}^2 \mathcal{B}_{+,3}^2 E_{2,3}^2
   E_4}{373248}\\
   &-\frac{643 \mathcal{A}_{+,3}^2 \mathcal{A}_{-,3}^2 \mathcal{B}_{+,3}^2 E_{2,3}
   E_6}{419904}-\frac{751 \mathcal{A}_{+,3}^2 \mathcal{A}_{-,3}^2 \mathcal{B}_{+,3}^2
   E_4^2}{3359232}+\frac{35 \mathcal{A}_{+,3}^2 \mathcal{A}_{-,3} \mathcal{B}_{+,3}^2 \mathcal{B}_{-,3}
   E_{2,3}^3}{2592}+\frac{85 \mathcal{A}_{+,3} \mathcal{A}_{-,3}^2 \mathcal{B}_{+,3}^3
   E_{2,3}^3}{15552}\\
   &-\frac{\mathcal{A}_{+,3}^2 \mathcal{A}_{-,3} \mathcal{B}_{+,3}^2
   \mathcal{B}_{-,3} E_6}{2592}-\frac{13 \mathcal{A}_{+,3}^2 \mathcal{B}_{+,3}^2 \mathcal{B}_{-,3}^2
   E_{2,3}^2}{3456}+\frac{\mathcal{A}_{+,3}^2 \mathcal{B}_{+,3}^2 \mathcal{B}_{-,3}^2
   E_4}{3456}-\frac{7 \mathcal{A}_{+,3}^2 \mathcal{A}_{-,3} \mathcal{B}_{+,3}^2 \mathcal{B}_{-,3}
   E_{2,3} E_4}{2592}\\
   &-\frac{31 \mathcal{A}_{+,3} \mathcal{A}_{-,3}^2 \mathcal{B}_{+,3}^3 E_{2,3}
   E_4}{31104}-\frac{\mathcal{A}_{+,3} \mathcal{A}_{-,3}^2 \mathcal{B}_{+,3}^3
   E_6}{7776}-\frac{47 \mathcal{A}_{+,3} \mathcal{A}_{-,3} \mathcal{B}_{+,3}^3 \mathcal{B}_{-,3}
   E_{2,3}^2}{5184}+\frac{5 \mathcal{A}_{+,3} \mathcal{A}_{-,3} \mathcal{B}_{+,3}^3 \mathcal{B}_{-,3}
   E_4}{5184}\\
   &-\frac{1}{864} \mathcal{A}_{+,3} \mathcal{B}_{+,3}^3 \mathcal{B}_{-,3}^2
   E_{2,3}-\frac{17 \mathcal{A}_{-,3}^2 \mathcal{B}_{+,3}^4
   E_{2,3}^2}{10368}+\frac{\mathcal{A}_{-,3}^2 \mathcal{B}_{+,3}^4 E_4}{5184}+\frac{1}{864}
   \mathcal{A}_{-,3} \mathcal{B}_{+,3}^4 \mathcal{B}_{-,3} E_{2,3}+\frac{\mathcal{B}_{+,3}^4
   \mathcal{B}_{-,3}^2}{288}\Bigg)\,.
\end{split}
\end{align}

\begin{align}
\begin{split}
\phi^{(4)}_2= &\frac{ \Delta_8^{\frac{1}{2}} }{\eta^{24}(4\tau)}(E_{2,2}-E_{2,4})^2\Bigg(\frac{7079 \mathcal{A}_{-,4}^2 E_4^3 \mathcal{A}_{+,4}^4}{14495514624}+\frac{179
   \mathcal{B}_{-,4}^2 E_4^2 \mathcal{A}_{+,4}^4}{63700992}+\frac{3287 \mathcal{A}_{-,4}^2
   E_{2,4}^2 E_4^2 \mathcal{A}_{+,4}^4}{1207959552}\\
   &-\frac{5263 \mathcal{A}_{-,4}
   \mathcal{B}_{-,4} E_{2,2} E_4^2 \mathcal{A}_{+,4}^4}{11777605632}+\frac{64067
   \mathcal{A}_{-,4} \mathcal{B}_{-,4} E_{2,4} E_4^2 \mathcal{A}_{+,4}^4}{4416602112}-\frac{1025
   \mathcal{A}_{-,4}^2 E_{2,2} E_{2,4} E_4^2 \mathcal{A}_{+,4}^4}{37748736}\\
   &+\frac{261481
   \mathcal{A}_{-,4}^2 E_6^2 \mathcal{A}_{+,4}^4}{5087925633024}+\frac{30527 \mathcal{A}_{-,4}
   \mathcal{B}_{-,4} E_{2,4}^3 E_4 \mathcal{A}_{+,4}^4}{368050176}+\frac{407 \mathcal{B}_{-,4}^2
   E_{2,4}^2 E_4 \mathcal{A}_{+,4}^4}{5308416}-\frac{1015 \mathcal{B}_{-,4}^2 E_{2,2}
   E_{2,4} E_4 \mathcal{A}_{+,4}^4}{7077888}\\
   &-\frac{61265 \mathcal{A}_{-,4}^2 E_{2,4}^3
   E_6 \mathcal{A}_{+,4}^4}{2944401408}+\frac{28385 \mathcal{A}_{-,4} \mathcal{B}_{-,4}
   E_{2,4}^2 E_6 \mathcal{A}_{+,4}^4}{509607936}+\frac{198677 \mathcal{A}_{-,4} \mathcal{B}_{-,4} E_4
   E_6 \mathcal{A}_{+,4}^4}{317995352064}\\
   &-\frac{649 \mathcal{B}_{-,4}^2 E_{2,2}
   E_6 \mathcal{A}_{+,4}^4}{191102976}-\frac{2197 \mathcal{B}_{-,4}^2 E_{2,4} E_6
   \mathcal{A}_{+,4}^4}{63700992}+\frac{4837 \mathcal{A}_{-,4} \mathcal{B}_{-,4} E_{2,2} E_{2,4}
   E_6 \mathcal{A}_{+,4}^4}{1019215872}\\
   &-\frac{156265 \mathcal{A}_{-,4}^2 E_{2,2}
   E_4 E_6 \mathcal{A}_{+,4}^4}{282662535168}-\frac{1177459 \mathcal{A}_{-,4}^2
   E_{2,4} E_4 E_6 \mathcal{A}_{+,4}^4}{141331267584}+\frac{215 \mathcal{B}_{+,4}
   \mathcal{B}_{-,4}^2 E_{2,4}^3 \mathcal{A}_{+,4}^3}{359424}\\
   &+\frac{1741 \mathcal{A}_{-,4}
   \mathcal{B}_{+,4} \mathcal{B}_{-,4} E_4^2 \mathcal{A}_{+,4}^3}{63700992}-\frac{7057
   \mathcal{A}_{-,4}^2 \mathcal{B}_{+,4} E_{2,2} E_4^2
   \mathcal{A}_{+,4}^3}{11777605632}+\frac{133669 \mathcal{A}_{-,4}^2 \mathcal{B}_{+,4} E_{2,4}
   E_4^2 \mathcal{A}_{+,4}^3}{4416602112}\\
   &+\frac{71113 \mathcal{A}_{-,4}^2 \mathcal{B}_{+,4}
   E_{2,4}^3 E_4 \mathcal{A}_{+,4}^3}{368050176}+\frac{3649 \mathcal{A}_{-,4} \mathcal{B}_{+,4}
   \mathcal{B}_{-,4} E_{2,4}^2 E_4 \mathcal{A}_{+,4}^3}{5308416}+\frac{1693 \mathcal{B}_{+,4}
   \mathcal{B}_{-,4}^2 E_{2,2} E_4 \mathcal{A}_{+,4}^3}{103514112}\\
   &-\frac{185 \mathcal{B}_{+,4}
   \mathcal{B}_{-,4}^2 E_{2,4} E_4 \mathcal{A}_{+,4}^3}{2156544}-\frac{175 \mathcal{A}_{-,4}
   \mathcal{B}_{+,4} \mathcal{B}_{-,4} E_{2,2} E_{2,4} E_4
   \mathcal{A}_{+,4}^3}{131072}-\frac{1127 \mathcal{B}_{+,4} \mathcal{B}_{-,4}^2 E_6
   \mathcal{A}_{+,4}^3}{310542336}\\
   &+\frac{58735 \mathcal{A}_{-,4}^2 \mathcal{B}_{+,4} E_{2,4}^2
   E_6 \mathcal{A}_{+,4}^3}{509607936}-\frac{6023 \mathcal{A}_{-,4} \mathcal{B}_{+,4} \mathcal{B}_{-,4}
   E_{2,2} E_6 \mathcal{A}_{+,4}^3}{191102976}-\frac{10297 \mathcal{A}_{-,4} \mathcal{B}_{+,4}
   \mathcal{B}_{-,4} E_{2,4} E_6 \mathcal{A}_{+,4}^3}{31850496}\\
   &+\frac{17603 \mathcal{A}_{-,4}^2
   \mathcal{B}_{+,4} E_{2,2} E_{2,4} E_6 \mathcal{A}_{+,4}^3}{1019215872}+\frac{293707
   \mathcal{A}_{-,4}^2 \mathcal{B}_{+,4} E_4 E_6 \mathcal{A}_{+,4}^3}{317995352064}+\frac{35
   \mathcal{A}_{-,4} \mathcal{B}_{+,4}^2 \mathcal{B}_{-,4} E_{2,4}^3 \mathcal{A}_{+,4}^2}{14976}\\
   &-\frac{25
   \mathcal{B}_{+,4}^2 \mathcal{B}_{-,4}^2 E_{2,4}^2 \mathcal{A}_{+,4}^2}{10368}+\frac{1025
   \mathcal{A}_{-,4}^2 \mathcal{B}_{+,4}^2 E_4^2 \mathcal{A}_{+,4}^2}{42467328}+\frac{37
   \mathcal{B}_{+,4}^2 \mathcal{B}_{-,4}^2 E_{2,2} E_{2,4} \mathcal{A}_{+,4}^2}{20736}-\frac{5
   \mathcal{B}_{+,4}^2 \mathcal{B}_{-,4}^2 E_4 \mathcal{A}_{+,4}^2}{62208}\\
   &+\frac{2021
   \mathcal{A}_{-,4}^2 \mathcal{B}_{+,4}^2 E_{2,4}^2 E_4 \mathcal{A}_{+,4}^2}{3538944}+\frac{35
   \mathcal{A}_{-,4} \mathcal{B}_{+,4}^2 \mathcal{B}_{-,4} E_{2,2} E_4
   \mathcal{A}_{+,4}^2}{718848}-\frac{7 \mathcal{A}_{-,4} \mathcal{B}_{+,4}^2 \mathcal{B}_{-,4} E_{2,4}
   E_4 \mathcal{A}_{+,4}^2}{22464}\\
   &-\frac{2695 \mathcal{A}_{-,4}^2 \mathcal{B}_{+,4}^2 E_{2,2}
   E_{2,4} E_4 \mathcal{A}_{+,4}^2}{2359296}-\frac{107 \mathcal{A}_{-,4} \mathcal{B}_{+,4}^2
   \mathcal{B}_{-,4} E_6 \mathcal{A}_{+,4}^2}{6469632}-\frac{3427 \mathcal{A}_{-,4}^2
   \mathcal{B}_{+,4}^2 E_{2,2} E_6 \mathcal{A}_{+,4}^2}{127401984}\\
   &-\frac{5903
   \mathcal{A}_{-,4}^2 \mathcal{B}_{+,4}^2 E_{2,4} E_6 \mathcal{A}_{+,4}^2}{21233664}+\frac{115
   \mathcal{A}_{-,4}^2 \mathcal{B}_{+,4}^3 E_{2,4}^3 \mathcal{A}_{+,4}}{119808}-\frac{65 \mathcal{A}_{-,4}
   \mathcal{B}_{+,4}^3 \mathcal{B}_{-,4} E_{2,4}^2 \mathcal{A}_{+,4}}{15552}+\frac{\mathcal{B}_{+,4}^3
   \mathcal{B}_{-,4}^2 E_{2,2} \mathcal{A}_{+,4}}{23328}\\
   &-\frac{\mathcal{B}_{+,4}^3 \mathcal{B}_{-,4}^2
   E_{2,4} \mathcal{A}_{+,4}}{3888}+\frac{83 \mathcal{A}_{-,4} \mathcal{B}_{+,4}^3 \mathcal{B}_{-,4}
   E_{2,2} E_{2,4} \mathcal{A}_{+,4}}{31104}-\frac{7 \mathcal{A}_{-,4} \mathcal{B}_{+,4}^3
   \mathcal{B}_{-,4} E_4 \mathcal{A}_{+,4}}{93312}\\
   &+\frac{1667 \mathcal{A}_{-,4}^2 \mathcal{B}_{+,4}^3
   E_{2,2} E_4 \mathcal{A}_{+,4}}{103514112}-\frac{263 \mathcal{A}_{-,4}^2 \mathcal{B}_{+,4}^3
   E_{2,4} E_4 \mathcal{A}_{+,4}}{2156544}-\frac{2297 \mathcal{A}_{-,4}^2 \mathcal{B}_{+,4}^3
   E_6 \mathcal{A}_{+,4}}{310542336}+\frac{\mathcal{B}_{+,4}^4 \mathcal{B}_{-,4}^2}{1458}\\
   &-\frac{5
   \mathcal{A}_{-,4}^2 \mathcal{B}_{+,4}^4 E_{2,4}^2}{7776}-\frac{\mathcal{A}_{-,4} \mathcal{B}_{+,4}^4
   \mathcal{B}_{-,4} E_{2,2}}{23328}+\frac{\mathcal{A}_{-,4} \mathcal{B}_{+,4}^4 \mathcal{B}_{-,4}
   E_{2,4}}{3888}+\frac{23 \mathcal{A}_{-,4}^2 \mathcal{B}_{+,4}^4 E_{2,2}
   E_{2,4}}{62208}-\frac{\mathcal{A}_{-,4}^2 \mathcal{B}_{+,4}^4 E_4}{186624}\Bigg)\,.
\end{split}
\end{align}

\end{tiny}

\section{Refined BPS invariants}
\label{app:refBPS}

\begin{table}[h!]
	\centering
	{\tiny
		\begin{tabular}{|r|c|}\hline
			$N_{j_-j_+}^{(1,0)}$ &$2j_+=$0\\ \hline
			$2j_-=$0&16\\ 
			\hline \end{tabular}
			\begin{tabular}{|r|c|}\hline
			$N_{j_-j_+}^{(1,1)}$ &$2j_+=$0\\ \hline
			$2j_-=$0&128\\ 
			\hline \end{tabular}
		\begin{tabular}{|r|cc|}\hline
			$N_{j_-j_+}^{(1,2)}$ &$2j_+=$0&1\\ \hline
			$2j_-=$0&576&\\ 
			1 &&16\\ 
			\hline \end{tabular}
		\begin{tabular}{|r|cc|}\hline
			$N_{j_-j_+}^{(1,3)}$ &$2j_+=$0&1\\ \hline
			$2j_-=$0&2048&\\ 
			1 &&128\\ 
			\hline \end{tabular}
		\begin{tabular}{|r|ccc|}\hline
			$N_{j_-j_+}^{(1,4)}$ &$2j_+=$0&1&2\\ \hline
			$2j_-=$0&6320&&\\ 
			1 &&592&\\ 
			2 &&&16\\  
			\hline \end{tabular}
		\begin{tabular}{|r|ccc|}\hline
			$N_{j_-j_+}^{(1,5)}$ &$2j_+=$0&1&2\\ \hline
			$2j_-=$0&17536&&\\ 
			1 &&2176&\\ 
			2 &&&128\\  
			\hline \end{tabular}
		\begin{tabular}{|r|cc|}\hline
			$N_{j_-j_+}^{(2,0)}$ &$2j_+=$0 & 1\\ \hline
			$2j_-=$0& & 1\\
			\hline \end{tabular}
			\begin{tabular}{|r|cc|}\hline
			$N_{j_-j_+}^{(2,1)}$ &$2j_+=$0 & 1\\ \hline
			$2j_-=$0& & 128\\ 
			\hline \end{tabular}	
		\begin{tabular}{|r|cccc|}\hline
			$N_{j_-j_+}^{(2,2)}$ &$2j_+=$0&1&2 & 3\\ \hline
			$2j_-=$0& & 1942 &&\\ 
			1 &1&& 121 &\\ 
			 2&&&&1\\  
			\hline \end{tabular}
		\begin{tabular}{|r|cccc|}\hline
			$N_{j_-j_+}^{(2,3)}$ &$2j_+=$0&1&2 & 3\\ \hline
			$2j_-=$0& & 15616 &&\\ 
			1 &128&& 2176 &\\ 
			 2&&&&128\\  
			\hline \end{tabular}
		\begin{tabular}{|r|cccccc|}\hline
			$N_{j_-j_+}^{(2,4)}$ &$2j_+=$0&1&2 &3& 4&5\\ \hline
			$2j_-=$0& & 93163 &&121 &&\\ 
			1 &2063&& 19408 &&1&\\ 
			 2&&122&&2199&&\\  
			 3&&&1&&121&\\
			 4&&&&&&1\\
			\hline \end{tabular}
		\begin{tabular}{|r|cccccc|}\hline
			$N_{j_-j_+}^{(2,5)}$ &$2j_+=$0&1&2 &3& 4&5\\ \hline
			$2j_-=$0& & 455808 &&2176 &&\\ 
			1 &17792&& 124416 &&128&\\ 
			 2&&2304&&19840&&\\  
			 3&&&128&&2176&\\
			 4&&&&&&128\\
			\hline \end{tabular}		
			}
			\caption{Refined BPS invariants $N_{j_-,j_+}^{(b,d)}$ of $M_2$, for base degree $b\leq2$ and fiber degree $d\leq 5$. }
	\label{tab:M2ref}
\end{table}

\begin{table}[h!]
	\centering
	{\tiny
		\begin{tabular}{|r|c|}\hline
			$N_{j_-j_+}^{(1,0)}$ &$2j_+=$0\\ \hline
			$2j_-=$0&9\\ 
			\hline \end{tabular}
			\begin{tabular}{|r|c|}\hline
			$N_{j_-j_+}^{(1,1)}$ &$2j_+=$0\\ \hline
			$2j_-=$0&36\\ 
			\hline \end{tabular}
			\begin{tabular}{|r|c|}\hline
			$N_{j_-j_+}^{(1,2)}$ &$2j_+=$0\\ \hline
			$2j_-=$0&126\\ 
			\hline \end{tabular}
		\begin{tabular}{|r|cc|}\hline
			$N_{j_-j_+}^{(1,3)}$ &$2j_+=$0&1\\ \hline
			$2j_-=$0&324&\\ 
			1 &&9\\ 
			\hline \end{tabular}
		\begin{tabular}{|r|cc|}\hline
			$N_{j_-j_+}^{(1,4)}$ &$2j_+=$0&1\\ \hline
			$2j_-=$0&801&\\ 
			1 &&36\\ 
			\hline \end{tabular}
				\begin{tabular}{|r|cc|}\hline
			$N_{j_-j_+}^{(1,5)}$ &$2j_+=$0&1\\ \hline
			$2j_-=$0&1764&\\ 
			1 &&126\\ 
			\hline \end{tabular}
		\begin{tabular}{|r|c|}\hline
			$N_{j_-j_+}^{(2,0)}$ &$2j_+=$0 \\ \hline
			$2j_-=$0&  \\
			\hline \end{tabular}
		\begin{tabular}{|r|cc|}\hline
			$N_{j_-j_+}^{(2,1)}$ &$2j_+=$0 & 1\\ \hline
			$2j_-=$0& & 9\\ 
			\hline \end{tabular}	
		\begin{tabular}{|r|ccc|}\hline
			$N_{j_-j_+}^{(2,2)}$ &$2j_+=$0 & 1&2\\ \hline
			$2j_-=$0&&& 126\\ 
			\hline \end{tabular}
		\\
		\begin{tabular}{|r|ccc|}\hline
			$N_{j_-j_+}^{(2,3)}$ &$2j_+=$0&1&2 \\ \hline
			$2j_-=$0& & 756 &\\ 
			1 &&& 36 \\ 
			\hline \end{tabular}
		\begin{tabular}{|r|cccc|}\hline
			$N_{j_-j_+}^{(2,4)}$ &$2j_+=$0&1&2&3 \\ \hline
			$2j_-=$0& & 3838 &&\\ 
			1 & 9 & &333&  \\
			2 &&&&9\\ 
			\hline \end{tabular}
		\begin{tabular}{|r|cccc|}\hline
			$N_{j_-j_+}^{(2,5)}$ &$2j_+=$0&1&2 & 3\\ \hline
			$2j_-=$0& & 12852 &&\\ 
			1 &126&& 1890 &\\ 
			 2&&&&9\\  
			\hline \end{tabular}

			}
			\caption{Refined BPS invariants $N_{j_-,j_+}^{(b,d)}$ of $M_3$, for base degree $b\leq2$ and fiber degree $d\leq 5$.}
\end{table}

\begin{table}[h!]
	\centering
	{\tiny
		\begin{tabular}{|r|c|}\hline
			$N_{j_-j_+}^{(1,0)}$ &$2j_+=$0\\ \hline
			$2j_-=$0&8\\ 
			\hline \end{tabular}
		\begin{tabular}{|r|c|}\hline
			$N_{j_-j_+}^{(1,1)}$ &$2j_+=$0\\ \hline
			$2j_-=$0&16\\ 
			\hline \end{tabular}
		\begin{tabular}{|r|c|}\hline
			$N_{j_-j_+}^{(1,2)}$ &$2j_+=$0\\ \hline
			$2j_-=$0&56\\ 
			\hline \end{tabular}
		\begin{tabular}{|r|c|}\hline
			$N_{j_-j_+}^{(1,3)}$ &$2j_+=$0\\ \hline
			$2j_-=$0&112\\ 
			\hline \end{tabular}
		\begin{tabular}{|r|cc|}\hline
			$N_{j_-j_+}^{(1,4)}$ &$2j_+=$0&1\\ \hline
			$2j_-=$0&248&\\ 
			1 &&8\\ 
			\hline \end{tabular}
		\begin{tabular}{|r|cc|}\hline
			$N_{j_-j_+}^{(1,5)}$ &$2j_+=$0&1\\ \hline
			$2j_-=$0&464&\\ 
			1 &&16\\ 
			\hline \end{tabular}
		\begin{tabular}{|r|c|}\hline
			$N_{j_-j_+}^{(2,0)}$ &$2j_+=$0 \\ \hline
			$2j_-=$0&  \\
			\hline \end{tabular}
		\begin{tabular}{|r|cc|}\hline
			$N_{j_-j_+}^{(2,1)}$ &$2j_+=$0 & 1\\ \hline
			$2j_-=$0& & 2\\ 
			\hline \end{tabular}	
		\begin{tabular}{|r|cc|}\hline
			$N_{j_-j_+}^{(2,2)}$ &$2j_+=$0 & 1\\ \hline
			$2j_-=$0&& 28\\ 
			\hline \end{tabular}
		\begin{tabular}{|r|cc|}\hline
			$N_{j_-j_+}^{(2,3)}$ &$2j_+=$0 & 1\\ \hline
			$2j_-=$0&& 140\\ 
			\hline \end{tabular}
		\begin{tabular}{|r|ccc|}\hline
			$N_{j_-j_+}^{(2,4)}$ &$2j_+=$0&1&2 \\ \hline
			$2j_-=$0& & 532 &\\ 
			1 &&& 28 \\ 
			\hline \end{tabular}
		\begin{tabular}{|r|cccc|}\hline
			$N_{j_-j_+}^{(2,5)}$ &$2j_+=$0&1&2&3 \\ \hline
			$2j_-=$0& & 1702 &&\\ 
			1 & 2 & &130&  \\
			2 &&&&2\\ 
			\hline \end{tabular}
			}
			\caption{Refined BPS invariants $N_{j_-,j_+}^{(b,d)}$ of $M_4$, for base degree $b\leq2$ and fiber degree $d\leq 5$.}
\end{table}

\begin{table}[h!]
	\centering
	{\tiny
		\begin{tabular}{|r|c|}\hline
			$N_{j_-j_+}^{(1,0)}$ &$2j_+=$0\\ \hline
			$2j_-=$0&2\\ 
			\hline \end{tabular}
			\begin{tabular}{|r|c|}\hline
			$N_{j_-j_+}^{(1,1)}$ &$2j_+=$0\\ \hline
			$2j_-=$0&56\\ 
			\hline \end{tabular}
		\begin{tabular}{|r|cc|}\hline
			$N_{j_-j_+}^{(1,2)}$ &$2j_+=$0&1\\ \hline
			$2j_-=$0&268&\\ 
			1 &&2\\ 
			\hline \end{tabular}
		\begin{tabular}{|r|cc|}\hline
			$N_{j_-j_+}^{(1,3)}$ &$2j_+=$0&1\\ \hline
			$2j_-=$0&1136&\\ 
			1 &&56\\ 
			\hline \end{tabular}
		\begin{tabular}{|r|ccc|}\hline
			$N_{j_-j_+}^{(1,4)}$ &$2j_+=$0&1&2\\ \hline
			$2j_-=$0&3620&&\\ 
			1 &&270&\\ 
			2 &&&2\\  
			\hline \end{tabular}
		\begin{tabular}{|r|ccc|}\hline
			$N_{j_-j_+}^{(1,5)}$ &$2j_+=$0&1&2\\ \hline
			$2j_-=$0&10688&&\\ 
			1 &&1192&\\ 
			2 &&&56\\  
			\hline \end{tabular}
		\\
		\begin{tabular}{|r|c|}\hline
			$N_{j_-j_+}^{(2,0)}$ &$2j_+=$0 \\ \hline
			$2j_-=$0& \\
			\hline \end{tabular}
			\begin{tabular}{|r|c|}\hline
			$N_{j_-j_+}^{(2,1)}$ &$2j_+=$0 \\ \hline
			$2j_-=$0&  \\ 
			\hline \end{tabular}	
		\begin{tabular}{|r|ccc|}\hline
			$N_{j_-j_+}^{(2,2)}$ &$2j_+=$0&1&2 \\ \hline
			$2j_-=$0& & 133 &\\ 
			1 &&& 1 \\ 
			\hline \end{tabular}
		\begin{tabular}{|r|ccc|}\hline
			$N_{j_-j_+}^{(2,3)}$ &$2j_+=$0&1&2 \\ \hline
			$2j_-=$0& & 1936 &\\ 
			1 &&& 112 \\ 
			\hline \end{tabular}			
		\begin{tabular}{|r|ccccc|}\hline
			$N_{j_-j_+}^{(2,4)}$ &$2j_+=$0&1&2 & 3&4\\ \hline
			$2j_-=$0& & 15607 &&1&\\ 
			1 &134&& 2210 &&\\ 
			 2&&1&&137&\\
			 3&&&&&1\\  
			\hline \end{tabular}		
		\begin{tabular}{|r|ccccc|}\hline
			$N_{j_-j_+}^{(2,5)}$ &$2j_+=$0&1&2 & 3&4\\ \hline
			$2j_-=$0& & 93200 &&112&\\ 
			1 &2048&& 19328 &&\\ 
			 2&&112&&2160&\\
			 3&&&&&112\\  
			\hline \end{tabular}		
			
			}
			\caption{Refined BPS invariants $N_{j_-,j_+}^{(b,d)}$ of $M_2'$, for base degree $b\leq2$ and fiber degree $d\leq 5$.}
	\label{tab:M2pref}
\end{table}

\begin{table}[h!]
	\centering
	{\tiny
		\begin{tabular}{|r|c|}\hline
			$N_{j_-j_+}^{(1,0)}$ &$2j_+=$0\\ \hline
			$2j_-=$0&3\\ 
			\hline \end{tabular}
			\begin{tabular}{|r|c|}\hline
			$N_{j_-j_+}^{(1,1)}$ &$2j_+=$0\\ \hline
			$2j_-=$0&27\\ 
			\hline \end{tabular}
			\begin{tabular}{|r|c|}\hline
			$N_{j_-j_+}^{(1,2)}$ &$2j_+=$0\\ \hline
			$2j_-=$0&81\\ 
			\hline \end{tabular}
		\begin{tabular}{|r|cc|}\hline
			$N_{j_-j_+}^{(1,3)}$ &$2j_+=$0&1\\ \hline
			$2j_-=$0&243&\\ 
			1 &&3\\ 
			\hline \end{tabular}
		\begin{tabular}{|r|cc|}\hline
			$N_{j_-j_+}^{(1,4)}$ &$2j_+=$0&1\\ \hline
			$2j_-=$0&594&\\ 
			1 &&27\\ 
			\hline \end{tabular}
				\begin{tabular}{|r|cc|}\hline
			$N_{j_-j_+}^{(1,5)}$ &$2j_+=$0&1\\ \hline
			$2j_-=$0&1377&\\ 
			1 &&81\\ 
			\hline \end{tabular}
		\begin{tabular}{|r|c|}\hline
			$N_{j_-j_+}^{(2,0)}$ &$2j_+=$0 \\ \hline
			$2j_-=$0&  \\
			\hline \end{tabular}
		\begin{tabular}{|r|cc|}\hline
			$N_{j_-j_+}^{(2,1)}$ &$2j_+=$0 & 1\\ \hline
			$2j_-=$0& & \\ 
			\hline \end{tabular}	
		\begin{tabular}{|r|ccc|}\hline
			$N_{j_-j_+}^{(2,2)}$ &$2j_+=$0 & 1&2\\ \hline
			$2j_-=$0&&& 27\\ 
			\hline \end{tabular}
		\\
		\begin{tabular}{|r|ccc|}\hline
			$N_{j_-j_+}^{(2,3)}$ &$2j_+=$0&1&2 \\ \hline
			$2j_-=$0& & 237 &\\ 
			1 &&& 3 \\ 
			\hline \end{tabular}
		\begin{tabular}{|r|ccc|}\hline
			$N_{j_-j_+}^{(2,4)}$ &$2j_+=$0&1&2 \\ \hline
			$2j_-=$0& & 1296 &\\ 
			1 & 9 & &81  \\
			\hline \end{tabular}
		\begin{tabular}{|r|cccc|}\hline
			$N_{j_-j_+}^{(2,5)}$ &$2j_+=$0&1&2 & 3\\ \hline
			$2j_-=$0& & 5400 &&\\ 
			1 &27&& 621 &\\ 
			 2&&&&27\\  
			\hline \end{tabular}

			}
			\caption{Refined BPS invariants $N_{j_-,j_+}^{(b,d)}$ of $M_3'$, for base degree $b\leq2$ and fiber degree $d\leq 5$.}
\end{table}

\begin{table}[h!]
	\centering
	{\tiny
		\begin{tabular}{|r|c|}\hline
			$N_{j_-j_+}^{(1,0)}$ &$2j_+=$0\\ \hline
			$2j_-=$0&4\\ 
			\hline \end{tabular}
		\begin{tabular}{|r|c|}\hline
			$N_{j_-j_+}^{(1,1)}$ &$2j_+=$0\\ \hline
			$2j_-=$0&16\\ 
			\hline \end{tabular}
		\begin{tabular}{|r|c|}\hline
			$N_{j_-j_+}^{(1,2)}$ &$2j_+=$0\\ \hline
			$2j_-=$0&40\\ 
			\hline \end{tabular}
		\begin{tabular}{|r|c|}\hline
			$N_{j_-j_+}^{(1,3)}$ &$2j_+=$0\\ \hline
			$2j_-=$0&96\\ 
			\hline \end{tabular}
		\begin{tabular}{|r|cc|}\hline
			$N_{j_-j_+}^{(1,4)}$ &$2j_+=$0&1\\ \hline
			$2j_-=$0&204&\\ 
			1 &&4\\ 
			\hline \end{tabular}
		\begin{tabular}{|r|cc|}\hline
			$N_{j_-j_+}^{(1,5)}$ &$2j_+=$0&1\\ \hline
			$2j_-=$0&400&\\ 
			1 &&16\\ 
			\hline \end{tabular}
		\begin{tabular}{|r|c|}\hline
			$N_{j_-j_+}^{(2,0)}$ &$2j_+=$0 \\ \hline
			$2j_-=$0&  \\
			\hline \end{tabular}
		\begin{tabular}{|r|c|}\hline
			$N_{j_-j_+}^{(2,1)}$ &$2j_+=$0  \\ \hline
			$2j_-=$0& \\ 
			\hline \end{tabular}	
		\begin{tabular}{|r|cc|}\hline
			$N_{j_-j_+}^{(2,2)}$ &$2j_+=$0 & 1\\ \hline
			$2j_-=$0&& 10\\ 
			\hline \end{tabular}
		\begin{tabular}{|r|cc|}\hline
			$N_{j_-j_+}^{(2,3)}$ &$2j_+=$0 & 1\\ \hline
			$2j_-=$0&& 64\\ 
			\hline \end{tabular}
		\begin{tabular}{|r|ccc|}\hline
			$N_{j_-j_+}^{(2,4)}$ &$2j_+=$0&1&2 \\ \hline
			$2j_-=$0& & 286 &\\ 
			1 &&& 6 \\ 
			\hline \end{tabular}
		\begin{tabular}{|r|ccc|}\hline
			$N_{j_-j_+}^{(2,4)}$ &$2j_+=$0&1&2 \\ \hline
			$2j_-=$0& & 960 &\\ 
			1 &&& 64 \\ 
			\hline \end{tabular}
			}
			\caption{Refined BPS invariants $N_{j_-,j_+}^{(b,d)}$ of $M_4'$, for base degree $b\leq2$ and fiber degree $d\leq 5$.}
\end{table}

\newpage
\addcontentsline{toc}{section}{References}
\bibliographystyle{utphys}
\bibliography{names}
\end{document}